\documentclass[aps,prd,twocolumn,groupedaddress,asmsymb,amsmath,amssymb,longbibliography]{revtex4-1}

\usepackage{graphicx}
\usepackage{calligra}
\usepackage{upgreek}
\usepackage{color}
\usepackage{inputenc,caption}
\usepackage{stackrel}
\usepackage[usenames,dvipsnames]{xcolor}
\usepackage{subcaption}

\numberwithin{equation}{section}

\DeclareMathAlphabet{\mathpzc}{OT1}{pzc}{m}{it}

\newcommand{\be}{\begin{equation}}
\newcommand{\ee}{\end{equation}}
\newcommand{\bea}{\begin{eqnarray}}
\newcommand{\eea}{\end{eqnarray}}
\newcommand{\lb}{\label}

\newcommand{\bF}{{\bf f}}

\newcommand{\bv}{{\bf v}}
\newcommand{\bu}{{\bf u}}

\newcommand{\bk}{{\bf k}}
\newcommand{\bp}{{\bf p}}
\newcommand{\bq}{{\bf q}}
\newcommand{\bx}{{\bf x}}

\newcommand{\br}{{\bf r}}

\newcommand{\bB}{{\bf B}}
\newcommand{\bI}{{\bf I}}

\newcommand{\bV}{{\bf V}}

\newcommand{\boeta}{{\mbox{\boldmath $\eta$}}}

\newcommand{\btau}{{\mbox{\boldmath $\tau$}}}

\newcommand{\grad}{{\mbox{\boldmath $\nabla$}}}
\newcommand{\bdot}{{\mbox{\boldmath $\cdot$}}}

\newcommand{\bzed}{{\mbox{\boldmath $0$}}}

\newcommand{\ttau}{\widetilde{\tau}}

\newcommand{\bttau}{\widetilde{{\mbox{\boldmath $\tau$}}}}

\newcommand{\hvline}{{$\!$\textbf{-}$\!$\textbf{-}$\!$\textbf{-}$\!$\textbf{-}$\!$\textbf{-}}}

\definecolor{forestgreen}{rgb}{0.13, 0.55, 0.13}
\definecolor{bondiblue}{rgb}{0.0, 0.58, 0.71}
\definecolor{paleblue}{rgb}{0.4, 0.7, 1}
\definecolor{palered}{rgb}{1, 0.48, 0.25}

\newcommand{\red}[1]{\textcolor{black}{#1}}
\newcommand{\black}[1]{\textcolor{black}{#1}}

\begin{document}

%\preprint{}

\title{Dissipation-Range Fluid Turbulence and Thermal Noise}
%and Dissipative Anomalies}


\author{Gregory Eyink}
%{\bf DRAFT VERSION - DO NOT DISTRIBUTE}}
\email[]{eyink@jhu.edu}
%\thanks{}
\affiliation{Department of Applied Mathematics \& Statistics, The Johns Hopkins University, Baltimore, MD, USA, 21218}
\altaffiliation{Department of Physics \& Astronomy, The Johns Hopkins University, Baltimore, MD, USA, 21218}
\author{Dmytro Bandak, Nigel Goldenfeld}
\affiliation{Department of Physics, University of Illinois at Urbana-Champaign, Urbana, IL, USA, 61801} 
\author{Alexei A. Mailybaev}
\affiliation{Instituto de Matem\'atica Pura e Aplicada -- IMPA, Rio de Janeiro, Brazil 
22460--320} 


%\date{\today}

\begin{abstract}
We revisit the issue of whether thermal fluctuations are relevant for incompressible fluid turbulence, 
and estimate the scale at which they become important.  As anticipated by Betchov in a prescient 
series of works more than six decades ago, this scale is about equal to the 
Kolmogorov length, even though that is several orders of magnitude above the mean free path.  
This result implies that the deterministic version of the incompressible Navier-Stokes equation is 
inadequate to describe the dissipation range of turbulence in molecular fluids.  Within this range, 
the fluctuating hydrodynamics equation of Landau and Lifschitz is more appropriate. In particular, 
our analysis implies that both the exponentially decaying energy spectrum and the far-dissipation 
range intermittency predicted by Kraichnan for deterministic Navier-Stokes will be generally 
replaced by Gaussian thermal equipartition at scales just below the Kolmogorov length. 
Stochastic shell model simulations at high Reynolds numbers verify our theoretical predictions and 
reveal furthermore that inertial-range intermittency can propagate deep into the dissipation range, 
leading to large fluctuations in the equipartition length scale. We explain the failure of previous scaling 
arguments for the validity of deterministic Navier-Stokes equations at any Reynolds number and we 
provide a mathematical interpretation and physical justification of the fluctuating Navier-Stokes equation 
as an ``effective field-theory'' valid below some high-wavenumber cutoff $\Lambda$,  rather than 
as a continuum stochastic partial differential equation. At Reynolds number around a million, comparable to that 
in Earth's atmospheric boundary layer, the strongest turbulent excitations observed in our simulation 
penetrate down to a length-scale of about eight microns, still two orders of magnitude greater than 
the mean-free-path of air. However, for longer observation times or for higher Reynolds numbers, 
more extreme turbulent events could lead to a local breakdown of fluctuating hydrodynamics. 
\end{abstract}

\pacs{?????}
%\keywords{}

\maketitle

% Put \label in argument of \section for cross-references
%\section{}
%\subsection{}

\section{Introduction}\label{sec:I}

The incompressible Navier-Stokes equation for the fluid velocity field $\bu(\bx,t)$ 
as a function of space $\bx$ and time $t$: 
\be \partial_t\bu + (\bu\cdot\grad)\bu = -\grad p + \nu \triangle \bu, \quad \grad\bdot\bu=0. 
\lb{NS} \ee
since their original introduction \cite{navier1823memoire},
have been accepted for more than 100 years as the mathematical model 
of turbulence in molecular fluids at low Mach numbers and arbitrarily high Reynolds-numbers.  
These equations have been used as the starting point for statistical 
theories of turbulence, such as that of Reynolds \cite{reynolds1894dynamical}. 
In particular, these equations were invoked by Kolmogorov in his celebrated 
1941 theory of turbulence (K41), which postulated a universal scaling behavior 
in the dissipation range of turbulent flow \cite{kolmogorov1941local} and yielded
the exact ``4/5th-law'' \cite{kolmogorov1941dissipation}. 
There are rigorous derivations of the Navier-Stokes equations at any fixed Reynolds number, no matter 
how large, in the limit of small Mach number and small Knudsen number, starting from the 
Boltzmann equation for low-density gases \cite{bardos1991fluid,bardos1993fluid} 
and from stochastic lattice-gas models with no restriction on density \cite{quastel1998lattice}.  
In fact, by a well-known argument of Corrsin using the K41 turbulence theory 
\cite{corrsin1959outline} (see also \cite{frisch1995turbulence}, section 7.5), 
the hydrodynamic approximation which underlies the incompressible Navier-Stokes 
equation becomes increasingly better the higher the Reynolds number, because the Knudsen 
number decreases as an inverse power of Reynolds number.  
Mathematically, Leray \cite{leray1934mouvement,leray2016motion}
has shown that dissipative weak solutions of incompressible 
Navier-Stokes equation exist globally in time for any initial data of locally finite energy
and also that these solutions remain smooth and unique locally in time for smooth initial data. 
The global  smoothness remains an open question \cite{fefferman2006existence} 
and singularities might possibly appear in finite time for 
certain initial data at sufficiently high Reynolds number, as conjectured by Leray 
himself \cite{leray1934mouvement,leray2016motion}. However, the existing rigorous derivations 
\cite{bardos1991fluid,bardos1993fluid,quastel1998lattice} show that, even if singularities 
occur so that Leray solutions are non-unique, nevertheless the empirical velocity field will satisfy 
{\it some} Leray solution of incompressible Navier-Stokes equations. The physical and 
mathematical foundations for basing a theory of turbulence on these equations seems 
thus quite secure. 

There are effects intrinsic to molecular fluids that are omitted, however, by the 
incompressible Navier-Stokes equations, most importantly thermal fluctuations
\cite{dezarate2006hydrodynamic,schmitz1988fluctuations}.  These effects are described 
instead by an extension  of the usual deterministic fluid equations known as 
{\it fluctuating hydrodynamics}, originally due to Landau and Lifschitz \cite{landau1959fluid},
Ch. XVII. For an incompressible fluid satisfying $\grad\bdot\bu=0$ these equations have 
the form \cite{forster1976long,forster1977large,usabiaga2012staggered,donev2014low,nonaka2015low} 
\black{\footnote{\black{To prevent confusion, we note that the fluctuating hydrodynamics equations
are essentially different, both physically and mathematically, from the Navier-Stokes equation 
with a spatially-smooth white-in-time force added to represent large-scale stirring
\cite{edwards1964statistical,novikov1965functionals,edwards1969statistical}.  
When the forcing spectrum decays rapidly at high wavenumbers, then the random stirring 
in the latter case inputs energy which cascades inertially to small scales. Mathematically,
one can then safely take any UV cutoff $\Lambda$ to infinity with fixed viscosity $\nu$
and obtain a well-defined limit defined by a stochastic partial differential equation (SPDE) 
\cite{albeverio2008spde,vishik1988mathematical}. On the contrary, the 
fluctuating hydrodynamic equations have stochastic forcing prescribed by the 
fluctuation-dissipation relation, so that the statistical steady-state  is thermal 
equilibrium with time-reversible dynamics. See Appendix \ref{appA}.
Because the forcing spectrum is now {\it growing} with wavenumber, it is not 
at all obvious how to define a limiting SPDE as $\Lambda\to\infty$ \cite{albeverio2008spde}. 
In fact, the bare viscosity is now renormalized by the thermal fluctuations and becomes 
a scale-dependent quantity $\nu_\Lambda$ \cite{forster1976long,forster1977large}.
These important differences are discussed at greater length in the text.}}}\\

\vspace{-30pt} 
\be \partial_t\bu + (\bu\cdot\grad)\bu = -\grad p + \nu \triangle \bu + 
\grad\cdot \bttau \lb{FNS} \ee
with $\ttau_{ij}(\bx,t)$ a fluctuating stress prescribed as a Gaussian random field with 
mean zero and covariance 
\begin{eqnarray}
\langle \ttau_{ij}(\bx,t) \ttau_{kl}(\bx',t')\rangle& = &\frac{2\nu k_BT}{\rho}
\left(\delta_{ik}\delta_{jl}+\delta_{il}\delta_{jk} -\frac{2}{3}\delta_{ij}\delta_{kl}\right)\cr
&& \hspace{30pt} 
\times \delta^3(\bx-\bx')\delta(t-t') \lb{FDR} \end{eqnarray}  
whose realizations are symmetric and traceless, and with Boltzmann's constant
$k_B\doteq 1.38\times 10^{-23}$ $m^2\, kg/sec^2\, K.$ These equations can be phenomenologically 
derived by requiring that the equations with the added stochastic terms have the Gibbs measure
\be P_G[\bu]= \frac{1}{Z} \exp\left(-\frac{\rho}{2k_BT } \int_\Omega d^3x\, |\bu(\bx)|^2 \right) \lb{Gibbs} 
\ee 
as an invariant measure, in which case \eqref{FDR} is known as the {\it fluctuation-dissipation 
relation}. It is very difficult to give the equation \eqref{FNS},\eqref{FDR}, as written, precise 
mathematical meaning as a stochastic partial differential equation and to show that it has \eqref{Gibbs} 
as an invariant measure (e.g. see \cite{albeverio2008spde}). One approach to make mathematical sense 
of this equation is as an equivalent Onsager-Machlup action or large-deviations 
rate function \cite{graham1981onset,eyink1990dissipation}, and in this form 
it has been rigorously derived for a stochastic lattice gas \cite{quastel1998lattice}. 
However, the common approach of statistical physicists is to regard \eqref{FNS},\eqref{FDR} as an effective, 
low-wavenumber field theory that should be truncated at some wave-number cutoff $\Lambda$ 
which is larger than an inverse gradient-length $\ell_\nabla^{-1}$ of the fluid but smaller than an 
inverse \red{microscopic length $\lambda_{micr}^{-1}$ (in a gas, the inverse mean-free-path
$\lambda_{mfp}^{-1}$)}.  In fact, there are formal physical derivations of
fluctuating hydrodynamics in this sense for compressible fluids starting from molecular 
dynamics \cite{zubarev1983statistical,morozov1984langevin,espanol2009microscopic} and 
taking the low Mach-number limit yields precisely \eqref{FNS},\eqref{FDR}. Incorporating such 
a cut-off $\Lambda,$ the fluctuating hydrodynamic equations become well-defined stochastic ODE's 
for a finite number of Fourier modes, and the corresponding Fourier truncated 
measure \eqref{Gibbs} is easily checked to be time-invariant. See Appendix \ref{appA}. \\

\vspace{-12pt} 
\textcolor{black}{There have been, however, only a relatively few studies of the effects of thermal fluctuations 
on turbulent flows and most of these have focused on the role of weak noise in selecting a unique 
invariant measure for deterministic Navier-Stokes \cite{hosokawa1976ensemble,ruelle1979microscopic,
machavcek1988role}.}
%, however, other than a pioneering work of Ruelle \cite{ruelle1979microscopic} on predictability of turbulence. 
The presence of a new dimensional parameter in the 
fluctuating hydrodynamics equations \eqref{FNS}-\eqref{FDR}, the thermal energy $k_BT,$ 
vitiates the similarity analysis of Kolmogorov \cite{kolmogorov1941local,kolmogorov1941dissipation},
who postulated that the only relevant dimensional parameters in the dissipation 
range of a turbulent flow are the kinematic viscosity $\nu$ and the mean energy dissipation-rate 
per mass $\varepsilon.$ This violation of Kolmogorov's 1941 analysis is in addition to the
defect that he later noted himself \cite{kolmogorov1962refinement}, which is that space-time
intermittency introduces dependence upon the outer length-scale $L$ of the flow. See 
\cite{frisch1995turbulence} for an extensive review. The interplay between these two additional 
dimensional parameters, $L$ and $k_BT,$ is one of the major issues addressed in this work. 
For high Reynolds-number turbulent flows this interplay raises new questions 
regarding the precise formulation of fluctuating hydrodynamics, even within the 
``effective field-theory" point of view. Arguing from K41 theory, the gradient 
length $\ell_\nabla,$ or largest length below which the velocity field 
is smooth, should be the Kolmogorov scale $\eta=\nu^{3/4}\varepsilon^{-1/4}.$
However, space-time intermittency makes the dissipative cut-off fluctuate 
\cite{paladin1987degrees} (or \cite{frisch1995turbulence}, section 8.5.5) 
so that at some points $\ell_\nabla\ll \eta.$ In that case,
does a cut-off length $\Lambda^{-1}$ exist which satisfies the necessary conditions 
$\ell_\nabla\gg\Lambda^{-1}\gg \lambda_{micr}$? If so, how should $\Lambda$ be chosen in practice?
And, most importantly, what significant effects, if any, does thermal noise have on the 
dynamics and statistics of incompressible turbulence? \\

\vspace{-12pt} 
We develop answers
to all of these questions in this work, and, in particular, argue that thermal noise has 
profound observable consequences for turbulence. 
In a following work we shall discuss 
the inertial range of scales, but here we deal with the dissipation range at scales smaller 
than the Kolmogorov length $\eta.$ This range has been the subject of much 
theoretical and rigorous mathematical work within the framework of the incompressible 
Navier-Stokes equation for decades, for example, on the rate of decay of the spectrum 
\cite{heisenberg1985statistischen,chandrasekhar1956theory,kraichnan1959structure,
frisch1981intermittency,foias1990empirical,sirovich1994energy} and 
on intermittency in the dissipation range
\cite{kraichnan1967intermittency,frisch1981intermittency,frisch1991prediction}. 
There have also been intensive recent efforts to study the dissipation range \red{energy spectrum}  
by direct numerical simulations (DNS) of the incompressible Navier-Stokes equations
\cite{khurshid2018energy,gorbunova2020analysis,buaria2020dissipation}.  \red{This is part 
of a larger program to determine the most extreme events and most singular, smallest-scale 
structures in a turbulent flow, at lengths far below the Kolmogorov scale 
\cite{yeung2015extreme,yeung2020advancing, farazmand2017variational,buaria2019extreme,
buaria2020self,buaria2020vortex,nguyen2020characterizing}. The underlying science question which  
drives this work is whether the hydrodynamic equations can remain valid during such extreme turbulent 
events or whether strong singularities can lead to breakdown of the macroscopic, hydrodynamic description.}
We shall argue that much of this prior theory and simulation work is \red{called into 
question for} turbulence in real molecular fluids \red{and may} require 
substantial modifications, because effects of thermal noise become significant 
already at length-scales scales right around the Kolmogorov scale. 
\red{Laboratory experiments are now attempting to probe these small length-scales 
\cite{debue2018experimental,gorbunova2020analysis,debue2021three} but, as we shall discuss 
at length, all current experimental methods lack both the space-time resolution and the sensitivity 
to measure turbulent velocity fields accurately at sub-Kolmogorov scales. We regard this 
state of affairs as a crisis in turbulence research, which calls for the development of completely 
novel experimental techniques.}

\black{After the work in this paper was completed, we became aware of a series of 
remarkable papers published by Robert Betchov starting in the late 1950's \cite{betchov1957fine,
betchov1961thermal,betchov1964measure}, which anticipated several of our key 
conclusions. Betchov not only recognized the significant effects that thermal 
noise could have on dissipation-range statistics and other phenomena 
in fluid turbulence, such as transition and predictability, but he also developed the 
framework of fluctuating hydrodynamics for incompressible fluids 
\cite{betchov1961thermal}, independent of Landau and Lifschitz \cite{landau1959fluid}. 
Betchov carried out a novel experimental investigation with a multi-jet flow created 
by a perforated box \cite{betchov1957fine}, designed to lower as much as possible the  
space resolution scale of extant hot-wire methods, in order to test his ideas.  
\black{Unfortunately, despite improving upon the resolution and especially the accuracy 
of prior experiments by a couple of orders of magnitude, Betchov's experiments 
nevertheless lacked the sensitivity required to verify his predicted results.}
Because he employed linearized equations for his theoretical analysis,  Betchov's 
predictions mainly regarded 2nd-order statistics, such as energy spectra and 
one-dimensional dissipation, but he studied also experimentally the velocity-derivative 
skewness and kurtosis. Our analysis goes well beyond that of Betchov, taking 
full account of the nonlinearity of the fluid equations of motion and associated 
phenomena such as inertial-range intermittency, which were unappreciated in 
his day. However, Betchov's pioneering work should be more widely known and many 
of his ideas are still highly relevant today. We shall therefore compare his 
conclusions with our own results, which serve to confirm and extend his early insights. 
\black{In the conclusion section \ref{sec:conclusion} of our paper we shall briefly review 
Betchov's experiments and place them in the context of current efforts}.  
}

We shall proceed in this paper by developing simple theoretical arguments which 
are then tested numerically in a reduced dynamical model of turbulence, the Sabra 
shell model \cite{lvov998improved,lvov1999hamiltonian}. \black{Our numerical results 
for this model provide, to our knowledge, the first empirical confirmation of Betchov's 
essential predictions anywhere in the literature. Furthermore,} based upon these simulations, 
we shall then formulate more refined theoretical predictions for the dissipation-range 
of real fluid turbulence. A short report of our most essential physical 
predictions has been submitted elsewhere \cite{bandak2021thermal}, but we provide 
here full details of our numerical study and, furthermore, address the quite subtle 
and complex issues surrounding the hydrodynamic description of turbulent flows.

\textcolor{black}{The detailed contents of our paper are as follows: In Section 
\ref{sec:FNS} we discuss the incompressible fluctuating hydrodynamics model \eqref{FNS-Lambda} 
and its physical and mathematical foundations for describing turbulent 
fluid flow. This includes the formulation of the basic equations (\ref{formulation})  
and an extended dimensional analysis of turbulence taking into account thermal effects
(\ref{DA}). In particular, we discuss how thermal noise breaks the scaling 
symmetry of deterministic incompressible Navier-Stokes \eqref{NS} and why standard arguments 
on its validity for molecular fluids thus fail in the dissipation range of turbulent flows 
(\ref{scale}). We make also a preliminary evaluation of the effects of inertial-range intermittency 
by a phenomenological multifractal approach, in order to assess possible limitations to 
a hydrodynamic description of high-Reynolds turbulence (\ref{intermittent:sec}). To test 
these theoretical ideas we develop in Section \ref{shell:sec} a stochastic Sabra shell model
of fluctuating hydrodyamics. We justify carefully the use of this ``zero-dimensional'' model despite  
some significant differences from fluctuating hydrodynamics in three space dimensions (\ref{shell-intro}), 
and we discuss its meaning and numerical solution as an effective theory for low wavenumbers (\ref{sec:num}).  
The main Section \ref{num:sec} of our paper presents numerical results on thermal noise effects 
in turbulence, obtained by simulating the shell model, which confirm our theoretical predictions 
and motivate additional conjectures for fluid turbulence. After describing the set-up of our turbulence simulations 
(\ref{setup}), we present results on thermal effects in the statistics of modal energies (\ref{sec:spectra}), 
a comparison of dissipation-range intermittency for the deterministic and stochastic model (\ref{diss-range}), 
and a similar comparison of shell-model structure function scaling in the inertial-range as well as the 
dissipation-range (\ref{sec:strfun}). Finally, Section \ref{sec:conclusion} summarizes our conclusions 
and outlines directions for future work.}

\section{Fluctuating Hydrodynamics of Turbulent Flow}\lb{sec:FNS}

\subsection{Formulation of the Equations}\lb{formulation}  

In order to keep things simple, we take as flow domain a periodic box or $3-$torus,
$\Omega=L{\mathbb T}^3,$ with volume $V=|\Omega|=L^3.$
Then, to discuss steady-state turbulent flow, we shall modify the equation \eqref{FNS}
in two ways. First, we shall employ the standard theoretical artifice of adding 
to the local momentum balance an external body force $\rho\bF$ to drive the 
turbulent flow, assuming that this force is supported in Fourier space at 
very low wavenumbers $\sim 1/L.$  Second, and very essentially, we shall assume that the velocity 
field is comprised of Fourier modes with wavenumbers $|\bk|<\Lambda$ only, a condition 
which may expressed in terms of a projection operator $P_\Lambda$: 
$$ \bu(\bx)=P_\Lambda\bu(\bx) :=\frac{1}{V}\sum_{|\bk|<\Lambda} e^{i\bk\bdot\bx} \hat{\bu}_\bk, $$ 
and then the fluctuating hydrodynamic equation \eqref{FNS} is modified to
\begin{eqnarray}
&& \partial_t\bu + P_\Lambda(\bu\cdot\grad)\bu = \cr
&& \hspace{30pt} -\grad p + \nu \triangle \bu + 
\left(\frac{2\nu k_BT}{\rho}\right)^{1/2} \grad\cdot \boeta_\Lambda +\bF, \cr
&& \hspace{60pt} \grad\bdot\bu=0.
\lb{FNS-Lambda} \end{eqnarray}
Separating out the covariance of the thermal noise facilitates our scaling 
analysis below.
Thus $\boeta_\Lambda(\bx,t)$ is a tensor spacetime Gaussian field with mean zero and 
covariance 
\begin{eqnarray}
\langle \eta_{\Lambda ij}(\bx,t) \eta_{\Lambda kl}(\bx',t')\rangle& = &
\left(\delta_{ik}\delta_{jl}+\delta_{il}\delta_{jk} -\frac{2}{3}\delta_{ij}\delta_{kl}\right)\cr
&& \hspace{15pt} 
\times \delta^3_\Lambda(\bx-\bx')\delta(t-t'), \lb{etacov} \end{eqnarray}  
for
\be \delta^3_\Lambda(\bx-\bx')=P_\Lambda\delta^3(\bx-\bx') 
= \frac{1}{V}\sum_{|\bk|<\Lambda} e^{i\bk\bdot\bx}. \ee 
The kinematic pressure $p$ in \eqref{FNS-Lambda} is determined by the requirement that $\bu$
be solenoidal and obviously satisfies the condition $p=P_\Lambda p.$ In contrast 
to the original equation \eqref{FNS} with no UV cut-off $\Lambda,$ which is mathematically 
ill-defined {\it a priori}, the equation \eqref{FNS-Lambda} is equivalent to a system of It$\bar{{\rm o}}$ 
stochastic differential equations for the Fourier modes $\hat{\bu}(\bk)$ of the velocity 
field and its solutions are stochastically well-posed. For example, see the lectures 
of Flandoli \cite{flandoli2008introduction}.  As usual in discussions of fluctuating hydrodynamics 
\cite{forster1976long,forster1977large,zubarev1983statistical,morozov1984langevin,espanol2009microscopic,
usabiaga2012staggered,donev2014low}, it is assumed that $\Lambda$ can be chosen to satisfy 
\red{
$1/\ell_\nabla\ll \Lambda\ll 1/\lambda_{micr}.$ For a liquid, $\lambda_{micr}$ is the mean interparticle 
distance $\ell_{intp}\equiv n^{-1/3}$ defined in terms of particle number density $n.$ The condition 
$\Lambda\ell_{intp}\ll 1$ is a minimal requirement that coarse-graining cells of size $1/\Lambda$
should contain many molecules. For a low-density gas, the mean-free-path length $\lambda_{mfp}\gg \ell_{intp}$ 
and the condition $\Lambda\lambda_{mfp}\ll 1$ guarantees that terms higher than 2nd-order in gradients 
can be neglected. We thus take $\lambda_{micr}=\max\{\ell_{intp},\lambda_{mfp}\}.$}
For a 
high Reynolds-number turbulent flow where intermittency effects may cause the gradient length $\ell_\nabla$ 
to be much smaller than the traditional Kolmogorov length $\eta,$ it is not trivial that 
the crucial condition $\ell_\nabla\gg \lambda_{micr}$ should be satisfied. Even if so, one must 
determine how to choose $\Lambda$ in this range so that predictions are independent of the 
choice. %We shall address all of these issues below.

The latter problem is highly non-trivial and currently lacks an analytical solution, even at the physical 
level of rigor \footnote{This problem is solved in principle by the Green-Kubo formulas 
for the transport coefficients obtained in the microscopic derivations of the fluctuating 
hydrodynamic equations, such as Eq.(2.56) of \cite{zubarev1983statistical} or 
Eq.(48) of \cite{espanol2009microscopic}. However, evaluation of those formulas itself 
requires microscopic molecular dynamics simulations.}. 
It is expected that the ``bare viscosity'' must be chosen to have a 
cut-off-dependent form $\nu_\Lambda$ so that predictions of the model are independent of $\Lambda.$ 
This is due physically to the fact that the effective viscosity is renormalized by 
hydrodynamic fluctuations from eliminated modes at wavenumbers $>\Lambda.$  A renormalization group 
analysis of the fluid in the thermal equilibrium state \eqref{Gibbs} shows that the dynamics becomes 
asymptotically free in the infrared for space dimensions $d\geq 2$ and described by a linear 
Langevin dynamics (the Onsager regression hypothesis for long-wavelength, low-frequency velocity 
fluctuations)\cite{forster1976long,forster1977large}. This result can be easily understood 
by estimating the r.m.s. velocity $u_\ell$ at length scale $\ell$ in an equilibrium fluid with 
temperature $T$ and mass density $\rho$ by the central limit theorem (cf. \cite{ruelle1979microscopic}) 
\red{as
%\be u_\ell \sim \left(\frac{k_BT}{\rho \ell^d}\right)^{1/2} \lb{vel-CLT}\ee 
$u_\ell \sim \left(k_BT/\rho \ell^d\right)^{1/2}.$  It follows that the ``thermal Reynolds numbers'' 
of such equilibrium velocity fluctuations at length-scale $\ell$
\be Re^{th}_\ell := \frac{\ell u^{th}_\ell}{\nu}=\left(\frac{k_BT}{\rho\nu^2\ell^{d-2}}\right)^{1/2}  \lb{Re-th} \ee
is small at sufficiently large lengths $\ell$ in space dimensions $d>2,$which implies that the nonlinear 
coupling is weak.  In principle, the nonlinear coupling becomes large at sufficiently small distances 
for $d>2,$ but we shall argue later that this length-scale is so small that it lies outside the regime of 
validity of a hydrodynamic description.}  
%Therefore there is no exact analytical theory which 
%prescribes how to choose the bare viscosity (and possibly additional parameters) so that 
%predictions of the model are $\Lambda$-independent. 
Current numerical practice in fluctuating 
hydrodynamics \cite{donev2014low} evaluates wavenumber-dependent viscosity $\nu(k)$
by fitting molecular dynamics results for velocity-velocity correlations  
at low frequency and at wavenumber $k$ to the Lorentzian form predicted by nonlinear fluctuating 
hydrodynamics (e.g. see eq.(3.37) in \cite{forster1977large}). \red{The viscosity of a 
hard-disk fluid with $1/k$ near scales of order the mean-free-path is found to be well-predicted 
by the Enskog kinetic theory for dense fluids and scale-dependence  due to renormalization by thermal fluctuations 
is weak. See \cite{donev2014low}, Fig. 8.} 

We remark finally that setting $\nu=0$ in \eqref{FNS-Lambda} recovers the truncated Euler dynamics, 
which has been previously much studied. This ideal system was shown to satisfy a Liouville 
theorem by Burgers \cite{burgers1933application} and much later independently by 
Lee \cite{lee1952some}. Quadratic invariants of the ideal incompressible Euler equations 
such as kinetic energy and helicity are also conserved in detail for 
individual wavevector triads, as noted by Onsager \cite{onsager1949statistical}
and Kraichnan \cite{kraichnan1959structure}, so that these remain invariants 
of the truncated Euler system. It follows that a Gibbs measure of the form 
\eqref{Gibbs} is an invariant measure  of the truncated Euler system, as 
observed by both Lee \cite{lee1952some} and Hopf \cite{hopf1952statistical}.
Furthermore, restoring a positive viscosity $\nu>0,$ the fluctuating
hydrodynamics equation preserves that Gibbsian invariant measure, as 
shown in detail in Appendix \ref{appA}. It is worth remarking that the 
truncated Euler dynamics possesses other Gibbs-type invariant measures associated 
to fixed values of helicity in addition to kinetic energy \cite{kraichnan1973helical}, 
but these measures are no longer invariant in the presence of thermal noise.  

\subsection{Dimensional Analysis}\lb{DA} 

For our study of the turbulent dissipation range, it is appropriate to 
non-dimensionalize the equation \eqref{FNS-Lambda} with Kolmogorov dissipation
length and velocity scales 
\be \eta=\nu^{3/4}\varepsilon^{-1/4}, \quad u_\eta=(\varepsilon\nu)^{1/4}, \ee 
by setting  
\begin{eqnarray} 
&& \hat{\bu}=\bu/u_\eta, \quad \hat{\bx}=\bx/\eta, \quad \hat{t} = (u_\eta/\eta)t, \cr
&& \hat{p}=p/u_\eta^2,\quad \hat{\bF}=\bF/F, \quad \hat{\Lambda}=\Lambda\eta 
\lb{K-non-dim} \end{eqnarray} 
where $F$ is the typical magnitude of the force per mass (e.g. an r.m.s. value).
The equations of motion  in dimensionless form become 
\be \partial_{\hat{t}}\hat{\bu} + P_{\hat{\Lambda}}(\hat{\bu}\cdot\hat{\grad})\hat{\bu} = -\hat{\grad}{\hat p} + 
\hat{\triangle} \hat{\bu} + (2\theta_\eta)^{1/2} 
\hat{\grad}\cdot \hat{\boeta}_{\hat{\Lambda}} + \digamma_\eta\hat{\bF}\ee
with dimensionless temperature and force magnitude: 
\be \theta_\eta=\frac{k_BT}{\rho u_\eta^2\eta^3}, \quad \digamma_\eta=\frac{F\eta}{u_\eta^2} 
\lb{K-groups} \ee
 However, 
it is more natural to non-dimensionalize the large-scale force with inertial-range units of 
length $L$ and velocity $U=(\varepsilon L)^{1/3}$ \footnote{There is no assumption made here 
that the r.m.s. velocoty $u_{rms}\sim U$ and their ratio could be $Re$-dependent}, so that 
$\digamma_\eta=\digamma_L/Re^{1/4}$ with 
\be \digamma_L=\frac{FL}{U^2}=\frac{F}{(\varepsilon^2/L)^{1/3}},
\quad Re=\frac{UL}{\nu}=\frac{\varepsilon^{1/3}L^{4/3}}{\nu} \lb{L-groups} \ee
the dimensionless force magnitude at the integral scale and the Reynolds number, respectively. 
Scaled in this manner and with hats omitted, the fluctuating hydrodynamic equation 
\eqref{FNS-Lambda} becomes 
\be \partial_t\bu + P_\Lambda(\bu\cdot\grad)\bu = -\grad p + \triangle \bu + \left(2\theta_\eta\right)^{1/2} 
\grad\cdot \boeta_\Lambda + \frac{\digamma_L}{Re^{1/4}} \bF. \ee
The Reynolds-dependence of the final term just reflects the expectation that the 
direct effect of large-scale forcing will be negligible in the dissipation range for $Re\gg 1.$

The thermal noise term is also expected to be small at the Kolmogorov scale. The crucial
parameter $\theta_\eta$ is the ratio of the thermal energy to the energy of the Kolmogorov-scale 
velocity fluctuations $u_\eta$ in a spatial region of diameter $\sim \eta.$ 
It is interesting to consider concrete numbers corresponding to typical values of physical constants 
for the turbulent atmospheric boundary layer (ABL), taken from the monograph \cite{garratt1994atmospheric}
\begin{eqnarray} 
&&\varepsilon=400 \ cm^2/sec^3, \quad \nu = 0.15\ cm^2/sec, \cr 
&&\rho=1.2 \times 10^{-3}\ g/cm^3, \quad T=300^\circ K, \lb{ABL-param} 
\end{eqnarray} 
which gives
\be \eta=0.54\ mm, \quad u_\eta=2.78\ cm/sec, \quad \theta_\eta=2.83 \times 10^{-8} 
\lb{ABL-K} \ee 
The very small value of $\theta_\eta$ arises from the small value of Boltzmann's constant in cgs units,
$k_B=1.38\times 10^{-16} \ erg/K.$ 

Although this number is very small, it however rises rapidly at length-scales 
$\ell<\eta.$ It can be estimated very crudely by assuming an exponential decay 
for $\ell<\eta,$ so that the fluid velocity fluctuation level becomes
\be u_\ell \sim  u_\eta \exp(-\eta/\ell). \ee 
Thus,
\be \theta_\ell =\frac{k_BT}{\rho u^2_\ell \ell^3} \sim  \theta_\eta 
\left(\frac{\eta}{\ell}\right)^3 \exp(\eta/\ell) \lb{theta-eff} \ee
Because $\theta_\ell$ increases so rapidly for decreasing $\ell,$ one can see that $\theta_\ell\sim 1$ 
already for $\ell_{eq}\sim \eta/11$. In the ABL, this length is of the order of $49\ \mu m.$ For comparison, 
the mean-free path length of air at room temperature and standard atmospheric pressure is $\lambda_{mfp}=68\ nm$. 
Thus, thermal noise becomes of the same order as nonlinear terms already at sufficiently large scales where 
a hydrodynamic description remains valid. Similar estimates hold for other natural turbulent flows, 
such as the upper ocean mixing layer, and in laboratory experiments performed with a variety of fluids. 
For example, in the water experiment of Debue et al. \cite{debue2018experimental} 
at \black{$Re=3\times 10^5$ and} $20^\circ$C, the Kolmogorov scale is 
\black{$\eta=0.016\ mm$} and $\theta_\eta=2.5\times 10^{-7},$ 
so that thermal fluctuations become relevant around length-scale 
\black{$\ell_{eq}\sim\eta/11=1.5\ \mu m$} which is still much larger than the mean-free-path length of water $\lambda_{mfp}=0.25\ nm.$ These cases 
are typical. We therefore argue that theories of the ``far dissipation range'' of turbulence which omit 
thermal noise are \red{of questionable relevance} to molecular fluids in Nature. 

The example of the ABL and others are reassuring that a cut-off $\Lambda$ should exist
satisfying the fundamental requirement $\ell_\nabla\gg \Lambda^{-1}\gg \ell_{micr}$
for validity of fluctuating hydrodynamics. Since the nonlinearity 
at length scales $\ell\lesssim \eta$ is weak compared with the thermal noise and 
the viscous damping, one expects that the velocity fluctuations at those scales 
should reach thermal equilibrium with a Maxwell-Boltzmann distribution and
an equipartition energy spectrum 
\be E(k)\sim \frac{k_BT}{\rho} \frac{4\pi k^2}{(2\pi)^3}. \lb{eq-spectrum} \ee
See Appendix \ref{appA}. Since this spectrum is growing in $k,$ it must always exceed 
the spectrum of the turbulent velocity fluctuations at sufficiently high wavenumber. 
The physical origin of these large velocity fluctuations is the high speeds of 
the constituent molecules of the fluid, which are of the order of magnitude of the sound speed
$c_{th},$ or about 343 $m/sec$ in air and 1481 $m/sec$ in water. While these large velocities 
almost completely cancel in macroscopic spatial averages by the law of large numbers, the central 
limit theorem fluctuations grow $\sim \ell^{-d/2}$ in space dimension $d$ as the resolved 
length-scale $\ell$ decreases toward molecular scales. 
The crossover wavenumber to see thermal effects can again be crudely estimated by equating the 
dissipation-range turbulence spectrum with the thermal spectrum 
\be u_\eta^2\eta \exp(-k\eta) \sim \frac{k_BT}{\rho} k^2 \lb{E-balance} \ee 
which implies
%\be 
$\theta_\eta (k\eta)^2\exp(k\eta)\sim 1$ %\ee
and yields an estimate of the crossover length $\ell_{eq}\sim 1/k_{eq}\sim \eta/12$ consistent with that found earlier. 

At much larger wavenumbers than this, the velocity fluctuations  
should be close to Gaussian, with statistical independence of modes instantaneously 
in non-overlapping wavenumber bands. We therefore propose that the UV truncation wavenumber $\Lambda$ of the 
fluctuating hydrodynamic equation \eqref{FNS-Lambda} may be chosen anywhere 
in the range of $k$ where the equipartition spectrum is achieved and where 
\red{the high-pass filtered velocity $\bu^{>k}(\bx,t)$ is a Gaussian random field}. 
Assuming that such a range of wavenumbers exists, we expect that the cut-off $\Lambda$
may be selected arbitrarily in this range and the predictions of the model 
will be insensitive to the particular choice, as long as the bare viscosity 
$\nu_\Lambda$ is chosen appropriately. No analytical prescription 
currently exists for this choice but the prescription will be the 
same as that for the fluid in thermal equilibrium with the given 
temperature $T,$ mass density $\rho$ and cutoff wavenumber $\Lambda.$
Thus, it should suffice to choose $\nu_\Lambda$ by matching with
equilibrium velocity-velocity correlations from molecular dynamics simulations, 
as in current practice \cite{donev2014low}. %In agreement with our earlier arguments, 
\red{Note that} the equipartition range which we predict slightly beyond the 
Kolmogorov wavenumber will be in the ``low-wavenumber'' weak-coupling 
range of the thermal equilibrium fluid. One can verify this by substituting 
the definition \eqref{K-groups} of $\theta_\eta$ \red{into \eqref{Re-th} 
for $Re^{th}_\eta,$ giving 
\be Re^{th}_\eta=\theta_\eta^{1/2}. \lb{k-coup-K} \ee
We see that for $\theta_\eta$ with realistic values, $Re^{th}_\eta \ll1$}
and thus the nonlinear coupling in the thermal equipartition range 
should remain negligible for several decades of wavenumber above $1/\eta.$ 

\begin{figure}[]
 \begin{center}
 \includegraphics[width=264pt]{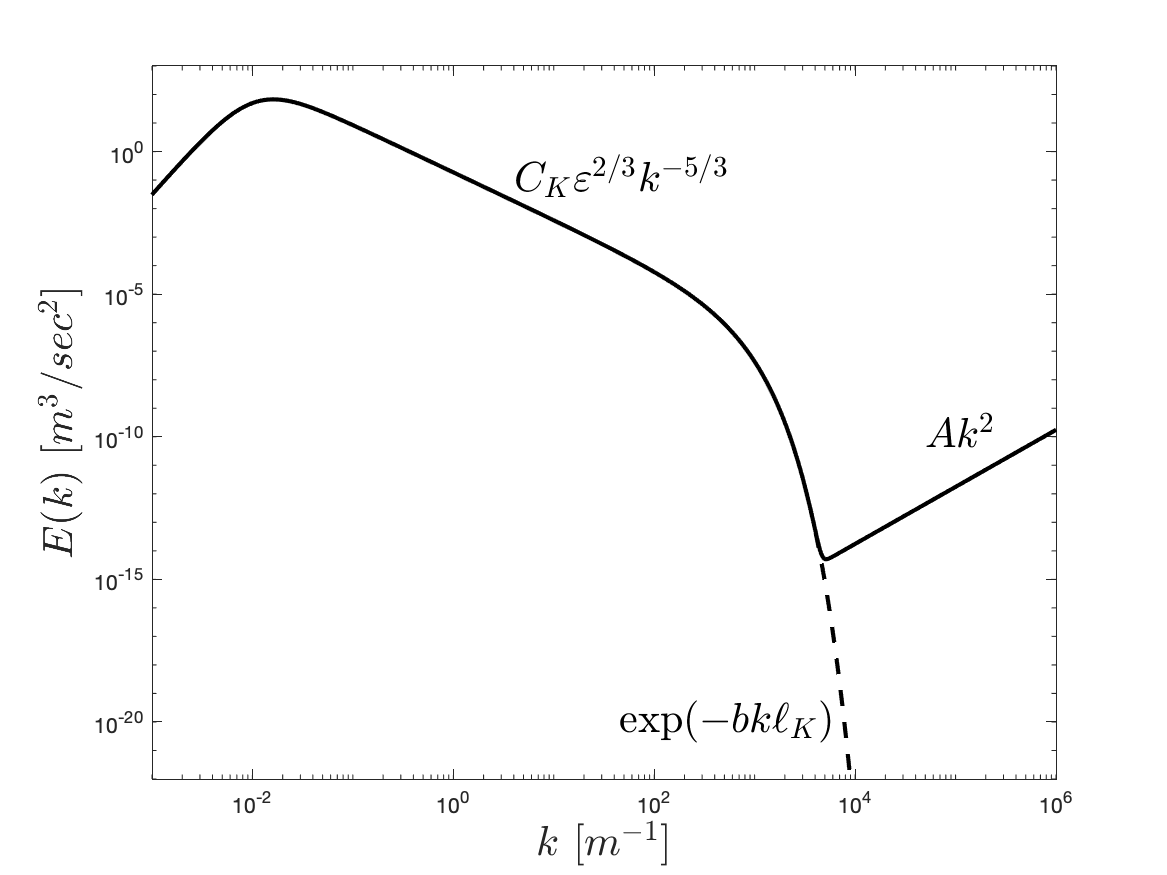}
 \end{center}
  \caption{Plot of the model turbulent energy spectrum \eqref{karman} 
  for the parameters \eqref{ABL-param} of the atmospheric boundary 
  layer and for a typical Reynolds number $Re=10^7$, as a solid line.
  The dashed line is the spectrum with no thermal noise.}\label{spectrum.fig}
\label{spectrum} \end{figure} 

The predicted energy spectrum which emerges from our arguments is illustrated 
in Figure \ref{spectrum.fig} with a model spectrum proposed by von K\'arm\'an 
\cite{vonkarman1948progress} 
\be E(k)=C_K (\varepsilon L)^{2/3}\frac{L^5k^4}{(1+(kL)^2)^{17/6}}\exp(-b k\eta)
+ Ak^2 \lb{karman} \ee 
which has been supplemented with an exponential factor $\exp(-b k\eta)$ to represent 
decay in the traditional ``far-dissipation range'' and with also an additive contribution 
from the thermal equipartition spectrum \eqref{eq-spectrum} for $A=\frac{k_BT}{\rho} 
\frac{4\pi}{(2\pi)^3}.$ In the plot we have used the 
parameters \eqref{ABL-param} and also a typical Reynolds number $Re=10^7$ for the ABL 
taken from  \cite{garratt1994atmospheric}. The exponential-decay factor is consistent
with asymptotic predictions \cite{kraichnan1959structure,sirovich1994energy}
and rigorous upper bounds \cite{frisch1981intermittency,foias1990empirical} for deterministic Navier-Stokes dynamics, with the coefficient $b=7$ 
chosen consistent with numerical observations \cite{khurshid2018energy} 
and with a conventional value $C_K=1.6$ of the Kolmogorov constant \cite{donzis2010bottleneck}. Plotted as well with a dashed line 
is the energy spectrum without thermal noise, obtained by formally setting $T=0$
in \eqref{karman}. Consistent with our earlier estimates, the two spectra agree
up to a wavenumber $k_{eq}\doteq 10/\eta$ where the equipartition spectrum begins 
to dominate. Below $k_{eq}$ lies only about half a decade of wavenumbers where the 
traditional exponential decay is manifested. The $k^2$ energy spectrum that appears 
at $k_{eq}$ will extend at least up to wavenumbers $k$ where the rms velocity $u_{1/k}$ 
approaches the sound speed $c_{th}.$ The incompressibilty assumption 
for the model \eqref{FNS-Lambda} then breaks down and 
density fluctuations become significant, so that fluctuating compressible 
equations \cite{zubarev1983statistical,morozov1984langevin,espanol2009microscopic}
must be employed. 

\black{We may observe at this point that our picture of the turbulent energy 
spectrum was almost entirely anticipated by Betchov in his first paper on 
the subject \cite{betchov1957fine}. Assuming that the hydrodynamic modes
at small scales would reach energy equipartition, he arrived at an expression for 
the thermal spectrum $E_{{\rm noise}}(k)$ identical to our Eq.\eqref{eq-spectrum}, except for an 
extra overall factor of $3/2.$ As previously noted by Hosokawa \cite{hosokawa1976ensemble},
Betchov did not take into account incompressibility and thus counted 3 degrees of freedom
for each independent wavenumber mode rather than 2. By an argument essentially identical 
to ours, Betchov then determined the wavenumber we call $k_{eq}$ by matching 
$E_{{\rm noise}}(k)$ with the turbulent energy spectrum $E(k)$ in the dissipation range. 
For the latter, he assumed a power-law form $E(k)\sim k^{-n},$ with $n=7$ predicted 
by the theory of Heisenberg \cite{heisenberg1948statistischen} but with $n=6$ more consistent 
with Betchov's own experimental results. Although he did not state a quantitative estimate
of $k_{eq},$ Betchov plotted his theoretical result for $E_{{\rm noise}}(k)$ together 
with his experimental data for $E(k)$ and, extrapolating the latter, they can be observed 
to cross at a wavenumber around 10 $\mu$m. In the final sentence of
his paper, Betchov concluded that ``the gap between turbulence and molecular agitation 
may not be as wide as is generally appreciated.'' In his later work \cite{betchov1964measure},
he arrived at the more quantitative estimate that $k_{eq}\eta\doteq 1$ by assuming that 
the turbulent spectrum drops essentially to zero for $k\eta>1.$ His schematic plot 
of the atmospheric spectrum, entirely analogous to our Fig.\ref{spectrum}, indicated 
an equilibration scale there of order 1 mm. Betchov ended the paper \cite{betchov1964measure}
by noting that, because $\eta\gg \lambda_{mfp},$ his conclusions differ from those of von Neumann
\cite{vonneumann1963recent}, who had argued that thermal noise becomes relevant 
only for length scales of order the mean free path and that separation of scales 
is thus ``unambiguous'' in turbulent flows.} 

\black{It should be emphasized that these conclusions are very robust and do not 
depend upon any particular model of the energy spectrum in the turbulent 
dissipation-range, as long as that spectrum decays rapidly in wave-number.
We made our estimate of the equilibration scale using the model spectrum 
\eqref{karman} with exponential decay, but Betchov \cite{betchov1957fine}
reached the same conclusion assuming a rapid power-law decay. In the 
Supplemental Materials we show similarly that the ``intermediate dissipation range''
predicted by Frisch \& Vergassola \cite{frisch1991prediction} using multifractal phenomenological models
\cite{kolmogorov1962refinement,she1994universal,yakhot2001mean}
implies also that the equipartition scale is only about an order of magnitude 
below the Kolmogorov scale. The stretched exponential decay predicted for 
the dissipation-range spectrum by functional renormalization group arguments 
\cite{tarpin2018breaking} will likewise lead to the same result. Of course, all of 
these arguments are phenomenological and the intuitive superposition of spectra
for turbulent and thermal fluctuations depends upon the hypothesis that small scales 
will achieve the same equilibrium distribution in a turbulent flow as in a laminar one. 
This hypothesis is supported by the standard presumption of weak turbulent 
fluctuations in the dissipation range, but it must be tested empirically and could 
even be false due to rare, intermittent bursts of turbulence that penetrate to very 
small scales \cite{paladin1987degrees}.}

The physical arguments \black{of Betchov and ourselves} 
which lead to our proposed picture of the turbulent energy 
spectrum will be corroborated in section \ref{shell:sec} by shell-model simulations.
Furthermore, the direct effects of thermal noise will be found to exist in those 
simulations at length-scales much larger than $1/k_{eq}$ in more refined statistical 
measures, such as negative-order structure functions or ``inverse structure functions'' 
\cite{jensen1999multiscaling}, which are more sensitive than the energy spectrum 
to rare low-amplitude events. Equipartition spectra similar to those predicted 
here have been observed in superfluid turbulence via numerical simulations 
%of Biot-Savart vortex filament models \cite{kivotides2006coherent},
%of the mesoscale HBVK equations \cite{salort2011mesoscale} and 
of the Gross-Pitaevskii equation \cite{shukla2019quantitative}, where they correspond to 
a thermal bath of phonons at high wave-numbers created by the forward energy cascade. 
Since superfluids have strictly zero viscosity, there is no exact analogue in these 
simulations of a dissipation range and equipartition spectra there join directly 
with the Kolmogorov cascade spectra. The same is true for equipartition spectra 
observed in numerical simulations of decaying turbulence for the truncated 
Euler system \cite{kraichnan1975remarks,cichowlas2005effective} where the $k^2$ spectrum corresponds to 
thermalized wave-number modes near the spectral cutoff $\Lambda$ of the model. 
In contrast to the fluctuating hydrodynamic equations \eqref{FNS}, which model 
molecular fluids at mesoscopic scales, the truncated Euler system does not correspond 
directly to any physical system in nature or in the laboratory. \red{For a more detailed 
comparison of  truncated Euler and related systems with our results in this paper, 
see section \ref{object}. 
} 

\subsection{Violation of Scale-Symmetry of Navier-Stokes}\lb{scale}  

A possible objection can be raised to our claim that the ``far dissipation range''
of deterministic Navier-Stokes turbulence is physically unachievable, based upon 
the well-known space-time scaling symmetry (\cite{frisch1995turbulence}, section 2.2;
\cite{majda2002vorticity}, section 1.2): 
\be \bu\to \bu'=\lambda\bu,\quad \bx\to\bx'=\lambda^{-1} \bx,
\quad t\to t'=\lambda^{-2}t \lb{INS-scale} \ee 
which maps an incompressible Navier-Stokes solution $\bu(\bx,t)$ to another solution
$\bu'(\bx',t')$ with the same Reynolds number $Re'=Re$
and with molecular viscosity $\nu$ also unchanged. This scale 
symmetry is equivalent to the familiar principle of hydrodynamic 
similarity. In the presence of an external body force $\bF$, one must also take
\be \bF\to \bF'=\lambda^3\bF. \lb{F-scale} \ee 
This symmetry of incompressible Navier-Stokes equation is the basis of its 
rigorous derivation from the Boltzmann equation \cite{bardos1991fluid,bardos1993fluid} 
or from lattice gas models \cite{quastel1998lattice} through a scaling limit with $\lambda\to 0$. 
Based on such mathematical treatments, one might conclude that the deterministic 
Navier-Stokes equation should hold in a turbulent flow to any desired accuracy 
by simply taking the outer length of the flow large enough and the r.m.s. velocity
small enough, at whatever Reynolds number desired. 

If, however, a solution of the fluctuating Navier-Stokes equation \eqref{FNS} in 3D 
is subjected to the same rescaling \eqref{INS-scale}, then 
\be \frac{k_B T}{\rho} \to \frac{k_B T'}{\rho'}=\lambda^{3}
\frac{k_B T}{\rho}. \lb{noise-scale} \ee 
Thus, thermal noise breaks the scaling symmetry of the deterministic  
Navier-Stokes equation, unless temperature can be decreased as  
$T\to T'=\lambda^a T$ and density increased as $\rho\to\rho'=\lambda^{-b}\rho$, 
so that thermal noise satisfies \eqref{noise-scale} with kinematic viscosity 
held fixed $\nu(\rho',T')=\nu(\rho,T).$ (For example, for an ideal 
gas $\nu\propto T^{1/2}/\rho$ and one must take $a=2,$ $b=1$). 
The presumed relation \eqref{noise-scale} is the reason that derivation of the 
incompressible Navier-Stokes equation through a hydrodynamic scaling limit 
corresponds to weak noise and leads to a large-deviations framework \cite{quastel1998lattice}. 
However, even this extended scaling symmetry of the fluctuating equation 
\eqref{FNS} is physically limited, since at low enough 
temperature and/or high enough density the fluid will undergo a phase transition 
to a binary gas-liquid mixture or to a solid state. If one instead keeps the temperature 
and density fixed, then the thermal noise does {\it not} become weaker and remains given by the 
fluctuation-dissipation relation \eqref{FDR} for both the solutions $\bu$ and $\bu',$ 
which are no longer statistically similar to each other. This violation of scale-symmetry is 
generally not observed in fluids because Boltzmann's constant $k_B$ is so small in macroscopic 
units (c.g.s. or m.k.s.). However, the breaking of hydrodynamic similarity 
due to thermal noise can be observed in microfluidics, for example in the 
mixing at fluid interfaces \cite{kadau2007importance}. 

Specializing these general considerations for the fluctuating hydrodynamic 
equation \eqref{FNS} to the case of forced turbulent flow with time-average power 
input per mass $\varepsilon=\langle \bu\bdot\bF\rangle,$ the scalings \eqref{INS-scale},
\eqref{F-scale} imply that 
\be \varepsilon \to\varepsilon'=\lambda^4\varepsilon. \lb{eps-scale} \ee 
In decaying turbulence, the same result holds by Taylor's relation $\varepsilon\sim U^3/L$
\cite{taylor1935statistical}. The result \eqref{eps-scale} then implies that 
the Kolmogorov velocity and length transform as 
\begin{eqnarray}
u_\eta&=&(\nu\varepsilon)^{1/4}\to u_\eta'=\lambda u_\eta, \cr 
\eta&=&\nu^{3/4}\varepsilon^{-1/4}  \to \eta'=\lambda^{-1}\eta 
\end{eqnarray} 
consistent with the scaling \eqref{INS-scale}. It is easy to check that 
along with the Reynolds number $Re,$ also the dimensionless force amplitudes 
$\digamma_\eta$ and $\digamma_L$ defined in \eqref{K-groups},\eqref{L-groups} 
are invariant under the rescaling \eqref{INS-scale}, but that 
\be \theta_\eta\to \theta_\eta'=\lambda \theta_\eta \lb{theta-scale} \ee 
when keeping $\rho$ and $k_BT$ fixed. This result holds because 
the kinetic energy of Kolmogorov-scale turbulent fluid eddies 
increases as $\lambda^{-1}.$ It is thus possible, in principle,
to observe deterministic Navier-Stokes predictions in the dissipation range 
by taking $\bu\to\bu'=\lambda \bu$ and $\bx\to\bx'=\lambda^{-1}\bx$
with $\lambda\ll 1.$ However, in practice, $\lambda$ must be chosen 
exponentially small, since the relations \eqref{theta-eff},\eqref{E-balance} 
show that deterministic Navier-Stokes predictions for the far-dissipation 
range will hold only up to a wavenumber $k\eta\sim \ln(1/\theta_\eta')$ 
growing as $\ln(1/\lambda).$ For example, in the case of the ABL we 
argued below \eqref{theta-eff} that noise would be relevant already at 
a length-scale $\ell\sim \eta/11.$ To make the deterministic predictions 
valid down to $\ell\sim \eta/22$ would require that the integral length be 
made $4e^{11}=240,000$ times larger and r.m.s. velocities 240,000 times weaker
\footnote{If relation $x^2e^x=1/\theta_\eta$ is solved for $x=k\eta,$
$\theta_\eta,$ then satisfying it for $x'=2x,$ $\theta'_\eta$ requires $1/\lambda=\theta_\eta/\theta_\eta'
=4e^x$}! 

Based on these considerations, we argue that an extended far-dissipation range  
spectrum described by deterministic Navier-Stokes will be practically 
unobservable both in natural flows and in laboratory experiments. The best hope 
of achieving a sizable exponential-decay range is probably with a highly viscous 
fluid at relatively low Reynolds numbers, so that both $\eta$ and $u_\eta$ are 
made as large as possible. In such a moderate Reynolds number turbulent flow, 
the dimensionless noise parameter $\theta_\eta$ will be as small as possible. This 
strategy should work best in liquids where viscosity can be increased by 
lowering temperature, whereas in gases large kinematic viscosity requires 
either high temperatures or low densities. In any case, there will be a fundamental 
difficulty in then, say, doubling the wavenumber extent of such a far-dissipation 
range spectrum, because this would require a further exponential decrease in $\theta_\eta,$ 
which will be unachievable.    

\subsection{Intermittency and Validity of Fluctuating Hydrodynamics}\lb{intermittent:sec}  

The previous considerations have ignored the phenomenon of small-scale 
turbulence intermittency, which leads to enhanced fluctuations 
in the inertial-range that may be described phenomenologically by
the Parisi-Frisch multifractal model \cite{frisch1985singularity,frisch1995turbulence}. 
This model postulates a dimension spectrum $D(h)$ of velocity H\"older exponents $h$ and 
suggests the existence of viscous cutoff lengths 
\be \eta_h\sim L Re^{-1/(1+h)} \lb{PVlength} \ee  depending 
upon $h$ which can be much smaller than the classical Kolmogorov length $\eta$, which corresponds 
to $h=1/3$ \cite{paladin1987degrees}. This model has been invoked to predict an ``intermediate 
dissipation range'' (IDR) of scales \cite{frisch1991prediction}, as an intermediate asymptotics at high-$Re$ 
between the inertial range where viscosity effects are negligible and the far-dissipation range 
where viscosity dominates to produce exponential decay of velocity fluctations.
The far-dissipation range is itself predicted to suffer extremely large fluctuations,
because the intermittency at lower wavenumbers is intensely magnified by the exponential 
drop-off in the spectrum \cite{kraichnan1967intermittency}. We should therefore 
discuss how such intermittency might modify the conclusions in the previous section 
about the choice of $\Lambda$ and indeed about the validity of a fluctuating hydrodynamic 
description of turbulent flow. 

We shall exploit here the Parisi-Frisch multifractal model to address these questions. 
We note that the multifractal concept of an IDR predicts a collapse of spectra at different 
$Re$ with a certain scaling \cite{frisch1991prediction}, and this ``multiscaling'' has been 
verified in the GOY shell model \cite{bowman2006links} and to some degree in direct numerical 
simulations (DNS) of the Navier-Stokes equations; see \cite{khurshid2018energy}, Fig.4. 
On the other hand, there are alternative theoretical ideas about the ``near-dissipation range'' 
of turbulent flows, for example, from functional renormalization group (FRG) analyses 
\cite{canet2017spatiotemporal,tarpin2018breaking}. Some recent experimental and DNS
studies have given stronger support to the FRG predictions than to the multifractal IDR prediction 
\cite{debue2018experimental,gorbunova2020analysis,buaria2020dissipation}. 
We therefore invoke the multifractal model only as a heuristic to draw qualitative 
conclusions. We note, however, that several of the results we obtain below can be confirmed 
independently by rigorous mathematical arguments. 

The considerations of Corrsin \cite{corrsin1959outline} on the validity of a hydrodynamic 
description within the K41 theory of turbulence can be easily extended to the multifractal 
model; see \cite{eyink2007turbulence}, section III(e). The essential point of Corrsin's 
analysis is that viscosity $\nu$ and microscopic length $\lambda_{micr}$ are not independent 
quantities, but are instead linked by the standard estimate from kinetic theory 
\be \nu \sim \lambda_{micr}c_{th} \lb{kinest} \ee
where $c_{th}$ is the thermal velocity/sound speed. Combining \eqref{kinest} with 
\eqref{PVlength} then easily yields
\be Kn_h:=\lambda_{micr}/\eta_h \sim Ma \, Re^{-h/(1+h)} \ee 
as an estimate of the ``local Knudsen number'' at a point with H\"older 
exponent $h,$ where $Ma=U/c_{th}$ is the Mach number based on the large-scale  
flow velocity $U.$ We see that the scale separation required for validity 
of a hydrodynamic description, as quantified by the fundamental condition $Kn_h\ll 1,$ will become 
progressively better with increasing $Re$ for $h>0$ and $Kn_h\ll 1$ will hold marginally 
even for $h=0$ if $Ma\ll 1.$ 

There is some evidence that the smallest H\"older exponent in incompressible 
fluid turbulence is $h_{\min}=0$ \cite{iyer2020scaling} (although this conclusion 
requires the assumption that negative $h$ cannot occur, e.g. \cite{frisch1995turbulence},
section 8.3, which can be called into question; see section \ref{validity}). If so, then 
there is a range of possible wavenumber cutoffs $\Lambda$ satisfying 
\be 1/\eta_h\ll \Lambda\ll 1/\lambda_{micr}, \quad \mbox{for all $h,$} \lb{cutcrit} \ee 
as long as $Ma\ll 1.$ 
Taking into account the thermal noise, we can expect for each $h$ that its effects 
will be manifested in the length scales just slightly below $\eta_h,$ where 
the local energy spectrum (defined e.g. by a wavelet transform) experiences 
exponential drop-off. In that case, we may further expect that Gaussian, thermal-equilibrium 
statistics will hold locally at any length-scale $\ell\ll \eta_h$. More precisely,
we may consider a locally coarse-grained velocity 
\be \bar{\bu}_\ell(\bx) =\sum_n \bv_n G_\ell(\bx-\bx_n)\Big/\sum_n G_\ell(\bx_n) 
\lb{vel-coarse} \ee 
% =\int d^3r\ G_\ell(\br)\, \bu(\bx+\br), 
where $G_\ell(\br)=\ell^{-3} G(\br/\ell)$ is a smooth, well-localized kernel function 
and where the sum $\sum_n$ is over molecules of the fluid with positions $\bx_n$
and velocities $\bv_n.$
%and its fluctuation field 
%\be \bu_\ell'(\bx)=\bu(\bx)-\bar{\bu}_\ell(\bx) \ee 
Because the microscopic velocity distributions are close to Maxwellian and because 
of the central limit theorem \cite{ruelle1979microscopic},  we expect to observe 
nearly Gaussian statistics \footnote{The meaning of statistics here is in the sense of ``local 
thermodynamic equilibrium.'' Thus, the region of diameter $\eta_h(\bx,t)$ centered 
at $\bx$ at time $t$ may be partitioned into $(\eta_h/\ell)^3$ subregions of diameter $\ell$ and the distribution of values $\bar{\bu}_\ell$ over the subregions should 
be the given Gaussian distribution \eqref{local-eq} for $\ell_{intp} \ll \ell\ll \eta_h$}
\be P(\bar{\bu}_\ell) \propto \exp\left(-C\frac{\rho\ell^3|\bar{\bu}_\ell-\bv|^2}{k_BT}\right),
\quad \ell_{intp} \ll \ell\ll \eta_h \lb{local-eq} \ee
with $\bv=\bar{\bu}_{\eta_h}$ locally at each point in space and time, with H\"older exponent $h=h(\bx,t).$
Thus, the primary effect of turbulent intermittency should be a strong fluctuation 
in space and time of the length scale $\eta_h$ below which Gaussian thermal equipartition 
sets in. Since the ratio $\eta_h/\lambda_{micr}$ grows with increasing 
$Re,$ as we have argued earlier, it should be possible to choose a cutoff satisfying 
\eqref{cutcrit} without any restriction on $Re$ as long as $Ma\ll 1.$

Our \red{tentative conclusion} is that fluctuating Navier-Stokes 
equation in the form \eqref{FNS-Lambda} should be valid with a suitable choice 
of wavenumber cutoff $1/\eta_\eta\ll \Lambda\ll 1/\lambda_{micr}$ for incompressible turbulent 
flows in the limit $Kn\ll 1$ and $Ma\ll 1$. \red{We shall return to this important issue after 
presenting and discussing our numerical results for the shell model in the following section 
(see section \ref{validity}). Most importantly, we have argued that} this model predicts a strikingly different 
behavior than does the deterministic Navier-Stokes equation in the far-dissipation 
range of turbulent flow, at length scales somewhat smaller than the Kolmogorov scale $\eta$. 
Whereas the deterministic equations predict an exponential decay of the 
energy spectrum \cite{kraichnan1959structure,frisch1981intermittency,foias1990empirical,
sirovich1994energy} and consequent enhanced intermittency 
\cite{kraichnan1967intermittency,frisch1981intermittency},
the fluctuating equations predict that the energy spectrum about only an order 
of magnitude above $1/\eta$ should rapidly bottom out and then rise again with increasing 
wavenumber $k$ in a thermal equipartition $k^2$ spectral range with Gaussian statistics. 
 
\section{Sabra Shell Model of Fluctuating Navier-Stokes Equation}\lb{shell:sec} 

In the preceding section we have made two fundamental claims: first, that turbulence 
in low Mach-number molecular fluids is described by the incompressible FNS equations 
\eqref{FNS} with a suitable high-wavenumber cut-off $\Lambda$ and, second, that those 
equations predict a thermal equipartition range a decade or so below the Kolmogorov scale 
rather than a ``far-dissipation range" with exponentially decaying energy spectrum. To check 
the first claim will require novel experimental investigations, which will be discussed later. 
The second claim is based on physically plausible reasoning, but can be verified  
by numerical simulations of the FNS equations. Such simulations are possible using existing 
numerical schemes such as the low Mach number FNS codes in 
\cite{usabiaga2012staggered,donev2014low,nonaka2015low}, which 
employ finite-volume spatial discretizations and explicit or semi-implicit 
stochastic Runge-Kutta integrators in time. \red{Motivated by the present study, such 
computations have been carried out and the results will be discussed briefly later.} 
An alternative numerical approach would be based on the lattice Boltzmann method
with thermal fluctuations incorporated at the kinetic level \cite{gross2011langevin,xue2020lattice}.
Neither of these numerical schemes even with the most powerful current computers can, however, 
reach Reynolds numbers comparable to those in the atmospheric boundary layer or even in 
many engineering flows. \red{Such high Reynolds numbers are particularly relevant to the 
issue whether a hydrodynamic description is valid for the most extreme turbulent events 
\cite{yeung2015extreme,yeung2020advancing, farazmand2017variational,buaria2019extreme,
buaria2020self,buaria2020vortex,nguyen2020characterizing}, since inertial-range intermittency 
increases with $Re.$} We shall therefore in this paper validate our physical arguments by numerical simulations 
of a simplified ``shell model'' of turbulence which can be solved at much higher Reynolds 
numbers. \black{These models long been used as surrogates of incompressible Navier-Stokes 
equations, both for physical investigation of high-$Re$ turbulence 
\cite{gledzer1973system,ohkitani1989temporal,biferale2003shell} and for 
comparative mathematical study \cite{vincenzi2021close}.}
Below we discuss the model, introduce a suitable numerical scheme, and then 
present numerical simulation results on the dissipation-range of the model in a statistical
steady state of high-Reynolds turbulence. 

\subsection{Introduction of the Model}\lb{shell-intro} 

The stochastic model that we consider is based upon the deterministic Sabra shell model 
\cite{lvov998improved,lvov1999hamiltonian,constantin2006analytic,constantin2007regularity} 
which describes the evolution of complex shell variables $u_n$  for 
discrete wavenumbers $k_n= k_0 2^n,$ $n=0,1,2,...,N$ via the coupled set of ODE's 
\be                                     du_n/dt = B_n[u]-\nu k_n^2 u_n +f_n      \lb{dSabra}  \ee  
with 
\begin{eqnarray}
B_n[u]  &=&  i [ k_{n+1} u_{n+2} u^*_{n+1}-(1/2) k_n u_{n+1} u^*_{n-1} \cr
&& \hspace{30pt} + (1/2) k_{n-1} u_{n-1} u_{n-2} ]. \lb{BS-def} \end{eqnarray}  
Here $u_n$ represents the ``velocity'' of an eddy of size $1/k_n,$ the parameter $\nu$
is ``kinematic viscosity'', and $f_n$ is an external body-force to stir the system. 
Shell models have long been studied as convenient surrogates for the Navier-Stokes 
equation in the numerical study of high-Reynolds-number 
turbulence \cite{ditlevsen2010turbulence,biferale2003shell} but the Sabra model has been 
especially popular because it enjoys symmetries most similar to those of the incompressible fluid equations, 
roughly analogous to translation-invariance and scale-invariance. There is no ``position space'' in the shell 
model on which a translation group can act, but one can shift phases by defining 
\be  u^\phi_n(t) = e^{i\phi_n} u_n(t) \lb{transl-symm} \ee
and if the constraint
$\phi_{n-2}+\phi_{n-1}=\phi_n $
is satisfied, then $u^\phi$ is a solution of the Sabra model whenever $u$ is a solution. This result
holds as well with a deterministic force, if the latter is also transformed 
as $f^\phi_n(t)=e^{i\phi_n} f_n(t).$ This symmetry is analogous 
to the action of space-translations by length displacement ${\bf a},$ acting on Fourier modes 
of the velocity field as 
\be \hat{{\bf u}}^{\bf a}({\bf k},t) = e^{i{\bf k}\cdot{\bf a}}\hat{{\bf u}}({\bf k},t).  \ee 
Other basic symmetries of the inviscid Sabra model are invariance under
continuous {\it time scaling} indexed by a real parameter $\btau>0$
\be u_n^{(\tau)}(t)=\tau u_n(\tau t) \ee
which maps solutions over time interval $[0,T]$ to solutions over the interval 
$[0,\tau^{-1} T]$ and also under discrete {\it space scaling} indexed by natural number $L$ 
\be u_n^{(L)}(t) = 2^L u_{n+L}(t) \ee   
where $N^{(L)}=N-L$ and likewise the lowest shell index is shifted from $M=0$ to $M^{(L)}=M-L.$
These are analogous to the scale symmetries of incompressible Euler, according to which for any 
$\lambda,\tau>0$ the transformation 
\be {\bf u}^{(\lambda,\tau)}(\bx,t) = \frac{\tau}{\lambda} 
{\bf u}\left(\lambda\bx,\tau t\right) \lb{scal-symm} \ee 
maps every Euler solution $\bu$ in the space-time domain $\Omega\times [0,T]$ 
to another solution $\bu^{(\lambda,\tau)}$ in the space-time domain $\lambda^{-1}\Omega\times [0,\tau^{-1} T].$ 
See \cite{majda2002vorticity}, section 1.2. Addition of viscous damping
in both shell models and real fluids leaves only the restricted symmetry group 
with the constraint $\tau=\lambda^2$ and this remaining scaling symmetry is that which leaves 
the Reynolds number invariant.  

Here we add to the deterministic model \eqref{dSabra} also {\it thermal noise} modeled by 
random Langevin forces 
\be    du_n/dt = B_n[u]  - \nu k_n^2 u_n  +  \left(\frac{2\nu k_BT}{\varrho}\right)^{1/2} k_n \eta_n(t) + f_n,   \lb{nSabra} \ee
where the complex white-noises $\eta_n(t)$ have covariance
\be                \langle  \eta_n(t) \eta^*_{n'}(t')\rangle  = 2 \delta_{nn'} \delta(t-t'),     \lb{white} \ee
for $n=0,1,...,N.$ Note that the ``translation-invariance" symmetry \eqref{transl-symm} still 
holds in the presence of thermal noise, but the remaining viscous scaling symmetry \eqref{scal-symm}
with $\tau=\lambda^2$ is broken. This {\it noisy Sabra model} is motivated as the shell-model 
caricature of the fluctuating Navier-Stokes equation \eqref{FNS-Lambda} at absolute temperature $T$ and 
mass density $\varrho$. Note that the ``density'' $\varrho$ has units of mass, because the shell model 
is zero-dimensional, describing a scale hierarchy of turbulent eddies at a single point. As usual in 
statistical physics, we do not attempt to define this stochastic model for infinitely many shells
$N=\infty,$ but instead truncate at some finite $N$ whose choice is discussed further below. 
The resulting system of stochastic ODE's is then well-posed globally in time (see Example 3.3
in \cite{flandoli2008introduction}). The specific form of the noise term can be justified as that required
to make the equilibrium {\it Gibbs distribution} 
\be P_G[u]= \frac{1}{Z} \exp(-\beta {\mathcal E}[u]) \lb{GibbsS} \ee 
the unique invariant measure for zero-forcing ($f=0$), where $\beta=1/k_BT$ and 
\be {\mathcal E}[u]= \sum_{n=0}^N \frac{1}{2}\varrho |u_n|^2  \lb{energy} \ee  
is the shell-model analogue of fluid kinetic energy. \red{In fact, this measure is in detailed
balance for the dynamics or time-reversible;} see Appendix A for the proof. 
It is not hard to verify by a modification of this 
argument that the Langevin forces in (\ref{nSabra}) are the only possible white-noise terms that 
make $P_G$ invariant, a result often called the ``Einstein relation'' or ``fluctuation-dissipation 
relation'' in statistical physics. These are well-known results in the literature 
on the fluctuating Navier-Stokes equations, here simply extended 
to the Sabra shell model. 

We should note that the infinite-$N$ limit of our model \eqref{nSabra},\eqref{white} in
the unforced case with $f_n=0$ has been been previously studied in \cite{bessaih2012invariant}. 
There it was shown that the stochastic dynamical equations with $N\leq +\infty$ define a time 
evolution which is strong in the probabilistic sense (i.e. for individual realizations of the noise) 
and that the corresponding probability measure \eqref{GibbsS} with $N=+\infty$ is time-invariant.
We do not have any need to consider such an infinite-$N$ limit in our study, although mathematical 
results of this type do bear upon the $N$-independence of the statistical predictions of our model, 
which will be discussed more below. The paper \cite{bessaih2012invariant} showed also that the 
unforced, deterministic, inviscid model with $\nu=0$ makes sense in the limit $N=+\infty$ with initial data chosen from 
the Gibbs measure \eqref{GibbsS}, which is again time-invariant. This result is likewise 
not of direct physical interest. The turbulent flows studied here will be described by (weak) solutions 
of the inviscid shell model dynamics in the limit $Re\to\infty,$ $N-M\to \infty,$ as 
mathematically studied in \cite{constantin2006analytic,constantin2007regularity},
but these weak solutions will dissipate kinetic energy and will not have $P_G$ as an invariant measure. 
There are additional invariant measures of the deterministic inviscid dynamics, associated 
for example to the ``helicity'' invariant  $H=\sum_n (-1)^n k_n|u_n|^2$ 
\cite{ditlevsen2010turbulence,biferale2003shell}. However, these measures are not invariant 
in the presence of viscosity and thermal noise, and are also not of any direct physical interest. 

To study our model in the dissipation range, we non-dimensionalize variables 
in analogy to \eqref{K-non-dim}, which brings the equation \eqref{nSabra} 
to the form
\be    du_n/dt= B_n[u]  -  k_n^2 u_n
+  \left(2\theta_\eta\right)^{1/2} k_n \eta_n(t) + f_n/Re^{1/4} ,   \lb{nSabra-K} \ee
where 
\be Re=\frac{U}{\nu k_0}, \quad \theta_\eta=\frac{k_BT}{\varrho u^2_\eta} \ee
and the second parameter is the ratio of thermal energies to kinetic energies of 
Kolmogorov-scale fluid fluctuations. In these units, the shell index now ranges over 
values $n=M,...,R$, where $M=-\left\lfloor\frac{3}{4}\log_2(Re)\right\rfloor$, $R=N-M,$
with $\lfloor x\rfloor$ denoting the integer part of $x,$ and now $n=0$ corresponds 
to the Kolmogorov wavenumber.  In a physically reasonable correspondence to real fluid 
turbulence, the parameter $\theta_\eta$ should be taken extremely small, but fixed independent of $Re.$ 
In our numerical studies in this work, we shall adopt the precise value in \eqref{ABL-K} appropriate to the ABL and which has magnitude $\theta_\eta\sim 10^{-8},$ but our results do not depend qualitatively upon the particular 
choice of this parameter. 

The existence of the $R\to\infty$ limit of the model \eqref{nSabra-K}, 
demonstrated in \cite{bessaih2012invariant} for $f_n=0$ and conjectured here 
with $f_n\neq 0$, is consistent with its much more benign UV-behavior 
than that of the 3D fluctuating Navier-Stokes model \eqref{FNS-Lambda}.
These differences in the two models arise both from the strictly local-in-wavenumber 
couplings of the shell model and also from its lower dimensionality, corresponding 
to a fluid in space dimension $d=0.$ If the RG analysis of 
\cite{forster1976long,forster1977large} is carried out for the unstirred shell 
model \eqref{nSabra} in thermal equilibrium, then it is found that that the dynamics is UV 
asymptotically-free for $k_n\gg u_{th}/\nu,$ with $u_{th}:=(2k_BT/\varrho)^{1/2}.$
The general result \eqref{k-coup-K} implies that \red{$Re^{th}_\eta=\theta_\eta^{1/2}\ll 1$} 
and thus the wavenumbers $k_n\eta\gtrsim 1$ are deep in the UV asymptotic-free 
regime of the thermal Gibbs state. In other words, the modes of the model \eqref{nSabra-K}
with shell numbers $n\gtrsim 0$ have dynamics given nearly by uncoupled linear Langevin 
equations and thus variables $u_n,$ $u_{n'}$ in that range with $n\neq n'$ are 
statistically independent not only instantaneously but also very nearly independent 
at unequal times. This statistical independence will be verified to hold in our numerical 
solutions to very good accuracy. It is therefore possible to increase $R\to R+1$ while keeping 
the bare-viscosity as a fixed constant $\nu_L=\nu$ and any resultant change in the stochastic 
dynamics is unobservable (see section \ref{N-independ}). As an aside, we remark  
that, conversely, the IR dynamics of the noisy shell model \eqref{nSabra} in the unstirred, 
thermal equilibrium state at wavenumbers $k_n\ll u_{th}/\nu$ will be strongly 
coupled and the time-correlations should exhibit nontrivial scaling analogous 
to that in the $d=1$ KPZ model \cite{forster1977large,kardar1986dynamic}.

It must be emphasized that, as a consequence, our shell model \eqref{nSabra-K} will 
{\it underestimate} the effects of thermal noise at high wavenumbers. This 
is due not only to the decrease of thermal noise effects for Sabra compared with 
increase for Navier-Stokes, but also due to the faster decay of the turbulent energy 
spectrum in Navier-Stokes without noise. As to the latter, the decay of 
the energy spectrum in the far-dissipation range of deterministic Navier-Stokes 
turbulence is expected to be exponential (up to a power-law prefactor), based upon 
physical theory \cite{kraichnan1959structure,sirovich1994energy}, rigorous 
mathematical arguments \cite{frisch1981intermittency,foias1990empirical}, and 
numerical simulations \cite{khurshid2018energy}. In the shell model, on the contrary, 
physical arguments \cite{schorghofer1995viscous,mailybaev2016spontaneous}, rigorous 
bounds \cite{constantin2006analytic} and numerical simulations \cite{lvov1998universal} 
support instead a stretched exponential decay: 
\be \langle |u_n|^2\rangle \sim \exp(-c(k_n\eta)^\gamma),
\quad \gamma=\log_2\left(\frac{1+\sqrt{5}}{2}\right) \lb{stretch-decay} \ee
This slower decay in the shell-model means that deterministic nonlinear 
effects will persist to higher wavenumbers. In addition, the Gibbs measure 
\eqref{GibbsS} for the shell model corresponds to an energy spectrum 
\be E_n:=\frac{\langle |u_n|^2\rangle}{2k_n} = A/k_n \lb{eq-spectrumS} \ee 
with $A=k_BT/\varrho.$ In contrast to the spectrum \eqref{eq-spectrum} for FNS in $d=3$,
which is growing $\sim k^2$ at high-$k,$ the corresponding equipartition spectrum 
\eqref{eq-spectrumS} for the noisy shell model is decaying $\sim 1/k_n.$ 
Because of these important differences of our model from reality, any thermal effects 
that we observe in our shell-model simulations should have appreciably greater 
analogues in real fluids. 

\subsection{Numerical Integration Method}\lb{sec:num}  

We now discuss the method for numerical solution of the noisy Sabra model 
\eqref{nSabra-K} which we shall use for our study of the steady-state 
statistics in this paper.  

\subsubsection{Slaved It$\bar{{\mathit o}}$-Taylor scheme}\lb{sec:ITscheme} 

We employ here a slaved 3/2-strong-order It$\bar{{\rm o}}$-Taylor scheme 
which was proposed by Lord and Rougemont for parabolic stochastic PDE's
(see \cite{lord2004numerical}, section 6). 
Because of the extreme stiffness of the shell model dynamics, it is desirable 
to use a scheme which explicitly solves the fast viscous dynamics by an 
integrating factor. The method we employ is a close analogue for stochastic 
equations of the slaved 2nd-order Adams-Bashforth method widely used for  
numerical simulation of deterministic shell models \cite{pisarenko1993further}.
The method in \cite{lord2004numerical} is based on stochastic It$\bar{{\rm o}}$-Taylor
expansions and is a slaved version of the 3/2-strong-order method of  
Platen \& Wagner \cite{platen1982taylor}; see \cite{kloeden2013numerical}, section 10.4
for a detailed discussion. The method of Lord-Rougemont \cite{lord2004numerical} 
used a Fourier-Galerkin method for spatial discretization that kept Fourier modes 
$|n|\leq N,$ which brings it very close to the formulation of the shell models.
It must be emphasized once again however that our view of the noisy Sabra 
model as an ``effective low-wavenumber theory'', although standard in the statistical physics 
literature, is quite different from the framework of stochastic PDE's, which requires 
a limit $N\to\infty.$ This difference will inform our discussion of convergence 
issues further below. 

To state the numerical scheme explicitly, we write the noisy Sabra model
\eqref{nSabra} in the form 
\be 
du_n = a_n dt +b_n dW_n, \quad n=0,1,...,N 
\ee 
with
\be 
a_n:= B_n[u]-\nu k_n^2u_n +f_n, \quad b_n:=\left(\frac{2k_BT}{\varrho}\right)^{1/2}k_n. 
\lb{anbn-def} \ee 
The method of \cite{lord2004numerical} solves this system approximately for a 
discrete set of times $t_k,$ $k=0,1,2,...$ by the iteration 
\bea 
&& u_n(t_{k+1})= e^{-\nu k_n^2 \Delta t}\Bigg\{ 
u_n(t_k) +\Delta t [B_n(t_k,u(t_k))+f_n(t_k)]\cr
&& +\frac{1}{2}(\Delta t)^2\left(\nu k_n^2 [B_n(t_k,u(t_k))+f_n(t_k)]+\dot{f}_n(t_k)\right)\cr
&& + \frac{i}{2}(\Delta t)^2 \Big[
 k_{n+1} (a_{n+2} u^*_{n+1}+u_{n+2} a^*_{n+1}) \cr 
&& \hspace{45pt}     -(1/2) k_n (a_{n+1} u^*_{n-1}+ u_{n+1} a^*_{n-1}) \cr
&& \hspace{45pt} + (1/2) k_{n-1} (a_{n-1} u_{n-2}+u_{n-1} a_{n-2})\Big] \cr    
&& + b_n \left[(1 +\nu k_n^2 \Delta t)\Delta W_n(t_k)-\Delta Z_n(t_k)\right] \cr
&& + i\Big[
 k_{n+1} (b_{n+2} \Delta Z_{n+2}(t_k) u^*_{n+1}+u_{n+2} b_{n+1}\Delta Z^*_{n+1}(t_k)) \cr 
&& \hspace{10pt}     -(1/2) k_n (b_{n+1}\Delta Z_{n+1}(t_k) u^*_{n-1}+ u_{n+1} b_{n-1}\Delta Z^*_{n-1}(t_k)) \cr
&& \hspace{-5pt} + (1/2) k_{n-1} (b_{n-1}\Delta Z_{n-1}(t_k) u_{n-2}+u_{n-1} b_{n-2}\Delta Z_{n-2}(t_k))\Big] \Bigg\}, \cr    
&& \lb{sabra-num} \eea
for $n=0,1,..,N,$ where 
\be \Delta W_n(t_k):=\int_{t_k}^{t_{k+1}} dW_n(t)=W_n(t_{k+1})-W_n(t_k), \lb{DW-def} \ee
\be  \Delta Z_n(t_k):= \int_{t_k}^{t_{k+1}}dt\ \left[W_n(t)-W_n(t_k)\right].  \lb{DZ-def} \ee 
The derivation of this scheme is sketched briefly in Appendix \ref{derive-slaved}, for 
completeness. In a practical implementation the pairs of complex random variables 
$\Delta W_m(t_k),$ $\Delta Z_m(t_k)$ have real and imaginary parts which can be generated 
from independent $N(0,1)$ random variables by the method in \cite{kloeden2013numerical}, 
section 10.4, eq.(4.3). For all of our tests of convergence of this scheme we used the same model 
parameters employed in the long-time steady-state simulation, discussed in detail in the next 
section \ref{setup}.

The convergence proofs in the mathematical literature for this method 
and related ones \cite{lord2004numerical,becker2016exponential,kloeden2011exponential} 
employ a joint limit $\Delta t\to 0$ and $N\to \infty$ and require a 
noise spectrum rapidly converging $b_n\to 0$ as $n\to\infty.$ By contrast, the spectrum 
defined in \eqref{anbn-def} that corresponds to thermal noise has in fact diverging $b_n,$
with $b_n\to\infty$ as $n\to\infty.$ However, we do not take the limit $N\to\infty$ but 
consider instead a fixed large $N,$ which makes our model a system of stochastic ODE's, for which 
standard convergence proofs in the limit $\Delta t\to 0$ apply \cite{kloeden2013numerical}. 
Our selection criterion for $N$ is that it must lie in the range of shell-numbers $n$ where the 
statistics are Gibbsian thermal equilibrium at temperature $T,$ with energy equipartition and 
independent Gaussian distributions of the shell variables $u_n$. We shall show in section
\ref{diss-range} for fixed individual realizations of the shell model state vector $u$ which are 
chosen from the long-time, turbulent steady-state that such an equipartition range in fact appears 
for all $n\geq N_e,$ with some $N_e,$ when these data $u$ are evolved under the stochastic 
dynamics \eqref{nSabra} for a very short time $\tau_e\sim 1/k_{N_e}u_{N_e}.$ Here $N_e=N_e(u)$ depends 
upon the state vector $u$ which is selected. We shall show furthermore that, for the range 
of integration times considered (300 large-eddy turnover times), there is a maximum value 
$N_e^\star=\max_t N_e(u(t))$ over all times $t.$ 
%which is $M_e=n_K+6,$ where $n_K$ is the Kolmogorov shell-number. 
In our convergence tests below we shall select as initial data $u$ a particular realization 
$u^\star$ such that $N_e^\star=N_e(u^\star),$ which corresponds to one these most intense ``bursts'' 
which we encountered in our long numerical run. See section \ref{diss-range} where this 
state $u^\star$ is more completely described and, also, \black{Supplemental Materials}. 
We only note here that this state $u^\star$ was found to have $N_e^\star=N_e(u^\star)=n_\eta+6,$ with $n_\eta$ the 
Kolmogorov shell-number, in a simulation with shell-number cut-off $N=n_\eta+7=22.$ Thus, only the two
highest wavenumber shells remained in thermal equilibrium for this intense bursting event $u^\star$. 
This is presumably the most stringent test case for convergence of our numerical scheme, which 
should require the smallest time-step $\Delta t$ and largest truncation shell-number $N$. However,
we have tested convergence of our numerical scheme as well for other initial data $u(0)$ selected from 
the turbulent steady-state and found very similar results. 

\subsubsection{Strong Convergence in $\Delta t$}\lb{dt-converge}  

To check strong (pathwise) convergence in the time-step $\Delta t$ in the model with 
$N=22$ shells, we must compare numerical solutions with different time-steps $\Delta t$
for the same realization of the complex Brownian motions $W_n(t)$.
This requires a statistically consistent choice of the random variables 
$\Delta W_n(t_k),$ $\Delta Z_n(t_k)$ for the different step-sizes $\Delta t.$
We create such a consistent set by first constructing these pairs 
by the method of \cite{kloeden2013numerical}, section 10.4, for the 
smallest time-step $\Delta t=\delta t.$ We then construct consistent values 
for $\Delta t=2\,\delta t,$ by the combinations \\
\be \Delta W_n^{(1)}(t_k^{(1)})=\Delta W(t_{2k+1})+\Delta W(t_{2k}) \lb{DW-comb} \ee
\be \Delta Z_n^{(1)}(t_k^{(1)})=\Delta Z(t_{2k+1})+\Delta Z(t_{2k}) 
+\Delta t\, \Delta W(t_{2k})\lb{DZ-comb} \ee   
where $t_k=k(\delta t)$ and $t_k^{(1)}=k(2\delta t),$ for $k=0,1,2,...$ 
The formulas \eqref{DW-comb},\eqref{DZ-comb} follow easily from the definitions 
\eqref{DW-def},\eqref{DZ-def}. This procedure may be iterated $p$ times to produce 
consistent sets of random increments $\Delta W_n^{(p)}(t_k^{(p)}),$
$\Delta Z_n^{(p)}(t_k^{(p)})$ for any time-step of the form $\Delta t=2^p\, \delta t,$
with $t_k^{(p)}=k(2^p\,\delta t)$ for $k=0,1,2,...$.

\begin{figure}[]
 \begin{center}
\includegraphics[width=240pt]{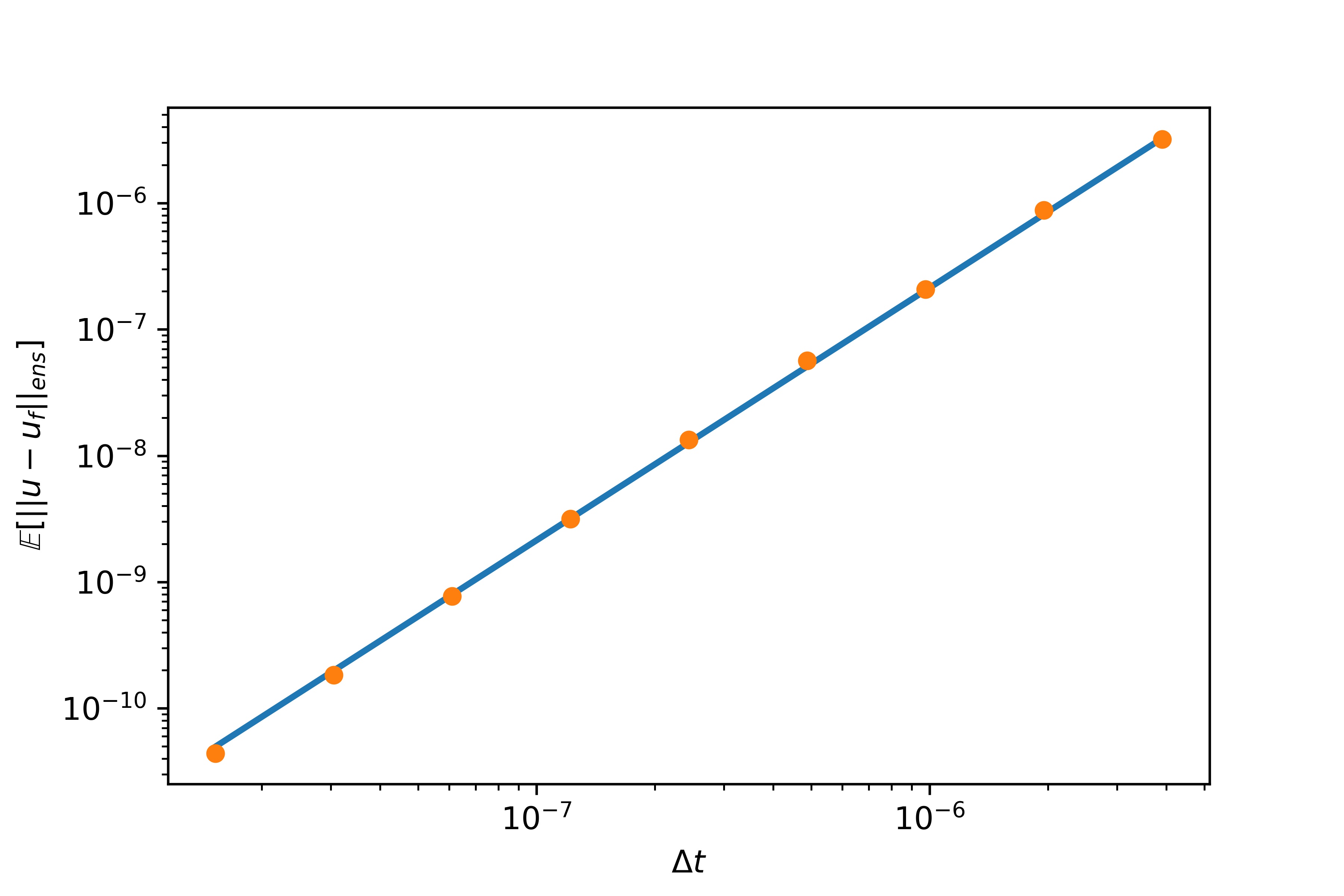}
 \end{center}
  \caption{Test of order of convergence of the numerical scheme, starting with a strong burst as the initial conditions. Solid orange dots (\textcolor{orange}{{\bf $\bullet$}}) correspond to expectation values of the enstrophy norm \eqref{ens-norm} of the error; solid cyan line ( \textcolor{bondiblue}{\hvline}) corresponds to least squares fit of a function $c \Delta t ^{2}$, where $c$ was determined to be $7.434$.}
 \label{fig-strong2}
 \end{figure}

As discussed in the previous section, we use as initial condition for our pathwise convergence
study the state $u^\star$ from the strong ``burst'' which propagated to the highest shell.  
We define error as the expectation over noise realizations ${\mathbb E}$ of the norm $\|\cdot\|$ of the 
difference $u^{\Delta t}(t) - u^{\delta t}(t)$, 
where $u^{\Delta t}(t)$ is the numerical solution starting from $u^\star$ using time-step $\Delta t$ 
and $u^{\delta t}(t)$ is the reference state obtained by integration with the finest time-step $\delta t$,  
which we take as the ``exact solution''. 
To make our convergence criterion most sensitive to the largest wavenumbers, we used the 
enstrophy norm associated to the space $h^1_2,$ or  
\be \|u\|_{ens}:=\sqrt{\sum_{n=0}^N k_n^2|u_n|^2}. \lb{ens-norm} \ee 
However, we obtained similar results with norms for other spaces such as energy norm 
for the space $l_2$  and sup-norm for the space $l_\infty$ and also for individual shells $n.$
We took $t=10^{-3}t_\eta,$ where $t_\eta=\eta/u_\eta$ is the Kolmogorov time, since this 
choice of time $t$ was large enough to obtain appreciable evolution at the highest shells. The results 
which are plotted in Figure \ref{fig-strong2} show convincingly that the strong order of convergence 
of the method is $2.$

\begin{figure}[h!]
 \begin{center}
\includegraphics[width=240pt]{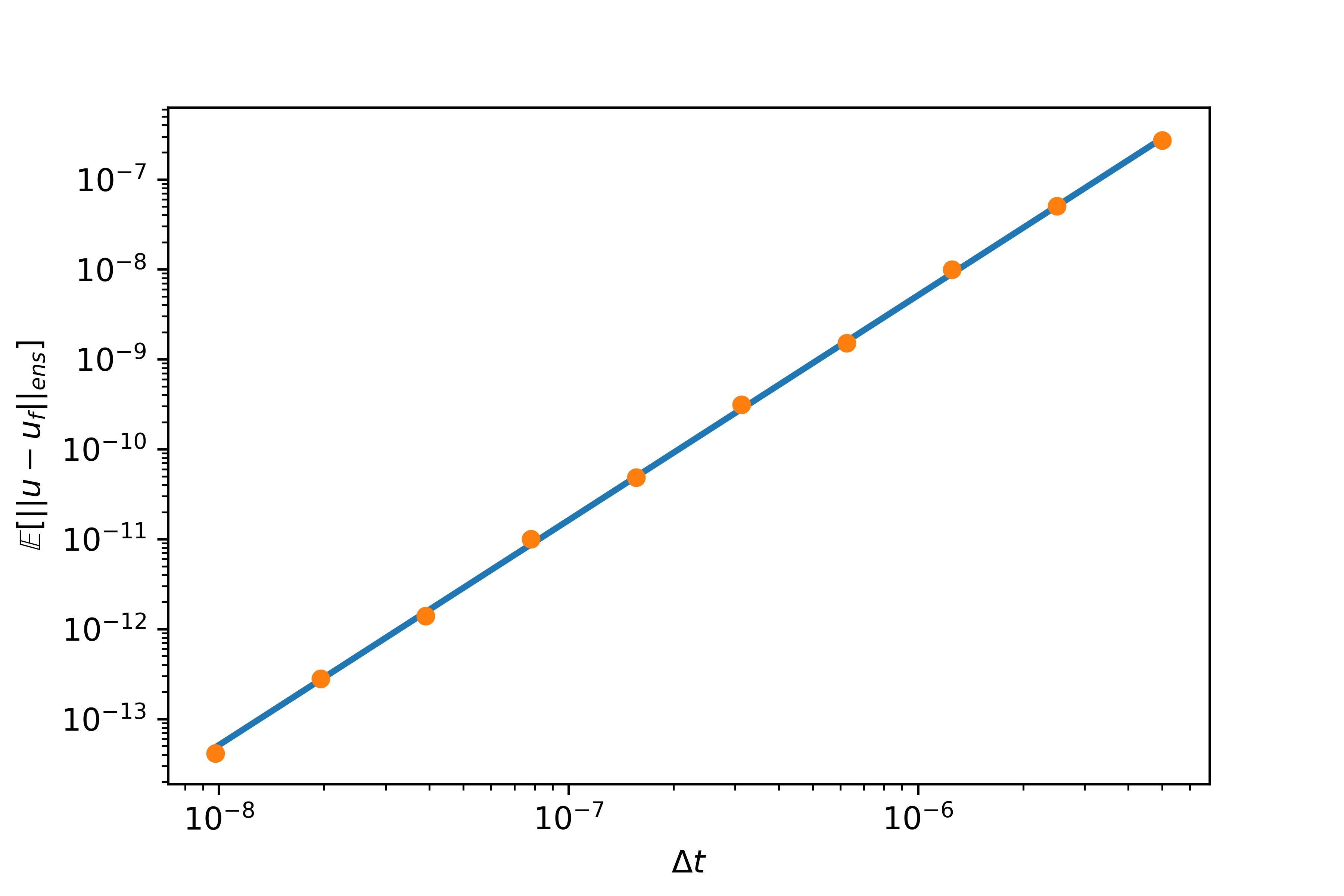}
 \end{center}
 \caption{Scaling $(\Delta t)^\frac{5}{2}$ for the local truncation error $T(\Delta t)$. Solid orange dots (\textcolor{orange}{{\bf $\bullet$}}) correspond to expectation values of the enstrophy norm \eqref{ens-norm} of the error; solid cyan line ( \textcolor{bondiblue}{\hvline}) corresponds to least squares fit of a function $c \Delta t ^{5/2}$, where $c$ was determined to be $6.710$.}
 \label{fig-strong1}
 \end{figure}

This is consistent with the mathematical theory \cite{lord2004numerical}, which 
establishes at least $\frac{3}{2}$-order (see also Appendix \ref{derive-slaved}), \\
but better than we had initially expected. We furthermore 
found the method to be strong order 2 for all other initial data that we considered. To illuminate this 
unexpectedly large convergence rate, we estimated the local truncation error 
$T(\Delta t)$ by calculating the error in the method with a single step 
${\mathbb E}[\| u^{\Delta t}(\Delta t)-u^{\delta t}(\Delta t)\|_{h^1_2}],$ for each stepsize $\Delta t.$ 
The results, plotted in Figure \ref{fig-strong1} show the scaling $T(\Delta t)\propto (\Delta t)^{5/2}$
that was expected. This would produce a global error scaling as $(\Delta t)^{3/2}$ 
in a number of steps $N_{\mathrm{steps}}\propto \frac{1}{\Delta t},$ if these errors accumulated without 
cancellation. The observed scaling of global error is explained if the errors at each step 
are in fact uncorrelated and of different signs, so that, by the central limit theorem, 
the total error then scales as $(\Delta t)^{5/2} \sqrt{N_{\mathrm{steps}}} \propto (\Delta t)^{2}$.

 In conclusion, our numerical scheme \eqref{sabra-num} converges pathwise with order 
at least $3/2$ (and effectively $2$) as $\Delta t\to 0$ for a choice of the cutoff $N>N^\star_e$.

\subsubsection{Independence of Wavenumber Truncation $N$}\lb{N-independ} 

We next verify that the statistical evolution of the shell modes $u_n$
for $n\leq N_e^\star$ is independent of the choice of cutoff $N\geq N_e^\star.$  
More precisely, we study the transition probability density 
\be P_N(u,t|u(0),0)={\mathbb E}[\delta(u-U_N(t;u(0),W)] \lb{trans-pdf} \ee    
where $U_N(t;u(0),W)$ is the (strong) solution of our stochastic shell model \eqref{nSabra} 
for number of shells $N\geq N_e(u(0))$ with initial data $u(0)$ and with the particular 
realization $W=(W_n:\ n=0,1,..,N)$ of the random Brownian motions over the time-interval $[0,t],$ 
and ${\mathbb E}[\cdot]$ again denotes expectation over those random Brownian motions. 
We present evidence that this transition probability for the set of modes 
$(u_n(0)=u_n^\star:\ n=0,1,..,N_e^\star)$ is independent of the choice of cut-off 
$N\geq N_e^\star$ and also independent of the initial data $(u_n(0):\ n=N^\star_e+1,...,N)$ 
of the modes with $n>N_e^\star$ when those are selected at random from the thermal Gibbs 
distribution for those shells. If this result holds, then we have a well-defined stochastic 
Markov evolution for the modes $(u_n(0):\ n=0,1,..,N_e^\star),$ which is independent of the 
cut-off $N.$ This gives a precise mathematical meaning to the stochastic shell-model 
\eqref{nSabra} as a ``low-wavenumber effective theory.'' 

The transition probability density \eqref{trans-pdf} of the entire state vector 
$(u_n:\ n=0,1,..,N_e^\star)$ is obviously too unwieldy to investigate in its entirety, so we consider 
instead reduced or marginal PDF's of $u_n$ for specific shell-numbers $n\leq N_e^\star.$ 
We focus our attention on the shell modes with $n$ near to $N^\star_e,$ since those must be 
most affected by the change in truncation $N.$ Here we present results for $n=N_e^\star$
itself, but comparable results are found also for $n<N_e^\star.$
%  see the Supplemental Materials.
For the time lapse $t$ in the transition PDF
\eqref{trans-pdf} we chose $t=t_\eta,$ one Kolmogorov time. This time should be sufficient 
for the influence of truncation $N$ to propagate through the entire dissipation range. 
Once the transition PDF \eqref{trans-pdf} has been verified to be independent of $N$
for times $t\leq t_\eta,$ then this independence of course extends to the transition PDF's 
for {\it all} times $t>0$ by the Chapman-Kolmogorov equation. Here we presume 
that $N$-independence of the transition probability for the most singular event, $u(0)=u^*,$ 
implies independence for any other choices of $u(0).$

\begin{figure}[t]
 \begin{center}
\includegraphics[width=240pt]{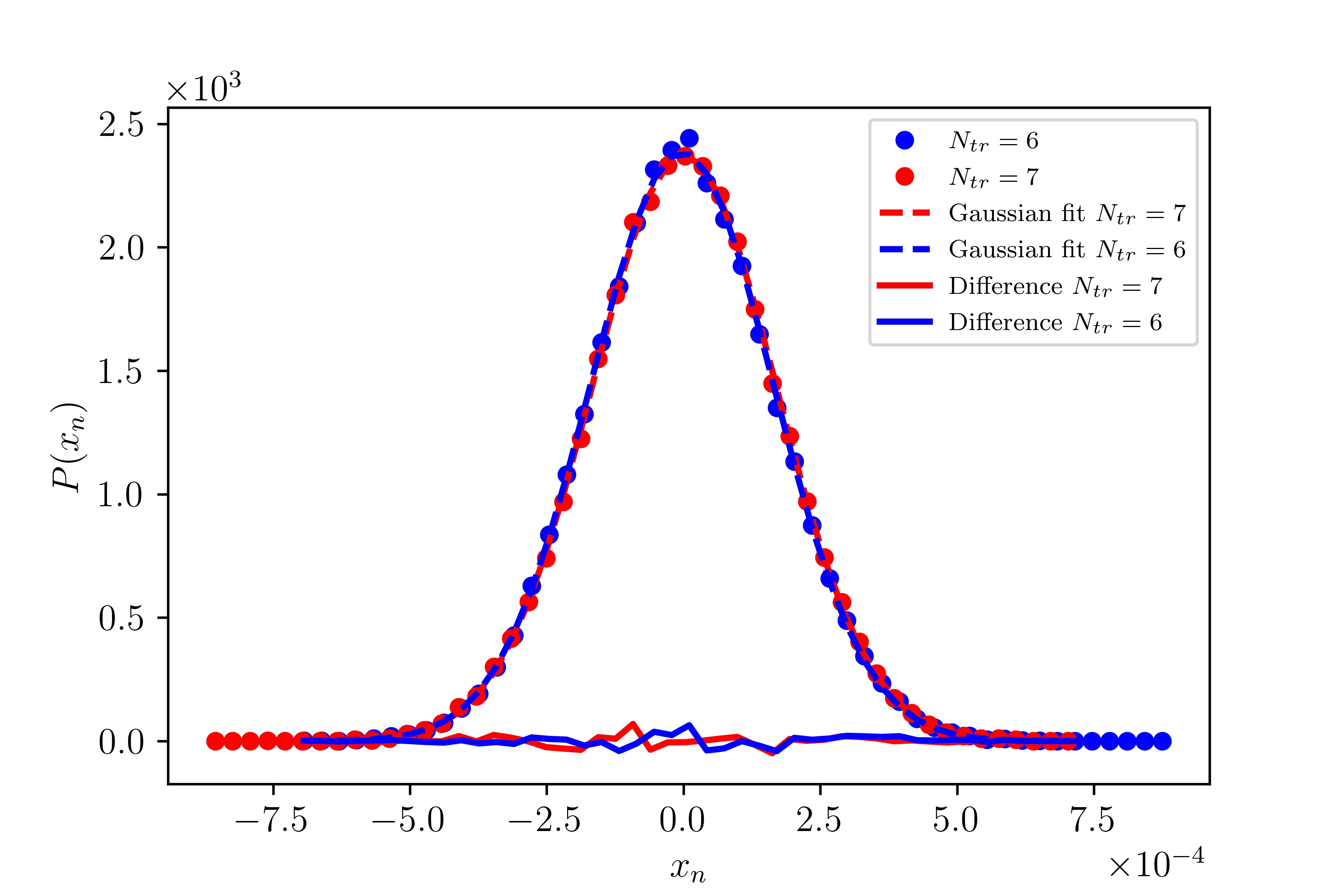}
 \end{center}
 \caption{Comparison of the two marginal transition PDFs 
 $P_N(x_{N^\star_e},t|u^\star,0)$ for the variable $x_{N^\star_e}={\rm Re}(u_{N^\star_e})$ 
 with the truncation wavenumbers $N=N^\star_e$ and $N=N^\star_e+1$.}
 \label{fig-trunc}
 \end{figure}

For the study of effects of truncation, we used the same model parameters and time step 
$\Delta t$ that were employed in the long-time steady-state simulation, discussed 
in detail in the next section \ref{setup}. Increasing $N$ from its original value 
$N=N^\star_e+1$ would require  a smaller time-step and this would have been numerically 
expensive. We therefore chose to make the much more demanding test of {\it reducing} 
the cut-off $N$ to the value $N=N^\star_e,$ taking $u_{N^\star_e+1}=0$ as boundary 
condition, and then comparing the transition PDF for the reduced 
value $N=N^\star_e$ with that for the original value $N=N^\star_e+1.$ We emphasize that for the 
intense burst event $u^\star,$ only the highest two modes with $n=N^\star_e,$ $N^\star_e+1$ remained 
in thermal equilibrium, so that our reduction to $N=N_e^\star$ leaves only a single mode
in equipartition. Plotted in Fig.\ref{fig-trunc} are the reduced transition PDF's 
for the real part $x_{N^\star_e}={\rm Re}(u_{N^\star_e})$ with both $N=N^\star_e$ and $N=N^\star_e+1$,
calculated by averaging over $8 \cdot 10^4$ independent samples of the Brownian motions 
$W=(W_n:\ n=0,1,..,N^\star_e+1).$ The two PDF's are identical within Monte Carlo error, 
a strong evidence for statistical decoupling of mode $u_{N^\star_e+1}$ from the 
dynamics of the modes $u_n$ with $n\leq N^\star_e.$ We obtained analogous results 
(not shown here) for the marginal transition PDF of the imaginary part $y_{N^\star_e}
={\rm Im}(u_{N^\star_e})$ and for the similar variables $x_n,$ $y_n$ with $n< N^\star_e.$ 

 These independence results support the claim that we have a well-defined stochastic  
dynamics whenever the cutoff $N$ is selected larger than $N^\star_e,$ in agreement with the 
idea that the shell modes in thermal equilibrium are UV asymptotically free. These results 
in fact suggest that a convergence result should hold for the idealized mathematical 
limit $N\to\infty.$ This is hinted also by the rigorous results in \cite{bessaih2012invariant} 
for the unforced, thermal equilibrium dynamics. Of course, even if such a limit result 
could be proved for the forced model, it would still not suffice to justify it physically, 
unless it could be shown even further that convergence is practically attained for a value 
of $N^\star_e=\log_2(\Lambda/k_0)$ corresponding to a length-scale $\Lambda^{-1}$
still much greater than $\lambda_{mfp}.$ Our results for the Sabra model are thus encouraging, 
because $N^\star_e=n_\eta+6$ corresponds to a length-scale only 64 times smaller than $\eta.$ 
In the case of the ABL with $\eta=0.54\ mm$ these events of most extreme intermittency 
would correspond to a length about $8.4\ \mu m,$ which is still 124 times greater than the 
mean-free path length of air, $\lambda_{mfp}=68\ nm.$ 
 
\section{Numerical Simulation Results}\lb{num:sec} 

\subsection{Set-up of the Simulations}\lb{setup}  

We undertook to perform our simulations of the noisy Sabra model in dimensionless form with 
the dissipation scaling \eqref{nSabra-K}, so that $\varepsilon=\nu=1$ and 
the Kolmogorov shellnumber is set to $n=0.$ We wanted dimensionless parameters in correspondence 
with the atmospheric boundary layer (ABL), and thus the dimensionless temperature was taken 
to have the value $\theta_\eta=2.83 \times 10^{-8}$ in \eqref{ABL-K}. The range of shell numbers 
in our simulation was chosen as $n = M,M+1,...,R$ with $M=-15$ and $R=7$ and with constant 
stirring forces applied at the first two shells, $M$ and $M+1.$ We aimed to achieve a Reynolds number 
$Re=u_{rms}2^M$ comparable to the value $Re\sim 10^7$ cited as typical in the ABL \cite{garratt1994atmospheric}.
However, with our choice of forcing, neither of the statistical quantities 
\be u^2_{rms}=\sum_{n=M}^R \langle |u_n|^2\rangle, \quad \varepsilon=\sum_{n=M}^R 
k_n^2 \langle |u_n|^2\rangle \ee 
was under our precise control. We therefore adjusted the forcing until we obtained 
$u_{rms}\sim O(10^2)$ and $\varepsilon\sim O(1),$ which was satisfied with the stirring forces 
\begin{multline*}
f_M = -0.008900918232183095 \\ -0.0305497603210104\,i,
\end{multline*}
\begin{multline}\lb{stirring}
f_{M+1} =  0.005116337459331228 \\ -0.018175040700335127\,i,
\end{multline}
which gave the precise value $\varepsilon\doteq 1.478$, close to our target value of unity. 
We then rescaled all quantities to correct dissipation-scale units by taking 
\be u_n\to u_n/\varepsilon^{1/4},\quad k_n\to k_n/\varepsilon^{1/4},\quad f_n\to f_n/\varepsilon^{3/4}, \ee
which yielded $u_{rms}= 56.48,$ $Re= 2.04 \times 10^6,$ and 
$\theta_\eta= 2.328 \times 10^{-8}.$ All results
presented below are in these Kolmogorov units.   

The time-step of our simulation was chosen (in original units) to be equal to $\Delta t=10^{-5}$ 
which was about an order of magnitude smaller than the viscous time at the highest wavenumber,  
$t_{vis}=1/k_R^2\doteq 6.1\times 10^{-5}.$ 
%\textcolor{black}{We checked that that our results for integration 
%of individual realizations of the shell model were negligibly changed by doubling this 
%time-step} and we also checked that $\Delta t=10^{-5}$ for the shell model with zero stirring  
%yielded a statistical steady-state distribution which was the Gaussian thermal equilibrium 
%\eqref{GibbsS} to high accuracy. 
Since one large-eddy turnover-time of the simulation was 
$T=1/k_M u_{rms}\doteq 640$ dimensionless time units, it was too time-consuming 
to calculate time-averages over many such times $T$ in a single run of the model on one 
computer. We therefore took advantage of parallel computing by using a strategy 
rst computing a long run of the 
model for $N_{samp}=300$ large-eddy turnover times with increased time-step $\Delta t=10^{-3}$
and then creating $N_{samp}$ independent initial-data by sampling that under-resolved 
solution in intervals of one turnover time. These 300 independent initial data were 
then distributed over the nodes of a computer cluster and each integrated again for 
time $T$ with the time-step $\Delta t=10^{-5}.$ To avoid possible early-time 
artefacts from under-resolution, we discarded the first 100 steps of these 
well-resolved simulations in calculating all long-time averages (and, in fact, 
the statistics were checked to change negligibly also by including those initial time-steps).

Both the noisy model \eqref{nSabra} and the deterministic model \eqref{dSabra} were 
solved with the same numerical Taylor-It$\bar{\rm o}$ scheme \eqref{sabra-num},
the latter simply by setting $\theta_\eta=0.$ The calculations were performed 
in double-precision arithmetic and only for the deterministic model at the 
highest wavenumbers did we approach the underflow level of double precision.
In fact, it is one of the numerical advantages of stochastic models that 
the requirements on arithmetic precision are considerably reduced, a fact which 
has been previously stressed for climate modelling \cite{palmer2019stochastic}.
The time-step $\Delta t$ and the total number $N_{samp}$ of large-eddy times 
used to calculate averages were the same for both noisy and deterministic cases. 
We discuss in Appendix \ref{converge-tests} the various tests that we have made 
that those parameter choices were adequate to yield well-converged statistics. 

\subsection{Energy Spectra with Thermal Noise} \lb{sec:spectra}

We first present numerical results on the simplest statistical quantity of interest,
the energy spectrum. This result was already reported in \cite{bandak2021thermal}, 
but it is included here also for completeness. In order to make thermal energy equipartition as evident 
as possible, we shall show the average $\bar{\epsilon}_n$ of the kinetic energy per mass in the mode 
with shell-number $n$ 
\be  \epsilon_n = \frac{1}{2} |u_n|^2, \lb{en-def} \ee 
plotted versus wavenumber $k_n.$ This differs slightly from the standard energy 
spectrum for the shell model which is conventionally defined by $E_n=\bar{\epsilon}_n/k_n,$
as in \eqref{eq-spectrumS}. The quantity \eqref{en-def} is more convenient
because thermal equipartition gives a constant value independent of $n$
\be e_n^{eq} = \theta_\eta, \lb{equip-val} \ee 
in Kolmogorov dissipation units. In Fig.~\ref{comp-spectrum} we plot the energy 
in mode $n$ given by \eqref{en-def} for both the deterministic and noisy Sabra models. 

\begin{figure}[t]
 \begin{center}
\includegraphics[width=240pt]{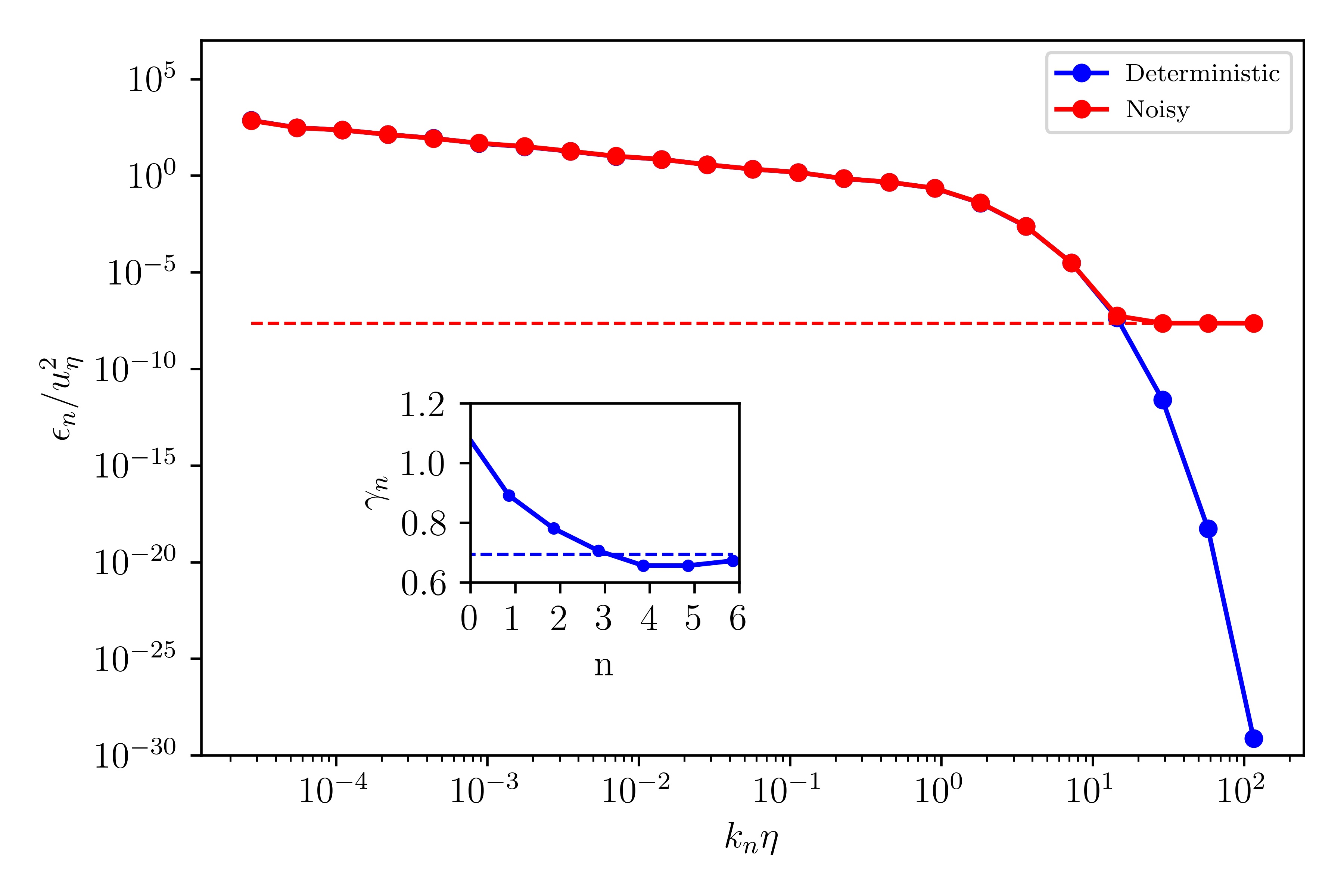}
 \end{center}
 \caption{The energy spectrum of the deterministic model (heavy blue line,
 \textcolor{blue}{\hvline}) and noisy model (heavy red line, \textcolor{red}{\hvline}).
 The dashed red line (\textcolor{red}{-$\,$-$\,$-}) shows the thermal equipartition 
 value in \eqref{equip-val}.  {\it Inset:} The local stretching exponents 
 \eqref{local-stretch} for the deterministic model spectrum 
 (blue solid circles, \textcolor{blue}{{\bf $\bullet$}}) and theoretical value
 in \eqref{stretch-decay} (dashed blue line, \textcolor{blue}{-$\,$-$\,$-}). }
\lb{comp-spectrum}  \end{figure} 

 The two spectra are indistinguishable in the inertial range, where both 
 exhibit a power-law decay $\propto k_n^{-\alpha}$ with exponent a bit larger than 
 the K41 value $\alpha=2/3.$ The increase of $\alpha$ above the K41 value is due to 
 standard inertial-range intermittency effects, which do not differ for the 
 deterministic and noisy models. The behavior of the spectrum in the dissipation-range 
 is, however, drastically different for the two models, with the spectrum 
 of the deterministic model exhibiting a very rapid exponential-type decay and 
 the spectrum of the noisy model saturating at the equipartition value \eqref{equip-val}. 
 Altogether, the results are very close to our predicted spectrum for 3D incompressible 
 fluid turbulence pictured in Fig.~\ref{spectrum}, except that the equipartition 
 wavenumber is slightly larger, $k_{eq}\eta\doteq 15,$ rather than $k_{eq}\eta\doteq 10$
 for 3D. The somewhat smaller equipartition wavenumber in 3D is due in part to the rising
 equipartition spectrum there and also to the faster exponential decay in the absence 
 of thermal noise. By contrast, the energy spectrum \eqref{en-def} of the shell 
 model becomes flat in thermal equipartition and exhibits the stretched-exponential 
 decay \eqref{stretch-decay} without thermal noise. We have verified this stretched-exponential 
 form of the spectrum from our numerical results, by calculating a local stretching exponent
 \be \gamma_n = \log_2 \left|\ln\langle|u_{n+1}|^2\rangle\right|-
 \log_2 \left|\ln\langle|u_{n}|^2\rangle\right| \lb{local-stretch} \ee 
 at dissipation-range shell numbers $n>0$ in the deterministic model, with the results 
 plotted in the inset of Fig.~\ref{comp-spectrum}. The local exponents indeed 
 agree well with the theoretically expected value $\gamma=\log_2\left(\frac{1+\sqrt{5}}{2}\right)$
 at $n\geq 4.$

 \begin{figure}[h!]
 \begin{center}
\includegraphics[width=240pt]{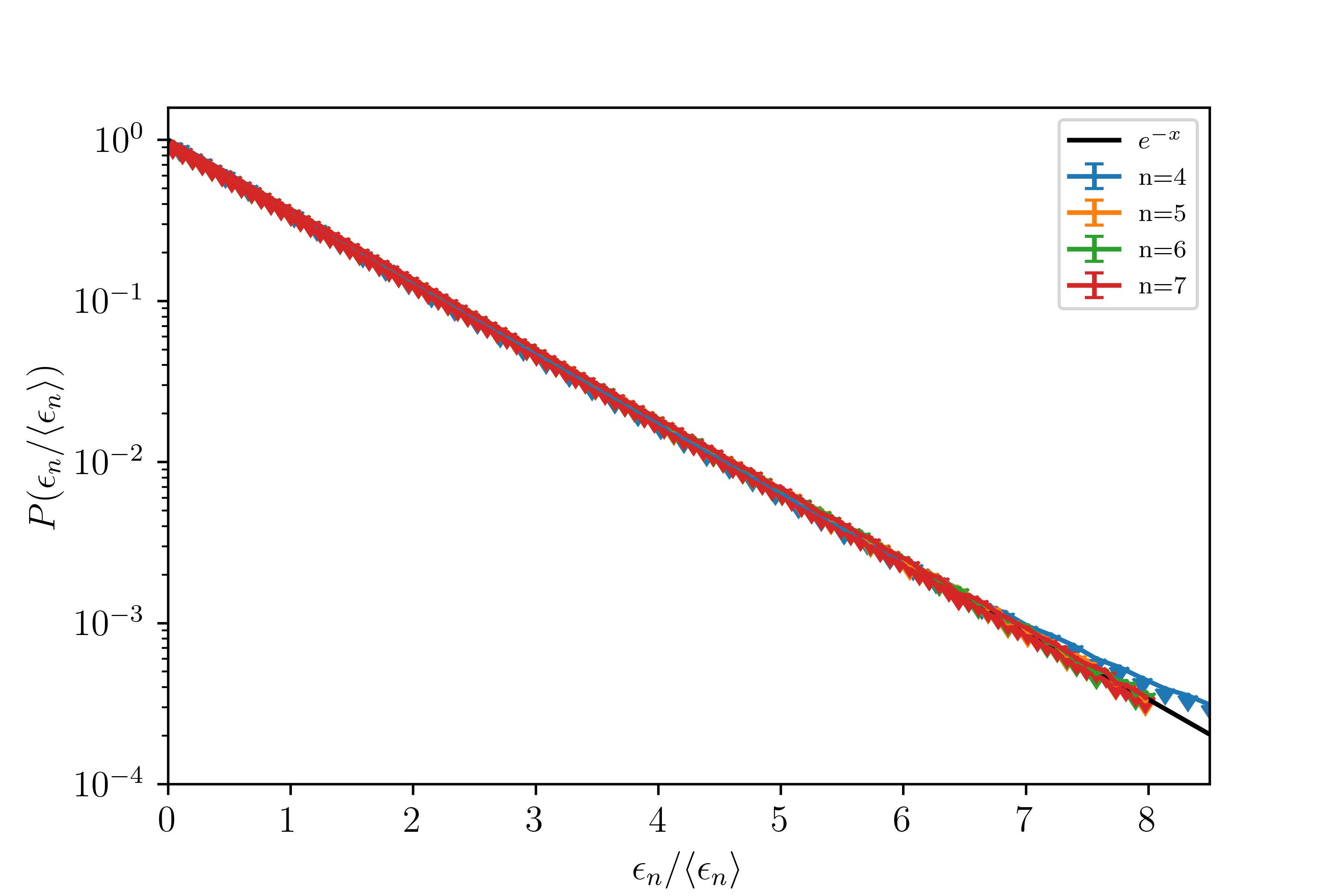}
 \end{center}
 \caption{PDFs of the normalized modal energy $\epsilon_n/\langle \epsilon_n \rangle$ 
 for shells $n = 4,5,6,7$ and an  exponential distribution with rate parameter 1. 
 The error bars depict standard error of the mean from the long-time average.}
\lb{energy-pdf} \end{figure} 

\begin{figure*}[t!]
  \centering
  \begin{subfigure}[b]{0.4\linewidth}
    \includegraphics[width=\linewidth]{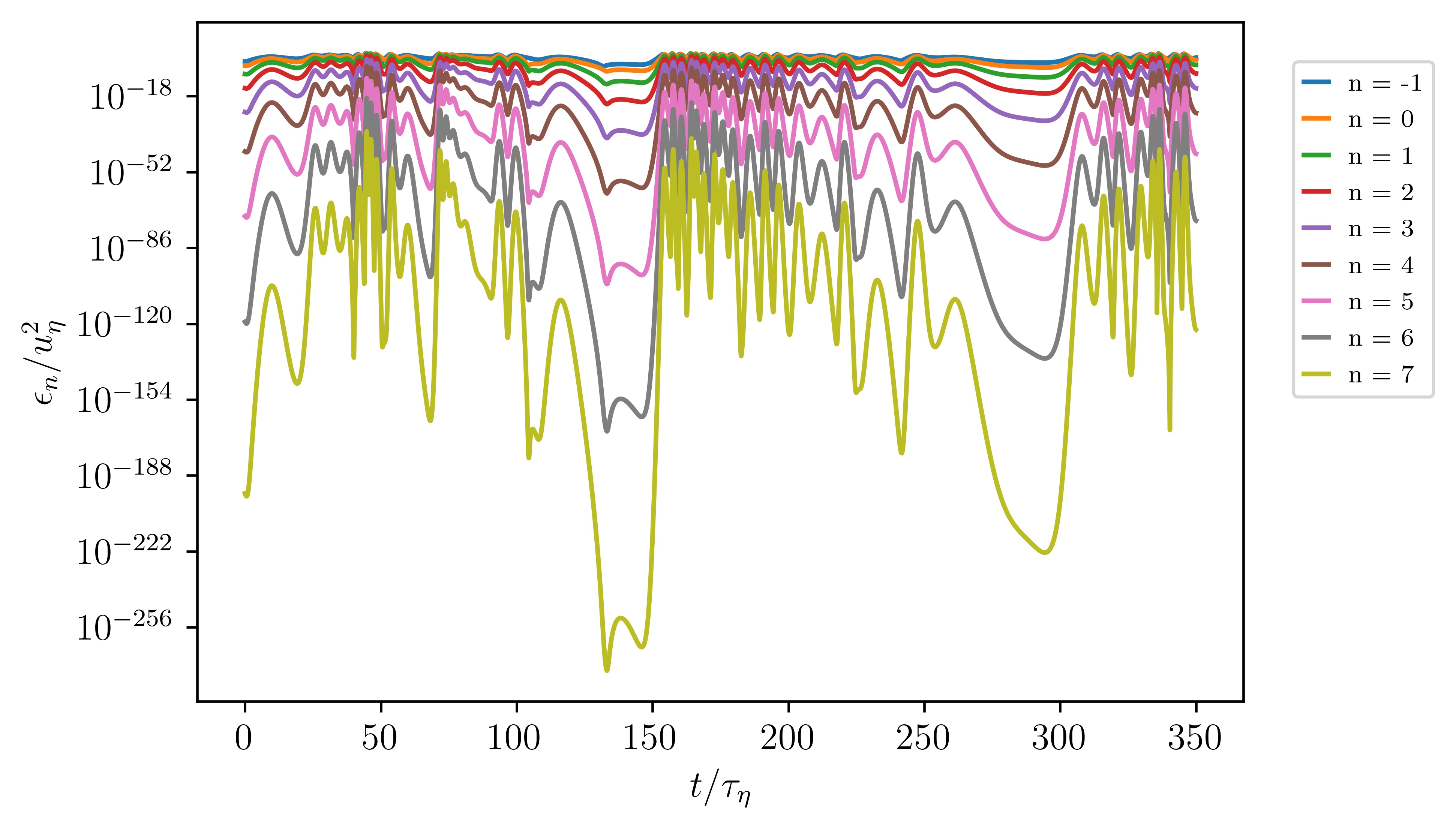}
    \caption{Deterministic Sabra}
  \end{subfigure}
  \begin{subfigure}[b]{0.4\linewidth}
    \includegraphics[width=\linewidth]{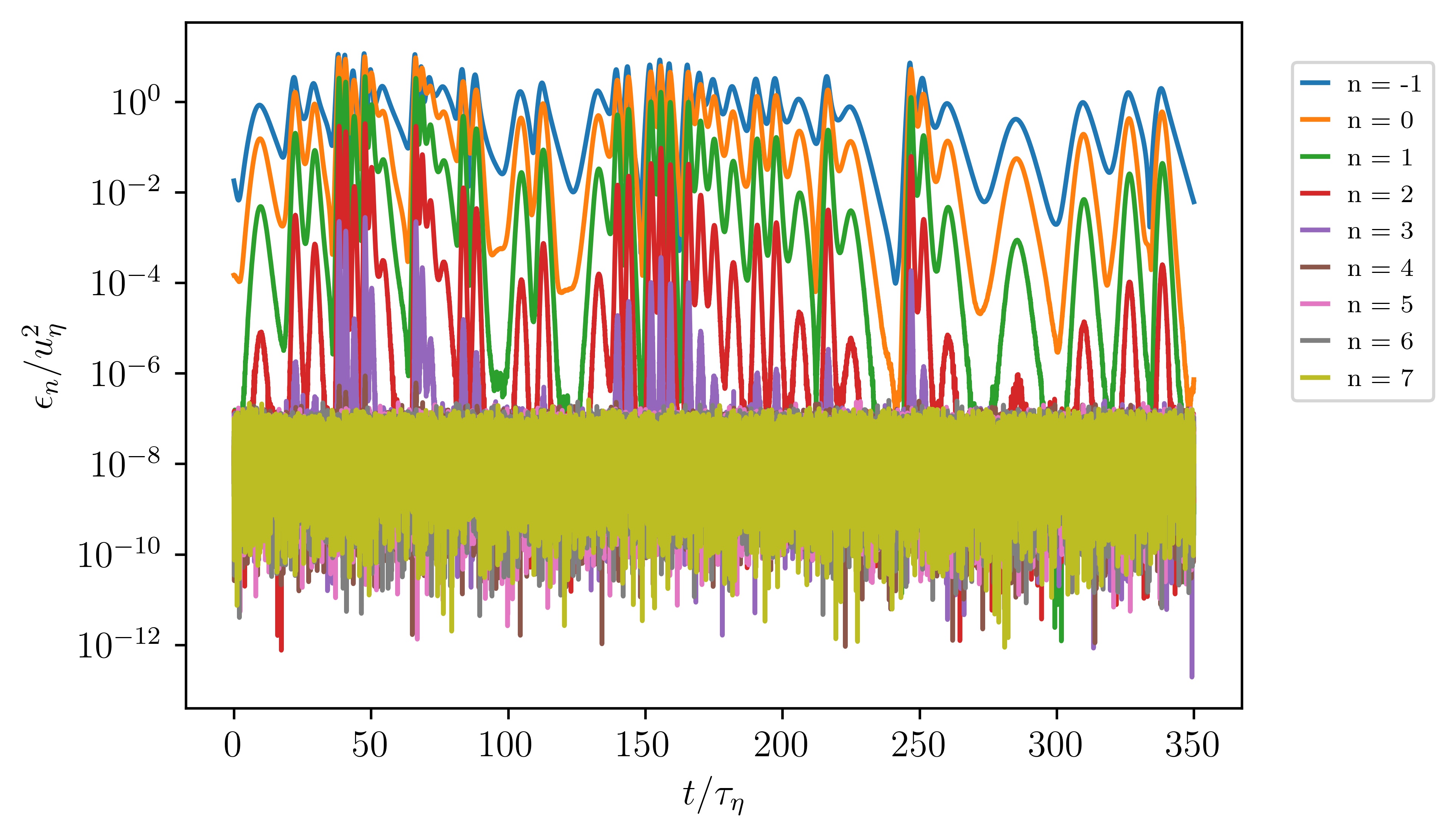}
    \caption{Noisy Sabra.}
  \end{subfigure}
 \caption{Plots of modal energies $\epsilon_n(t)$ versus time $t$ for shells
 in the dissipation range with $n=-1$ to $n=7$. (a) The deterministic Sabra model results, 
 exhibiting the extreme intermittency predicted by Kraichnan \cite{kraichnan1967intermittency},
 and (b) the stochastic model results, with the large intermittent fluctuations completely washed 
 out by Gaussian thermal fluctuations.} \label{kraichnan67}
\end{figure*}

 With thermal noise present, the modes at the highest wavenumber not only have energy 
 spectrum in thermal equilibrium, but in fact have all statistics quite accurately described 
 by the gaussian Gibbs measure \eqref{GibbsS}. Plotted in Fig.~\ref{energy-pdf}  
 are the PDF's of the modal energies $\epsilon_n=\frac{1}{2}|u_n|^2$ for the 
 four highest shells $n=4,5,6,7$ compared with the exponential PDF $p(\epsilon)=
 \beta e^{-\beta \epsilon}$ that follows from  \eqref{GibbsS}.  The energy for the lowest 
 of these modes, $n=4,$ agrees well with the exponential PDF out to about 7 standard 
 deviations, but has a distinctly broader tail past this point. The three highest 
modes with $n=5,6,7$ have energy PDF's that are indistinguishable from the 
thermal-equilibrium exponential PDF to within numerical errors (which arise mostly 
from the finite number of samples).

\subsection{Dissipation-Range Intermittency}\lb{diss-range}  

The dissipation range of 3D incompressible fluid turbulence is widely expected 
to exhibit extreme spatiotemporal intermittency because of the super-algebraic 
decay of the energy spectrum at high wavenumbeers $k.$ This was first 
predicted by Kraichnan \cite{kraichnan1967intermittency}, who argued heuristically 
that most of the contributions to moments of velocity-gradients at high-$k$
would come ``from the few exceptional regions'' where a large but still 
modest fluctuation around the Kolmogorov scale had been exponentially
magnified in relative magnitude by the rapid fall-off in the spectrum. 
This argument was later given in a more mathematically precise form
related to singular solutions of the dynamics in the complex time plane 
and extended also to temporal intermittency of simple nonlinear dynamics 
without spatial structure \cite{frisch1981intermittency}. This type of intermittency
in the ``far-dissipation range'' is expected to occur even in relatively low Reynolds 
number turbulent flows and has been observed in numerical simulations of the 
Navier-Stokes equation \cite{chen1993far,schumacher2007asymptotic}.  

The deterministic Sabra model \eqref{dSabra} possesses all of the ingredients for such 
extreme intermittency in its far-dissipation range and our numerics 
give vivid confirmation of the expected effects. For example, plotted 
in Fig.~\ref{kraichnan67}(a) are the modal energies $\epsilon_n(t)$ as 
a function of time $t$ for the dissipation-range shells $n=-1,...,R$ in a 
simulation with $M=-15$ and $R=7.$ It can be seen there that mild 
fluctuations for $n=-1,0$ appear with a slight time delay at larger $n$ 
but also hugely enhanced, so that the energies at distinct instants
differ by hundreds of orders of magnitude. The fluctuations are 
so severe that our computations are brought near to the underflow 
level of the double-precision arithmetic. An average of $\epsilon_n(t)$ 
over the pictured time interval for a dissipation-range mode with 
$n>0$ would thus be completely dominated by the single exceptional 
event around $t\simeq 50.$ However, the same quantities plotted in 
Fig.~\ref{kraichnan67}(b) for the noisy Sabra model \eqref{nSabra}, which includes 
physically realistic thermal noise, has these extreme fluctuations replaced 
by Gaussian velocity fluctuations at the level of energy equipartition. This thermal velocity $u_{th}$ 
is small compared with the Kolmogorov-scale velocity $u_\eta$ (by a factor of $\theta_\eta^{1/2}$)
but much larger than the extremely tiny velocities attained in the deterministic theory. We expect 
likewise for incompressible fluid turbulence occurring in the laboratory or in Nature that 
the type of strong intermittency predicted by Kraichnan \cite{kraichnan1967intermittency} to appear 
in the far-dissipation range is, in fact, thoroughly erased by thermal noise.

Sizable dissipation-range intermittency remains in the stochastic shell 
model, as made clear by the results in Fig.~\ref{kraichnan67}(b), but we argue 
that it is a high-$Re$ effect of inertial-range intermittency propagating into 
the dissipation range. This ``near-dissipation range intermittency'' is present also 
in the deterministic Sabra model and, in a spatiotemporal form, in turbulence 
modelled by the deterministic Navier-Stokes equation. \red{This is the type of 
near-singularity due to extreme events intensively studied in recent works 
\cite{yeung2015extreme,yeung2020advancing, farazmand2017variational,buaria2019extreme,
buaria2020self,buaria2020vortex,nguyen2020characterizing}.} In fluid turbulence such 
intermittency is known to lead to strong fluctuations in the ``local viscous/cutoff length'' 
$\eta(\bx,t),$ which is conventionally defined \cite{paladin1987degrees,schumacher2007sub} 
by the condition that the local Reynolds number at that scale be order unity:
\be \frac{\eta\delta_\eta u(\bx,t)}{\nu}\simeq 1, \lb{eta-loc} \ee
where $\delta_\eta u(\bx,t)=|\bu(\bx+\eta,t)-\bu(\bx,t)|.$ While typically 
$\eta(\bx,t)\simeq \eta,$ much larger and much smaller cutoff lengths appear. 
As already discussed in section \ref{intermittent:sec}, this fluctuating viscous 
length has been represented phenomenologically \cite{paladin1987degrees} by 
$\eta_h\sim L Re^{-1/(1+h)},$  where $h=h(\bx,t)$ is the local H\"older exponent 
of the velocity in the Parisi-Frisch multifractal model 
\cite{frisch1985singularity,frisch1995turbulence}.  Below the length-scale $\eta(\bx,t)$
the velocity field is expected to be smooth, with a local energy spectrum 
defined by a suitable band-pass filter that is exponentially decaying. These 
are the considerations that led Frisch and Vergassola to predict  for velocity structure 
functions an ``intermediate dissipation range'' \cite{frisch1991prediction},
bridging the inertial-range and the far-dissipation range. 
A definition of  a ``local viscous shellnumber'' analogous to \eqref{eta-loc} 
may be made also in shell models, both deterministic and noisy, by setting 
\be N_{vis}(u):=\min\left\{n:\, \frac{|u_n|}{\nu k_n}\leq 1\right\}. \lb{Nvisc} \ee
and the predictions of \cite{frisch1991prediction} concerning the intermediate dissipation 
range were previously verified in a numerical simulation of the GOY shell model 
\cite{bowman2006links}.  

We expect that a similar physics lies behind the intermittency displayed 
in Fig.~\ref{kraichnan67}(b) for our noisy shell model, but with the rapid 
exponential decay of amplitudes below the viscous cutoff replaced in 
the noisy model by a thermal equipartition, like that exhibited by the 
average energy spectrum for shellnumbers $n\geq 5$ in Fig.~\ref{comp-spectrum}.
As a consequence of the inertial-range intermittency, however, the shell-number at which 
this equipartition first sets in must fluctuate greatly from realization to realization. 
It is actually a somewhat subtle issue how to define precisely ``energy equipartition" 
for an individual realization, because equipartition is a statistical concept. 
Even realizations selected from the thermal Gibbs state \eqref{GibbsS} 
show considerable fluctuations in energy from the equipartition value 
and some averaging in time is typically required to bring the modal 
energy close to the ensemble mean value even for such an equilibrium realization. 
\black{As one possible measure of the ``equilibration shell-number'' $N_e(u)$
for an individual realization $u$ from our turbulent simulation, we can average 
the modal energy $\epsilon_n=(1/2)|u_n|^2$ over one Kolmogorov time $t_\eta=\eta/u_\eta$ 
and we then identify $N_e(u)$ as the smallest integer such that this local time-average 
$(1/2)\langle|u_n|^2\rangle_\eta$ is below $2 \theta_\eta$ for all $n\geq N_e(u).$
For examples of such locally-time averaged realizations, see Figure 3 in \cite{bandak2021thermal}.
We have plotted in Fig.~\ref{multi-pdf} the PDF of the local equipartition shellnumber 
$N_e(u)$ obtained from our DNS with the above definition, together 
with the PDF of the viscous cutoff shell-number $N_{vis}(u)$ defined in \eqref{Nvisc}.}   

Before examining the simulation results, we must first acknowledge that our definition 
of the ``equilibration shell-number'' $N_e(u)$ suffers from a good bit of arbitrariness. 
We have therefore explored as well alternative definitions. For example,
if averaging over time $t_{avg}$ produces an equipartition spectrum down to shell 
$N(t_{avg}),$ then averaging over a longer time might extend that range. Since the 
natural viscous time-scale of the stochastic dynamics at one lower shell increases by 4, 
we have considered another possible definition by successively increasing the averaging 
time $t_{avg}$ from $t_\eta$ by factors of 4 and by redefining $N_e(u)=N(4t_{avg})$ as long 
as $N(4t_{avg})<N(t_{avg})$ \footnote{Note that it is possible that $N(4t_{avg})>N(t_{avg})$
because a new strong ``burst'' may enter the dissipation range between the times 
$t_{avg}$ and $4t_{avg}.$}.  This alternative definition yielded slightly lower 
estimates of $N_e(u)$ for some realizations $u,$ but with the same qualitative features.
A completely different approach to defining $N_e(u)$ would be to apply standard 
distribution tests from mathematical statistics, such as $p$-values \cite{marden2000hypothesis}, 
to the hypothesis that the shell variables $u_\eta,u_{K+1},...,u_N$ are drawn from the 
multivariate distribution \eqref{GibbsS} and then define $N_e(u)$ to be the smallest 
$K\leq N$ for which that hypothesis is accepted. However, since any definition of the 
``equilibration shell-number'' seems to involve various subjective choices and 
since all definitions that we have considered exhibit qualitatively similar intermittency, 
we shall only discuss here the quantity $N_e(u)$ defined in the previous paragraph. 

\begin{figure}[t!]
 \begin{center}
\includegraphics[width=260pt]{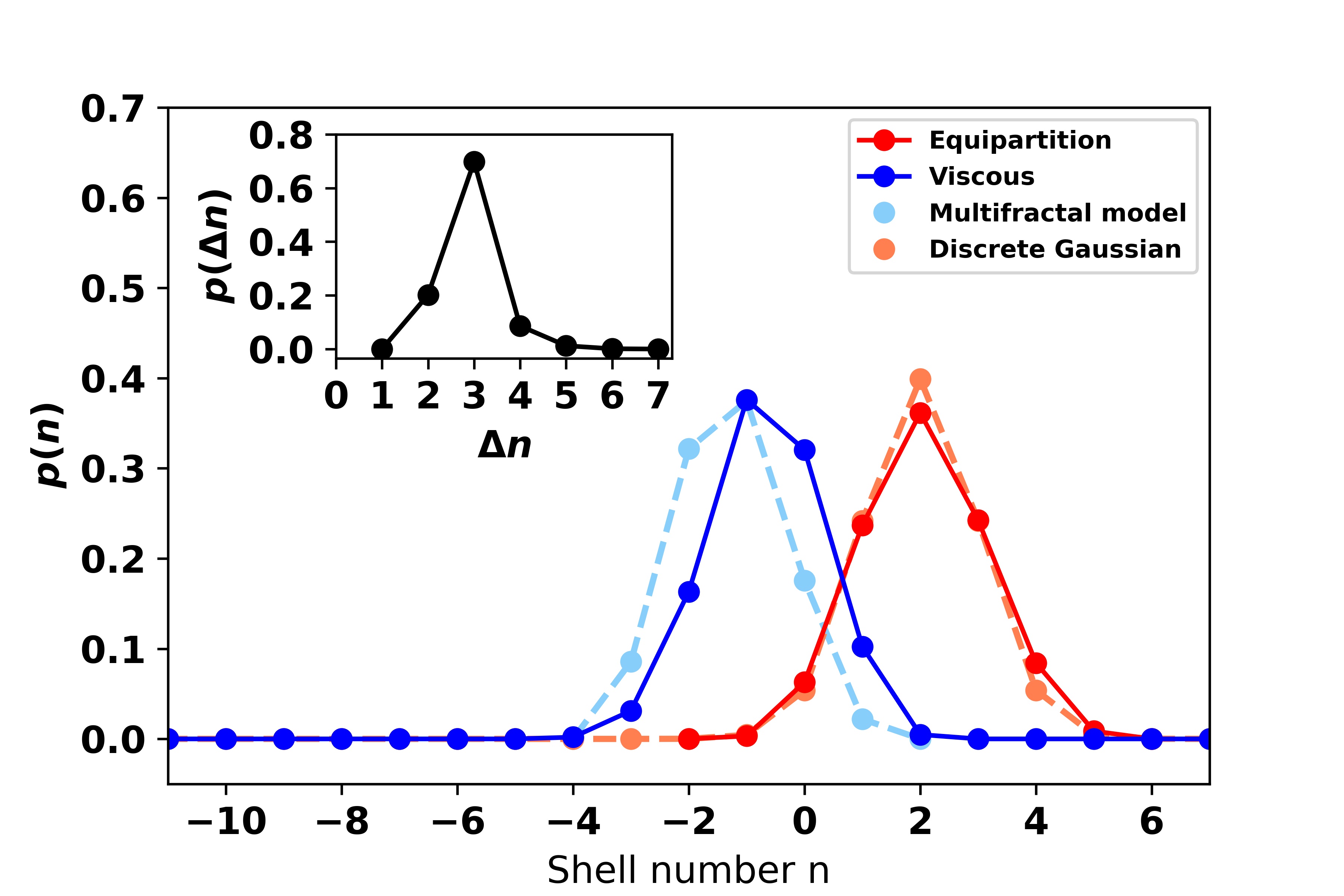}
 \end{center}
 \caption{PDF's of $N_{vis}(u)$ (\textcolor{blue}{{\bf blue}}) and $N_{e}(u)$ 
 (\textcolor{red}{{\bf red}}), along with a simple multifractal model  
 for the PDF of $N_{vis}(u)$ (\textcolor{paleblue}{{\bf pale blue}})
 and a fit by a discrete Gaussian distribution for the PDF of $N_e(u)$ 
 (\textcolor{palered}{{\bf pale red}}). {\it Inset:} The PDF of the 
 shift $\Delta N(u)$ in each realization. Standard errors of the mean for $N_{vis}(u)$ and $N_{e}(u)$ are smaller than the marker size.}
\lb{multi-pdf} \end{figure}

We first observe in Fig.\ref{multi-pdf} that the equilibration shellnumber $N_{e}(u)$ 
does fluctuate substantially from realization to realization, with non-vanishing 
probability to take values from -2 to 6 in our ensemble of events gathered 
from a simulation of 300 large-eddy turnover times. We have no theoretical 
prediction for the PDF of  $N_{eq}(u)$, but we observe that it is fit fairly well 
in this core range by a standard discrete Gaussian PDF \cite{agostini2019discrete} 
of the form $p(n)=e^{-(n-2)^2/2}/\Theta,$ plotted as well in Fig.~\ref{multi-pdf}. 
Most interestingly, the PDF's of $N_{vis}(u)$ and $N_e(u)$ are roughly 
similar in form, but that for $N_e(u)$ is shifted to the right by $2-3$
shells. This observation suggests that the same dynamics is responsible 
for the fluctuations in both $N_{vis}(u)$ and $N_e(u).$ This supports the 
picture we have proposed that inertial-range intermittency leads to a fluctuating 
viscous cutoff shellnumber, above which the amplitude of $u_n$ drops drastically 
in $n$ with the stretched-exponential decay \eqref{stretch-decay} which is the same 
for all realizations. In that case, within a small number of shells which 
is nearly independent of $u$ the amplitude of the shell velocity drops 
to the level $u_{th}$ where thermal equipartition is achieved. As a further 
test of this explanation, we have also calculated with our DNS the PDF 
of the shift $\Delta N(u)=N_e(u)-N_{th}(u)$ in each realization, with 
the result plotted in the inset of Fig.~\ref{multi-pdf}. This data 
shows a probability of about $2/3$ for $\Delta N=3,$ and considerably 
smaller probabilities $\doteq 2/9$ for $\Delta N=2$ and $\doteq 1/9$ 
for $\Delta N=4.$ We conclude that the shell velocity amplitude indeed 
drops to the equipartition level $u_{th}$ about 3 shells above the 
viscous cutoff shellnumber $N_e(u)$ in every realization. The intermittency
that appears in the dissipation range of the noisy model thus appears 
to be imprinted by small- and large-amplitude events that propagate down
from the inertial range.  The few events observed with $N_e(u)=-2$ 
may be described as very low-intensity ``lulls'' and the handful of events 
with $N_e(u)=6$ as extreme amplitude ``bursts''.  

The burst event $u$ that we considered in our convergence studies 
in sections \ref{dt-converge}-\ref{N-independ} is 
\textcolor{black}{the only realization} with $N_e(u)=6$ that we encountered in our long 
numerical simulation over 300 large-eddy turnover times. As discussed there, even 
this largest value $N_e(u)=6$ would correspond in the ABL 
to an ``equilibration length'' about 124 times larger than the mean-free path of air. 
Such a significant separation in scales suffices to justify the validity of a hydrodynamic 
description even for such extreme events. Of course, we cannot rule out that running 
our shell model dynamics for much longer times would produce much more intense 
events still, with $N_e(u)>6.$ To properly identify the most extreme event at 
our given Reynolds number $Re$ and dimensionless temperature $\theta_\eta$ which 
achieves the largest value $N^\star_e$ (possibly =$\infty$) would require specialized tools 
of rare-event sampling beyond the scope of the current work. However, more
systematic investigation of such extreme behaviors is important work for the future. 

The results of the present section demonstrate that sizable intermittency 
remains in the dissipation range of our stochastic shell model with thermal noise.
Our numerical results are consistent with the hypothesis that these strong fluctuations 
are due to intermittent events, ranging from ``lulls'' to ``bursts'', that propagate in 
from the inertial-range. However, a quantitative relationship remains to be established
with inertial-range scaling.  Furthermore, it remains to be understood how such intermittency 
manifests in standard statistical averages and at what wavenumber scale. 
All of these issues are addressed in the following section. 

\subsection{Structure Functions}\lb{sec:strfun} 

\begin{figure*}
    \centering
    \begin{subfigure}[b]{0.475\textwidth}
        \centering
        \includegraphics[width=\textwidth]{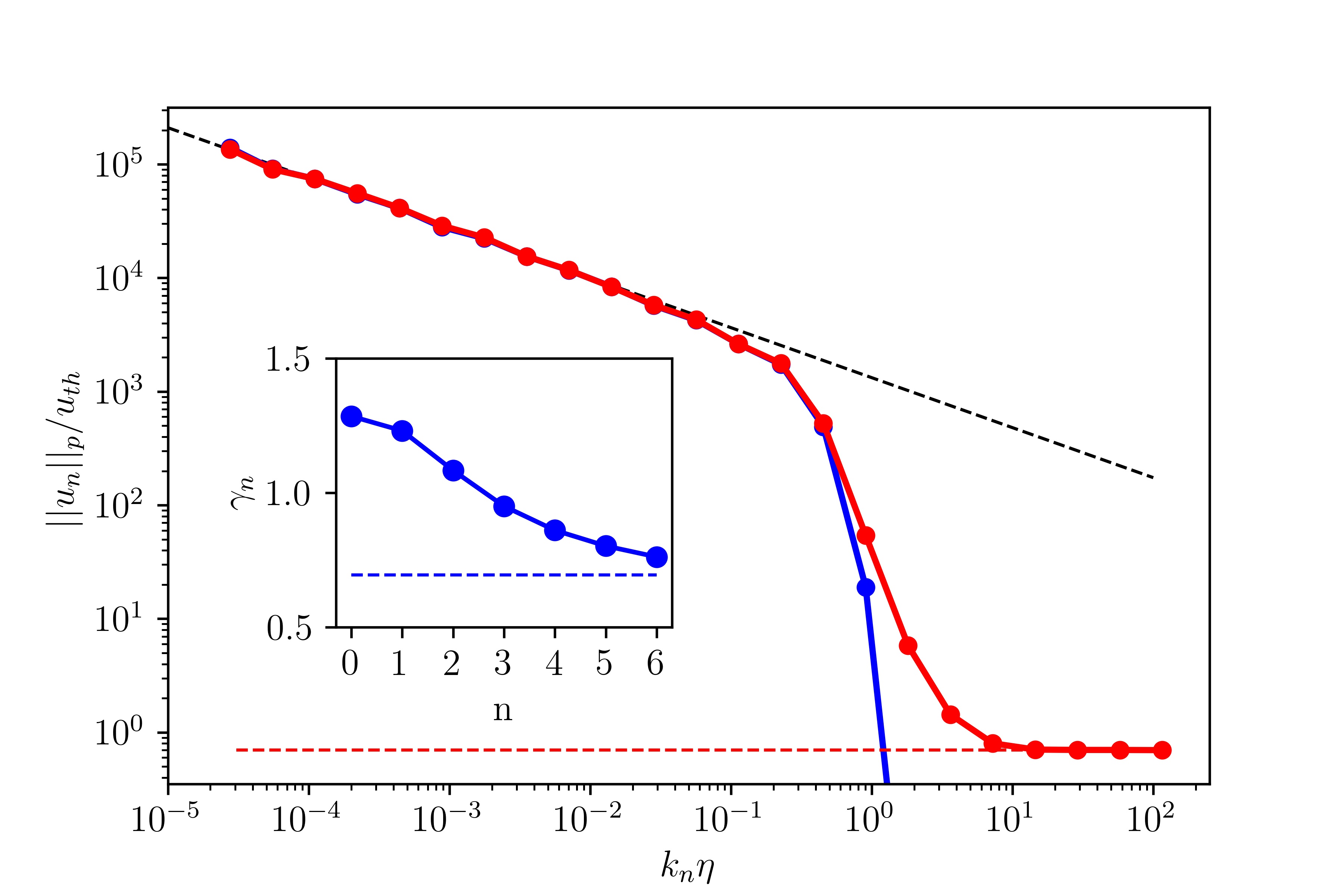}
        \caption{} 
    \end{subfigure}
    \hfill
    \begin{subfigure}[b]{0.475\textwidth}  
        \centering 
        \includegraphics[width=\textwidth]{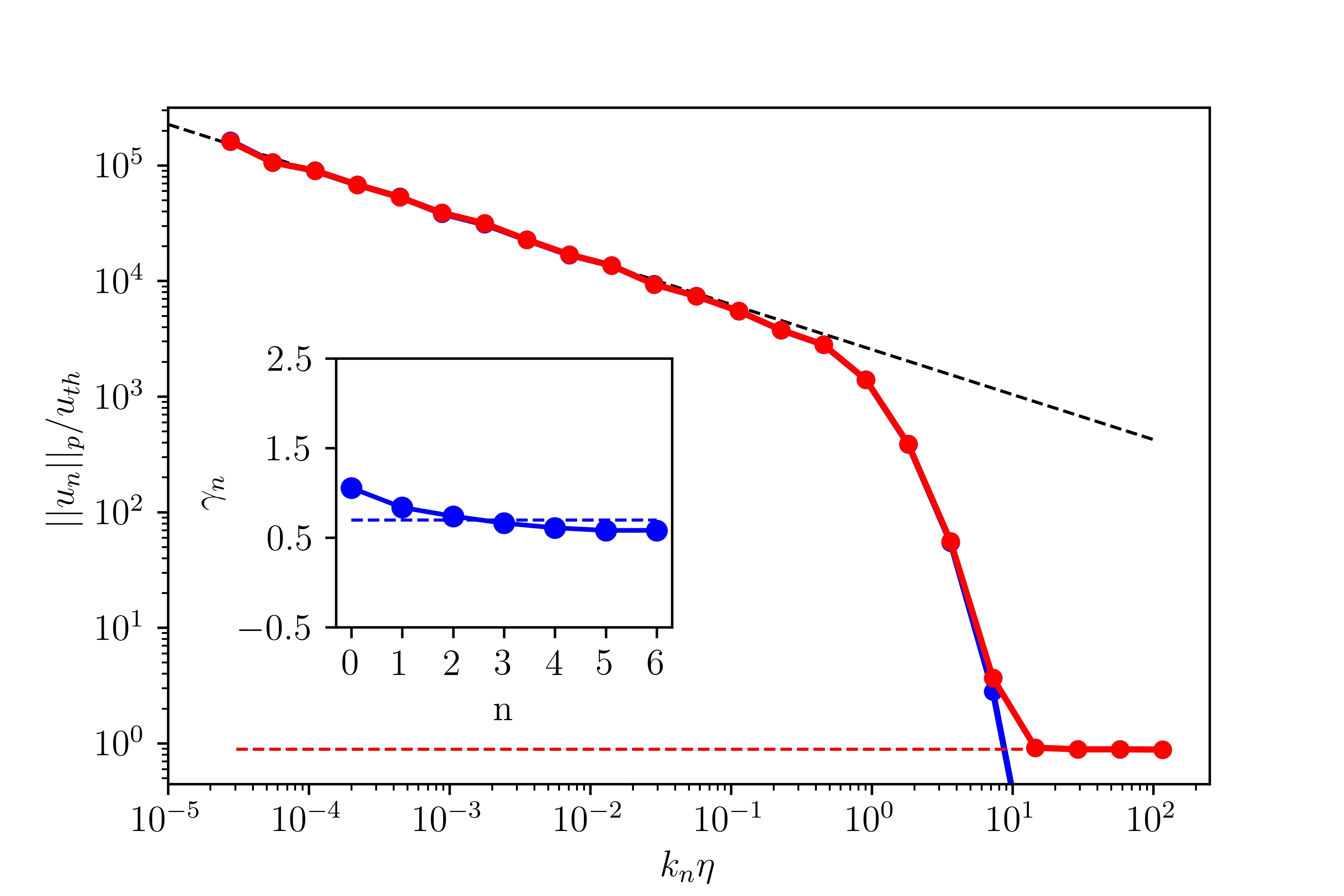}
        \caption{} 
    \end{subfigure}
    \vskip\baselineskip
    \begin{subfigure}[b]{0.475\textwidth}   
        \centering 
        \includegraphics[width=\textwidth]{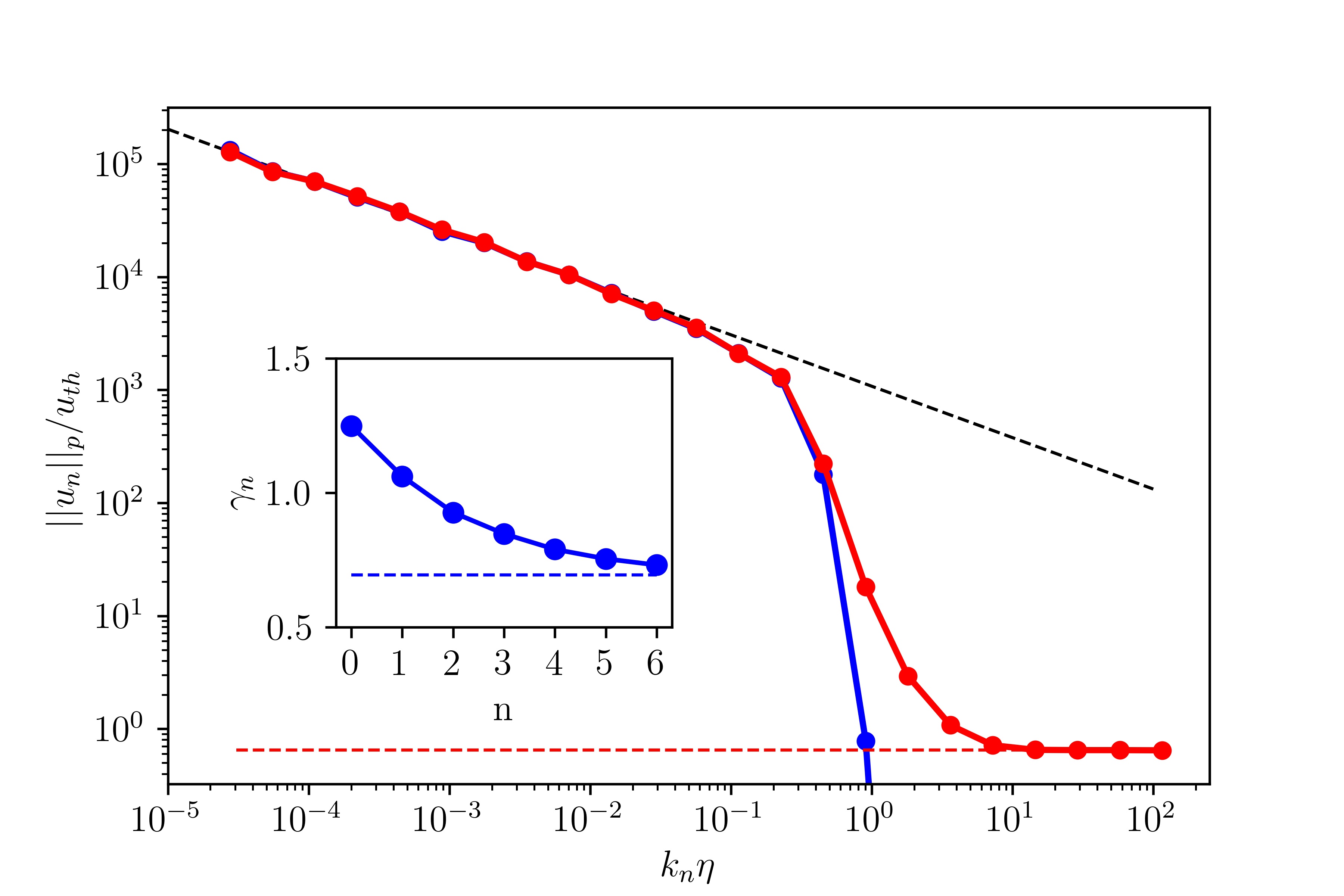}
        \caption{} 
    \end{subfigure}
    \hfill
    \begin{subfigure}[b]{0.475\textwidth}   
        \centering 
       \includegraphics[width=\textwidth]{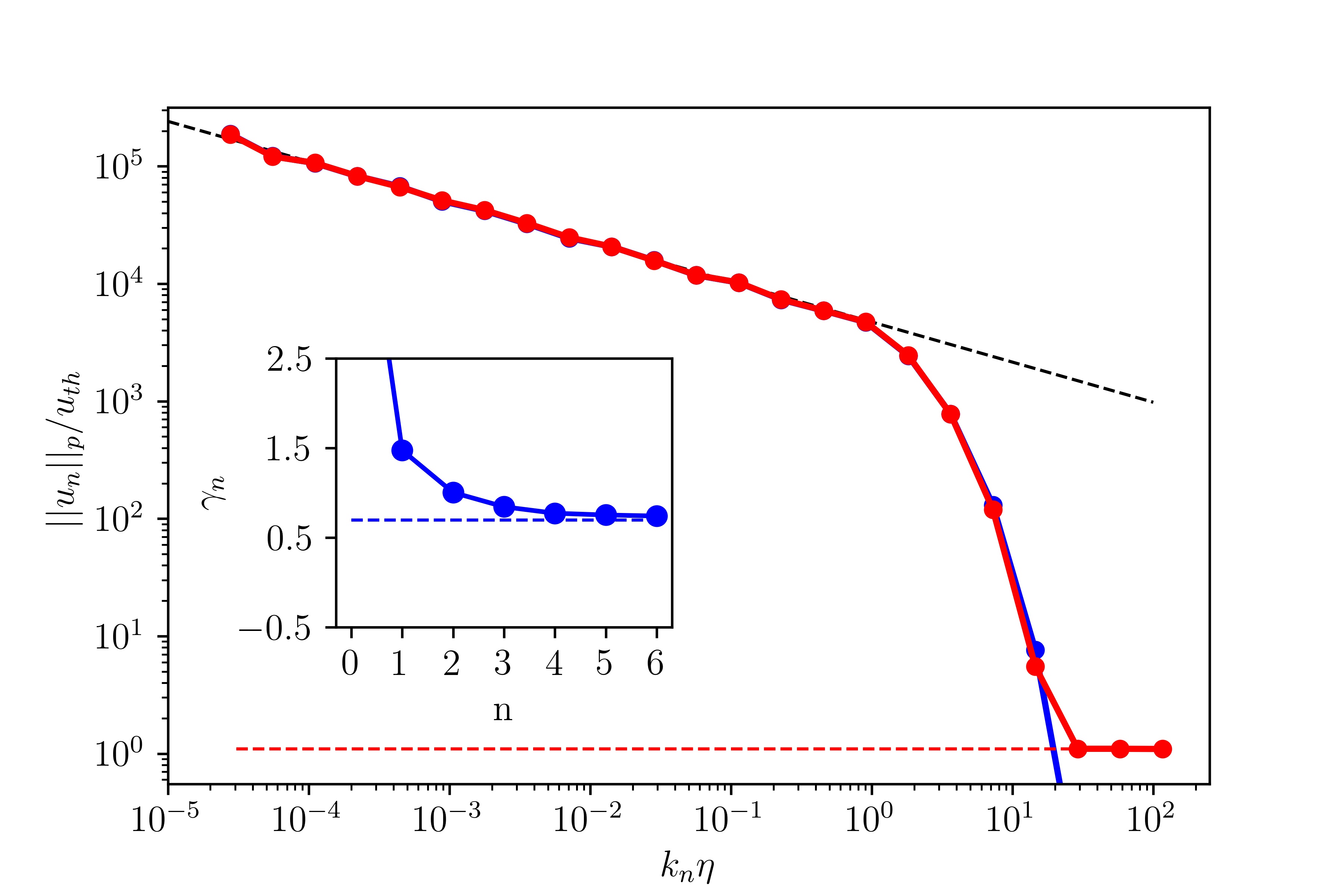}
        \caption{} 
    \end{subfigure}
        \vskip\baselineskip
    \begin{subfigure}[b]{0.475\textwidth}   
        \centering 
        \includegraphics[width=\textwidth]{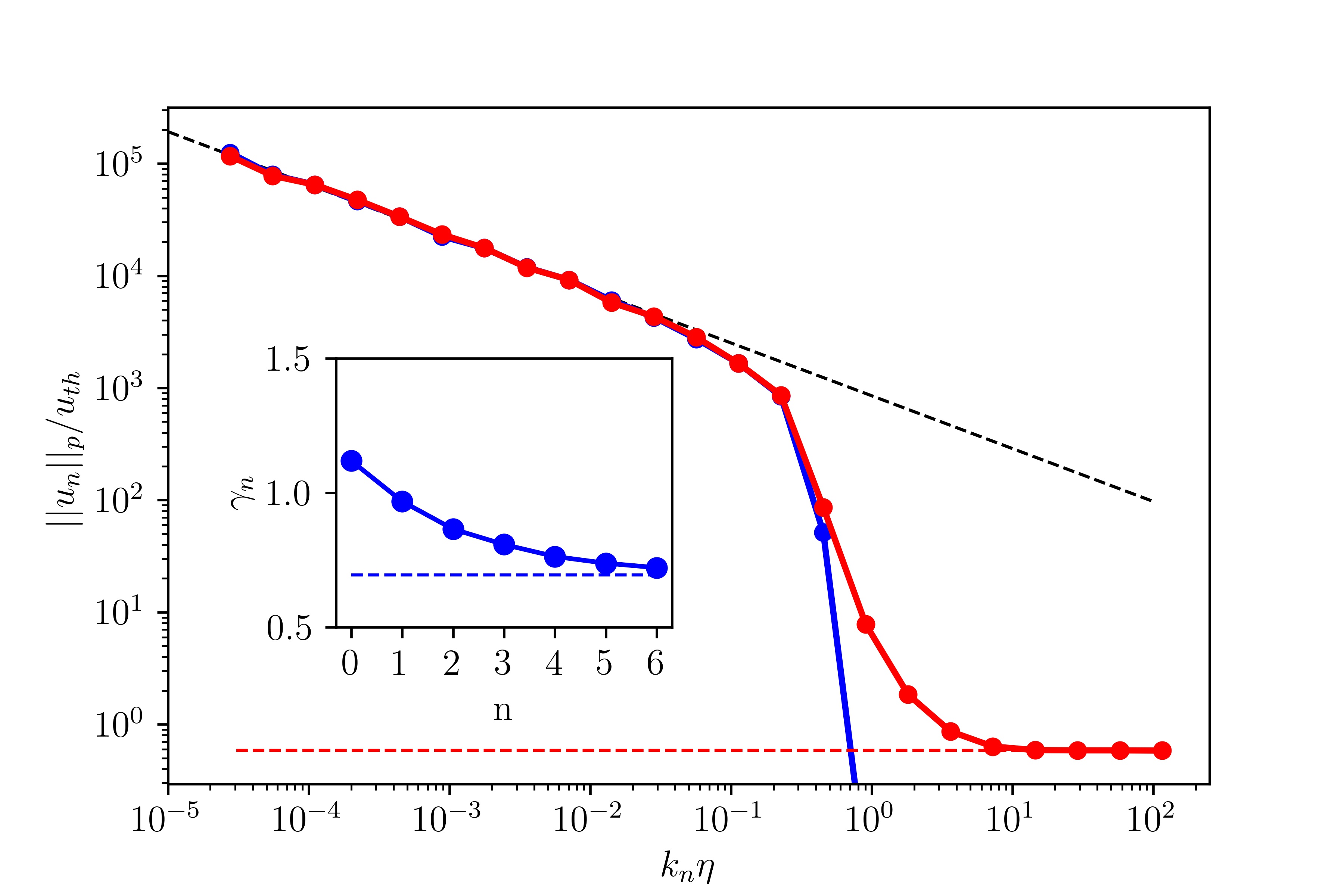}
        \caption{} 
    \end{subfigure}
    \hfill
    \begin{subfigure}[b]{0.475\textwidth}   
        \centering 
        \includegraphics[width=\textwidth]{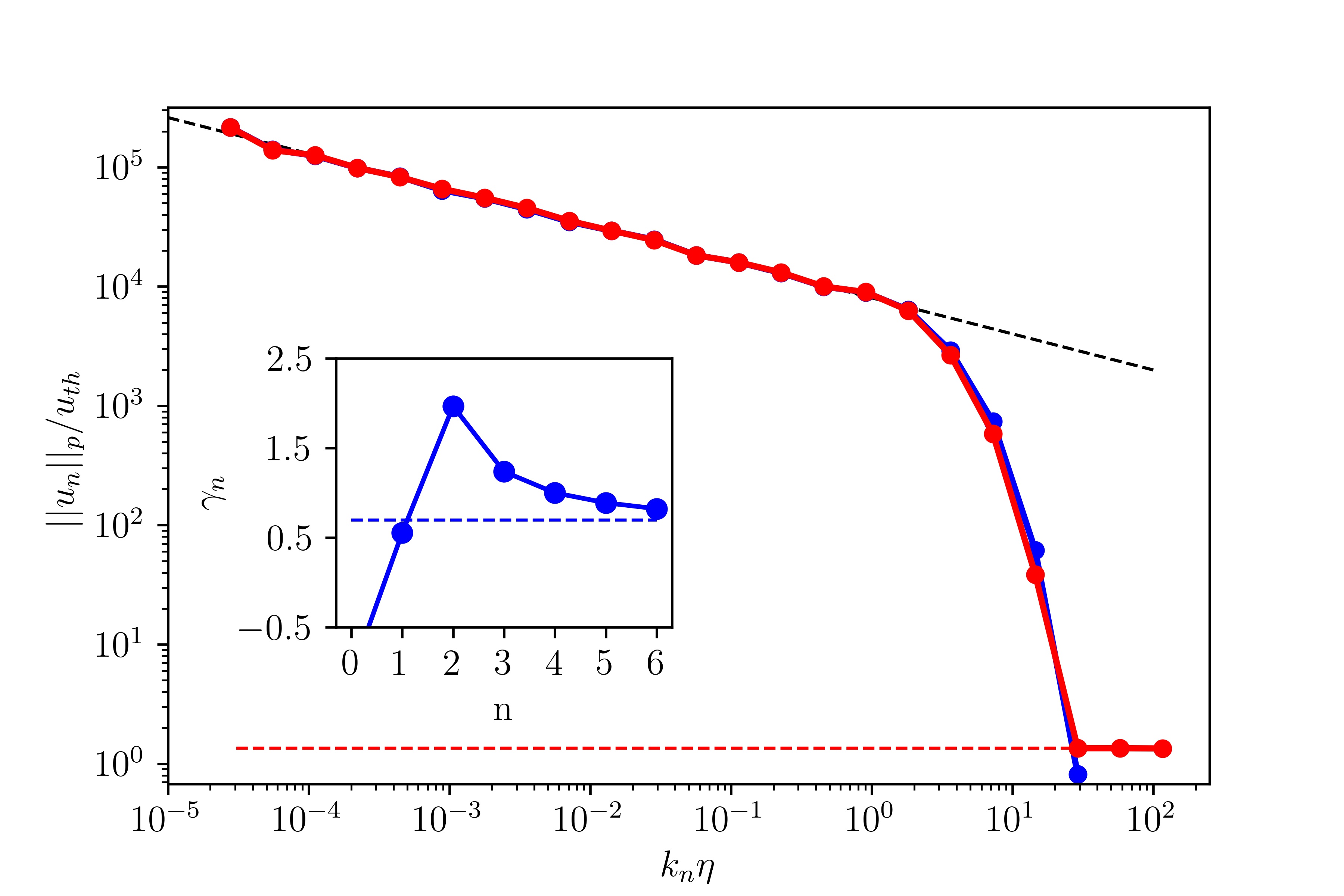}
        \caption{} 
    \end{subfigure}
    \caption{Structure functions \eqref{p-norm} for the deterministic model
    (heavy blue lines, \textcolor{blue}{\hvline}) and for the noisy model
    (heavy red lines, \textcolor{red}{\hvline}). The predicted equipartition levels $[\Gamma(1+p/2)]^{1/p}$
    are indicated by the horizontal dashed red lines (\textcolor{red}{-$\,$-$\,$-}) and power-law fits in the 
    inertial range are plotted as black dashed lines (-$\,$-$\,$-). The left panels show 
    negative values {\bf (a)} $p=-0.3$, {\bf (c)} $p=-0.6$, {\bf (e)} $p=-0.9$, 
    and the right panels show positive values {\bf (b)} $p=1$, {\bf (d)} $p=3$, {\bf (f)} $p=6.$ 
    Standard errors of the mean for structure functions are smaller than the marker size.
    The insets show with heavy blue line ( \textcolor{blue}{\hvline}) the 
    local stretching exponent \eqref{local-stretch-p} and with horizontal dashed blue line (\textcolor{blue}{-$\,$-$\,$-})theoretical prediction $\gamma=\log_2\left(\frac{1+\sqrt{5}}{2}\right).$}
\lb{strucfun:fig} \end{figure*}

In order to study intermittency effects across both inertial- and dissipation-ranges in our simulations, 
we use $p$-th-order structure functions, which we define here as statistical ``$p$-norms'' 
\be \|u_n\|_p=\langle |u_n|^p\rangle^{1/p} \lb{p-norm} \ee 
of the shell node $u_n.$ The additional $p$th-root in \eqref{p-norm} compared with the standard 
definition makes comparing results for different choices of $p$ more transparent. Although 
these are norms literally only for $p\geq 1,$ we consider all real values of $p>-1,$ because
negative $p$ values give information about rare events which are smoother or more regular 
than typical. Similar information could be obtained also from so-called ``inverse 
structure functions'' \cite{roux2004dual} but we confine ourselves here to the more 
traditional direct structure functions. Note that the shell-model quantities \eqref{p-norm} 
correspond only in the inertial-range to the standard structure-functions defined by moments 
of velocity-increments in Navier-Stokes turbulence, but more generally they 
correspond to structure-functions of a suitably band-passed velocity field $\bu_n(\bx,t)$
at wavenumber magnitudes $k_n\sim 2^n k_0.$ This distinction is crucial in the turbulent 
dissipation range, where first-order velocity-increments are completely dominated 
by the linear term in their Taylor series and are an inadequate tool to probe 
the energy spectrum and intermittency effects. An empirical study of incompressible 
fluid turbulence which aimed to investigate the same physics issues that we do here 
for shell models would need to employ a band-pass filter kernel $f_n(\bk)$ with 
very rapid decay for $|\bk|<k_n,$ optimally vanishing identically for $|\bk|<c k_n$
with some constant $c>0$.  

In Fig.~\ref{strucfun:fig} we plot the structure functions $\|u_n\|_p/u_{th}$ 
versus $n$ for the deterministic and noisy models, both normalized by the thermal 
velocity $u_{th}=(2\theta_\eta)^{1/2}$ in Kolmogorov units. With this normalization, 
the thermal equipartition value of the $p$th-order function is $\left[\Gamma\left(1+\frac{p}{2}\right)\right]^{1/p}$
in terms of the Euler $\Gamma$-function. We plot the structure functions for six representative 
values of $p$, both positive and negative, over the entire range of $n.$ 
The first important observation is that the structure functions for the deterministic 
and noisy models are identical in the inertial-range, within numerical accuracy.
This is not unexpected, because the thermal noise is extremely weak in the inertial
range and the direct effect on the dynamics should be negligible. The $p$th-order
functions here exhibit power-law scaling $\propto k_n^{-\sigma_p}$, which is indicated 
by the straight-line fits in the log-log plots in Fig.~\ref{strucfun:fig}. 
These exponents are related by $\sigma_p=\zeta_p/p$ to the standard structure-function
scaling exponents $\zeta_p,$ and the values obtained from our linear fits agree 
for $p>0$ with those previously appearing in the literature. See Fig.~\ref{sigexp:fig} 
where we plot our values together with those reported in \cite{lvov998improved}, 
showing agreement within our error bars \footnote{Whereas we averaged over 300
large-eddy turnover times, \cite{lvov998improved} averaged over several thousand such 
times. We could not afford to do likewise, because our much better-resolved dissipation
range forced us to use a much smaller time-step.}. 
The decrease of $\sigma_p$ with the order $p$ is a reflection of temporal 
intermittency in the shell model dynamics, which has been long understood to
be associated with strong ``bursts'' that propagate from low to high wavenumbers
\cite{siggia1978model,pisarenko1993further,mailybaev2013blowup}. This inertial-range  
dynamics is essentially unaltered by the presence of weak thermal noise. 

 \begin{figure}[t!]
 \begin{center}
 \includegraphics[width=260pt]{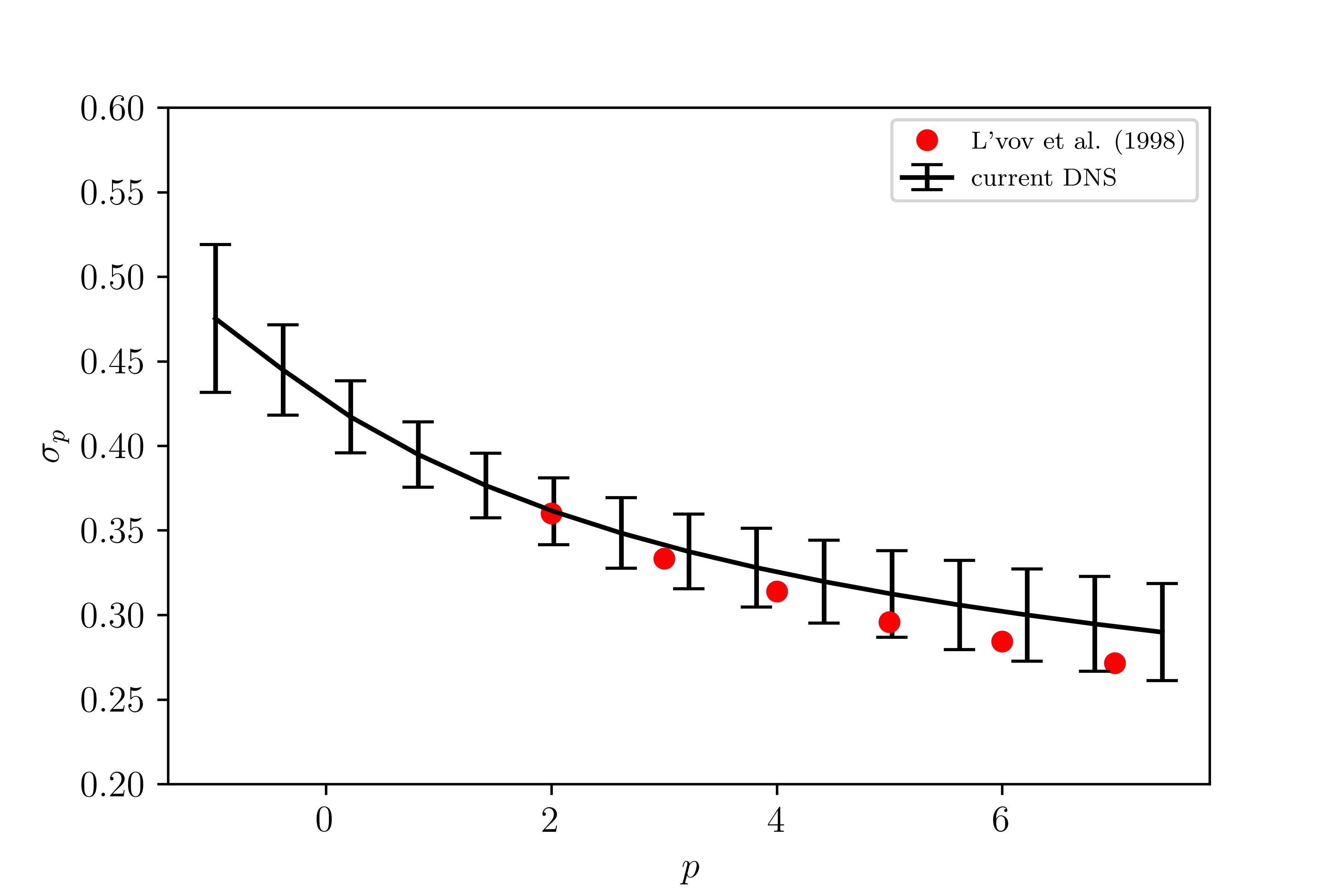}
 \end{center}
 \caption{The scaling exponents $\sigma_p=\zeta_p/p$ of structure functions $\|u_n\|_p$
 for 210 values of $p$ between $p=-1$ and $p=8,$ from our DNS 
 (solid black line, \hvline). The values for $\sigma_p$ from \cite{lvov998improved} 
 for $p=1-7$ are plotted as solid red circles (\textcolor{red}{$\bullet$).} 
 \textcolor{black}{The errors are calculated as a sum of two sources, 
 presumed independent. The first source of error is calculated by restricting the range 
 of the linear fit from 10 to 6 shells, shifting it within the original 10 shells, and taking the 
 standard deviation of all the obtained slope values as the dominant source of the error. 
 The second source of error is calculated by sub-sampling the data set into 10 parts, 
 calculating the scaling exponents separately for each, and finding 
 the standard deviation of the obtained exponents. The systematically smaller 
 results obtained in \cite{lvov998improved} compared with our mean values
 are presumably due to extremely rare, intense events missed in our 
 sample of 300 large-eddy turnover times but encountered in the longer runs of \cite{lvov998improved}.}}
\lb{sigexp:fig} \end{figure} 

In the dissipation range, however, the $p$th-order structure functions of the 
deterministic and noisy models are entirely different. Analogous to what is observed 
in the pair of energy spectra ($p=2)$ plotted in Fig.~\ref{comp-spectrum}, the structure 
functions for the noisy model approach thermal equipartition at high wavenumbers 
whereas the same functions for the deterministic model exhibit a super-algebraic 
decay. In fact, the $p$th-order structure functions of the deterministic model for all 
$p$-values exhibit the same stretched-exponential decay \eqref{stretch-decay} as 
does the energy spectrum. This is verified in the insets of Fig.~\ref{strucfun:fig}, which 
plot the local stretching exponents: 
 \be \gamma_n^{(p)} = \log_2 \left|\ln\|u_{n+1}\|_p\right|- \log_2 \left|\ln\|u_{n}\|_p\right| \lb{local-stretch-p} \ee 
and the theoretical prediction $\gamma=\log_2\left(\frac{1+\sqrt{5}}{2}\right).$ Especially 
the negative $p$-values and $p=1$ exhibit the predicted stretched-exponential, whereas 
$p=3$ and $p=6$ only approach this behaviour for large $n.$ The plausible explanation 
is that the intense ``bursts'' which dominate the structure functions for $p=3$ and $6$
penetrate to very high wavenumbers, so that for our deterministic simulation the cutoff 
$R=7$ is too small to capture well the stretched-exponential. For $-0.9<p<1,$ 
on the other hand,  the ``lulls'' or low-intensity events which dominate those $p$-values 
end at relatively low wavenumbers, so that the stretched-exponential above that wavenumber 
is well-resolved. The wavenumbers where the asymptotic behaviors first appear, equipartition or 
stretched-exponential, likewise strongly depend on the order $p$ and, from Fig.~\ref{strucfun:fig}, 
clearly increase with $p.$ This is directly associated with the strong 
fluctuations in the equipartition shellnumber $N_e(u),$ observed in the statistics 
in Fig.~\ref{multi-pdf}. In consequence, the negative-order structure functions which get
dominant contributions from ``lulls' with $N_e(u)\leq 0$ show direct effects 
of thermal noise at the Kolmogorov wavenumber and even lower wavenumbers.
This should be true also for three-dimensional hydrodynamic turbulence, but obtaining 
those structure functions from a laboratory experiment would require extremely 
accurate velocity measurements, since the thermal velocity $u_{th}$ which sets 
the floor will be $3-4$ orders of magnitude smaller than the Kolmogorov velocity. 

To establish a quantitative connection with inertial-range scaling, we have made 
also a very simple multifractal model of the PDF of the viscous cutoff shellnumber 
$N_{vis}(h),$ with the ansatz (in Kolmogorov units) 
\be N_{vis}(h)={\rm Round}\left[\log_2(Re)\left(\frac{1}{1+h}-\frac{3}{4}\right)\right], 
\lb{Nvis-MF} \ee 
where ``Round'' denotes rounding to the nearest integer, and where the 
PDF of the H\"older exponent $h$ is taken to be $P(h)\propto Re^{D(h)/(1+h)}$
for multifractal dimension spectrum $D(h).$ Using the Legendre transform relation 
$D(h)=\inf_p\{ph-\zeta_p\},$ we have determined the dimension spectrum from 
our numerical results for $\zeta_p,$ with the resulting model PDF of $N_{vis}$ also 
plotted in Fig.~\ref{multi-pdf}. Although our model is less sophisticated than 
the corresponding multifractal model developed for the PDF of $\eta(\bx,t)$
in 3D fluid turbulence \cite{biferale2008note}, it matches reasonably well the 
PDF of $N_{vis}(u)$ from our DNS. This is consistent with an earlier study 
\cite{bowman2006links} verifying in the GOY shell model the predictions 
of the multifractal model for an ``intermediate dissipation range'' 
\cite{frisch1991prediction}, since those predictions are based on the same 
assumption \eqref{Nvis-MF}. We conclude that the fluctuating viscous 
cutoff $N_{vis}(u)$ in our noisy Sabra model has statistics plausibly 
associated to inertial-range intermittency of ``lulls'' and ``bursts'', and 
the equipartition shellnumber $N_{eq}(u)$ follows suit, occurring generally 
about just three shells higher.

\section{Discussion and Conclusions} \lb{sec:conclusion}

The main claim of this paper is that thermal noise fundamentally modifies the 
far-dissipation range of turbulent flow, leading to a thermal equipartition 
range in the turbulent energy spectrum at length-scales about $\eta/10,$ 
one-tenth of the Kolmogorov length $\eta$, or even larger. If so, the correct equation 
to describe low-Mach-number fluid turbulence down to sub-Kolmogorov scales 
is not incompressible Navier-Stokes, but is instead the fluctuating hydrodynamics 
of Landau-Lifschitz \cite{landau1959fluid} in its incompressible limiting form 
\cite{forster1976long,forster1977large,usabiaga2012staggered,donev2014low,nonaka2015low}. 
\black{This conclusion was already anticipated in the pioneering papers by Betchov 
\cite{betchov1957fine,betchov1961thermal}.}  
We have \black{further} explained why standard scaling arguments \cite{bardos1991fluid,bardos1993fluid,
quastel1998lattice} for validity of the deterministic Navier-Stokes equation 
at arbitrarily high Reynolds numbers do not contradict our conclusions. 
We have also discussed interactions of turbulent intermittency with thermal noise
effects that should lead to large spatiotemporal fluctuations in the length-scale 
at which thermal equipartition occurs in individual realizations. Finally, we have 
verified our various theoretical conclusions by simulations with a Sabra shell model 
of fluctuating hydrodynamics. 

More questions are certainly raised by our results in this 
paper than can be currently answered definitively, and there is an urgent need for new 
computations, laboratory experiments, physical theory, and rigorous mathematical analysis 
to address them. \black{Many of these questions were raised already by Betchov in his 
early works \cite{betchov1957fine,betchov1961thermal,betchov1964measure}, as we discuss further
below, and remain completely open to the present day.} 

\subsection{\red{Relations to Prior Theory}}\lb{object}  

\red{First, however, it is important to discuss relations of our work to earlier studies.
There is a large body of work attempting to define fluctuating hydrodynamics equations 
of the form \eqref{FNS} as continuum stochastic partial-differential equations (SPDE's), with 
the view that this is necessary for the understanding of turbulence. For example, the excellent 
book on SPDE's in hydrodynamics \cite{albeverio2008spde} states in its preface that 
\begin{quotation}
``In a sentence, one of the purposes of the course was to understand the link between the 
Euler and Navier-Stokes equations or their stochastic versions and the phenomenological 
laws of turbulence.''
\end{quotation} 
and the chapters of this book attempt to make mathematical sense of equations like 
\eqref{FNS} as continuum SPDE's. This was also the point of view of the earlier paper on the 
stochastic shell model \cite{bessaih2012invariant} which showed how to make sense of equations 
\eqref{nSabra} in the limit $N\to\infty$ (for thermal equilibrium).} An exact mathematical solution 
of this problem \red{for fluctuating hydrodynamics} would allow the limit $\Lambda\to\infty$ formally 
to be taken so that the equation \eqref{FNS} would make sense as an SPDE with the invariant 
measure \eqref{Gibbs}. This is a problem of the same nature as the non-perturbative construction 
of a renormalized quantum field-theory and it has only been solved for simpler models such as the
KPZ equation \cite{kardar1986dynamic}, where it is already extremely difficult \cite{hairer2013solving}. 
\red{We have argued that this point of view is, in fact, physically incorrect for fluctuating hydrodynamics 
and that at any finite Reynolds number, however large, the equations \eqref{FNS} must be regarded 
as low-wavenumber ``effective field theories'' with an explicit UV cut-off $\Lambda$, somewhat arbitrary,  
but chosen to satisfy the constraints $1/\ell_\nabla\ll \Lambda\ll 1/\lambda_{micr}.$ This point 
of view is not novel, of course, but is standard in the field of fluctuating hydrodynamics
\cite{forster1976long,forster1977large,zubarev1983statistical,morozov1984langevin,espanol2009microscopic,
usabiaga2012staggered,donev2014low}} 

\red{There have also been renormalization group analyses of stochastically forced 
Navier-Stokes fluids to study systematically the effect of changing $\Lambda,$ 
notably by Forster et al. \cite{forster1976long,forster1977large},
who included the case with stochastic force representing thermal noise as their ``Model A''.  This paper 
carried out a Wilson Fourier-slicing RG analysis of the thermal fluid at equilibrium obtaining 
scale-dependent viscosity $\nu_\ell$ and temperature $T_\ell$ with effective dimensionless 
nonlinear coupling constant of the velocity fluctuations at scale $\ell$ given in space dimension $d$ by 
\be g_\ell :=\left(\frac{k_BT_\ell}{\rho\nu_\ell^2\ell^{d-2}}\right)^{1/2}  \lb{g-def} \ee
These authors reached the conclusion that the thermal fluid is IR asymptotically 
free for $d>2$, with coupling $g_\ell\to 0$ for $\ell\to\infty,$ corresponding to a linear Langevin 
model for relaxation of long-wavelength fluctuations (Onsager regression hypothesis). 
Conversely, they concluded that the thermal fluid is UV strongly coupled,
with the constant $g_\ell$ becoming large for $\ell\to\infty,$ and they equated this strong-coupling 
regime with turbulence:   
\begin{quotation} 
``we will not attempt to treat the formidable and probably more interesting problem of the ultraviolet 
(short-distance, short-time) correlations described by (2.1), i.e., of fully developed turbulence...''
 \cite{forster1977large}
\end{quotation} 
Exactly the opposite situation was shown to hold for $d<2$ by \cite{forster1976long,forster1977large},
with now thermal velocity fluctuations UV asymptotically free and IR strongly coupled. As discussed 
in section \ref{shell-intro},} the shell model that we study numerically in this work 
is effectively a model in 0 space dimension \red{ and, according to the RG analysis of 
Forster et al., it is UV asymptotically free, so that the shell variables $u_n$ are described 
by independent linear Langevin models for very large $n.$ Because 3D FNS is instead 
UV strongly coupled, it might be argued that thermal fluctuations in our shell model are 
qualitatively different at high wavenumbers than those for 3D fluids and that our shell model 
is thus unsuitable to test the effects of thermal noise in the turbulent dissipation range.} 

\red{In fact, it is not hard to see that thermal velocity fluctuations both in 3D fluids and in our 
stochastic shell model are  weakly coupled at the Kolmogorov dissipation length and down 
to much smaller scales. Note that the coupling constant $g_\ell$ of Forster et al. in \eqref{g-def} 
coincides for $\ell=\eta$ with the thermal Reynolds number $Re^{th}_\eta$ which we defined 
in \eqref{Re-th} and thus $g_\eta=\theta_\eta^{1/2}$ by \eqref{k-coup-K}. 
It follows that the coupling constant $g_\eta$
at the Kolmogorov scale is tiny for realistic magnitudes of $\theta_\eta.$ The na\"\i ve belief 
expressed in  \cite{forster1977large} that turbulent flows must correspond to large coupling 
constant $g_\ell$ is not necessarily true.} 

\red{It is true that this coupling will increase as $\ell$ is further decreased, both in 
turbulent flows and even in a thermal equilibrium fluid at rest. One can estimate the 
wavenumber $k_{coup}$ where coupling becomes strong by setting $g_\ell\sim 1$ 
in \eqref{g-def} and solving for $k\sim 1/\ell,$ yielding} 
\footnote{Here we neglect, however, scale-dependence of the renormalized viscosity 
$\nu(k).$ If this viscosity decreases as $k$ grows, then $k_{coup}$ is correspondingly 
lowered. \black{A kinematic viscosity on order of magnitude $\nu\sim \lambda_{micr} c_{th}$ 
follows, as is well-known, from the kinetic theories of Boltzmann and Enskog. This is a reasonable 
estimate of the ``bare viscosity" at length scales of order the mean-free-path, before it is 
dressed or renormalized by thermal fluctuations. Indeed, it is well-known that the 
Boltzmann-type kinetic equations neglect the effects of thermal fluctuation, which 
must be incorporated by additional Langevin terms. See \cite{bixon1969boltzmann,
fox1970contributions,spohn1983fluctuation} and, for more recent work, 
\cite{bouchet2020boltzmann,bodineau2020statistical}. Furthermore, this estimate 
of the ``bare viscosity'' is in agreement with the study of \cite{donev2014low}.
See their Appendix C, Figure 8.}}
\be k_{coup}=\left(\frac{k_BT}{\rho\nu^2}\right)^{\frac{1}{d-2}}. \lb{k-coup}  \ee
Simple estimates then imply that $k_{coup}$ in \eqref{k-coup} is so large that it is strictly 
outside the regime of validity of a hydrodynamic description! To see this, we can substitute 
the standard kinetic theory estimate for kinematic viscosity $\nu\sim \lambda_{micr} c_{th}$, 
with $c_{th}$ the thermal velocity/sound speed, into \eqref{k-coup}, 
which yields 
\be k_{coup}\sim (n\lambda_{micr}^2)^{\frac{1}{d-2}} 
\sim \left(\frac{\lambda_{micr}^2}{\ell_{intp}^d}\right)^{\frac{1}{d-2}} \ee 
where $n$ is the particle number density of the fluid and $\ell_{intp}:=n^{-1/d}$
is the mean interparticle distance. In a liquid, $\lambda_{micr}\sim \ell_{intp}\sim R,$
where $R$ is the radius of the molecule, and we see that $k_{coup}R\sim 1$. In a 
gas described by the Boltzmann kinetic equation, $k_{coup}$ is even larger.
The low-density limit for validity of the Boltzmann equation first identified 
by Bogolyubov \cite{bogolyubov1946problem} and Grad \cite{grad1949kinetic} 
is that in which $\lambda_{mfp}/\ell_{intp}\gg 1$ and $\ell_{intp}/R\gg 1$
with $\lambda_{mfp}\sim1/nR^{d-1}$ held fixed. It is then easy to see that 
for such a Boltzmann gas $k_{coup}R\sim (\ell_{intp}/R)^{\frac{d}{d-2}}\gg 1.$
These considerations suggest that the strong-coupling problem encountered
in the limit $\Lambda\to\infty$ is of only academic mathematical interest and 
is not relevant to the physical description of molecular fluids in thermal 
equilibrium. 

\red{It is worth pointing out that the limit $\Lambda\to\infty$ reappears in a different guise
when taking the infinite Reynolds-number limit, $Re=UL/\nu\to\infty$ with 
mean dissipation $\varepsilon=U^3/L$  and fluid parameters $\rho,$ $T$ all held fixed.
In that case, $\widehat{\Lambda}:=L\Lambda>L/\eta\to\infty$ so that the UV cutoff 
diverges to infinity when the fluctuating hydrodynamic equations are non-dimensionalized 
with the integral-scale quantities $L$ and $U.$ This limit will be the subject of our following 
work \cite{bandak2020spontaneous}. Here we just note that, assuming the validity 
of the Landau-Lifschitz equations \eqref{FNS} for arbitrarily large values of $Re,$
we find in \cite{bandak2020spontaneous} that the limiting velocity fields are singular (weak) 
solutions of the {\it deterministic} Euler equation. Thus, both the molecular noise and 
the molecular viscosity vanish in this limit. It is not entirely clear, however, that the 
fluctuating hydrodynamic equations \eqref{FNS} do remain valid for very large $Re,$
because increasing intermittency could allow extreme turbulent singularities to 
reach down to the microscopic length-scale $\lambda_{micr}.$ This possibility will 
be discussed more in section \ref{validity} below.}

\red{Finally, there has been much 
work on thermalization and equipartition spectra in various mathematical model fluid problems, 
especially truncated Euler 
\cite{kraichnan1975remarks,cichowlas2005effective,krstulovic2008two,ray2011resonance,murugan2021many}, 
but also truncated Burgers 
\cite{majda2000remarkable,venkataraman2017onset,dileoni2018dynamics,murugan2020suppressing}, 
hyperviscous Navier-Stokes \& Burgers 
\cite{frisch2008hyperviscosity,frisch2013real,banerjee2014transition,agrawal2020turbulent}, 
and shell models \cite{ditlevsen1996cascades,gilbert2002inverse,levant2010statistical,tom2018revisiting}. 
For a good overview of this large literature, see \cite{ray2015thermalized}.  
In these various deterministic models, equipartition energy spectra ($\sim k^{d-1}$ for dimension $d$)
and Gaussian thermal statistics have been observed over certain ranges of wavenumber, rather similar to our 
observations. None of these works, however, have included stochastic terms to model the effects of thermal noise. 
There have been a few prior works on stochastic shell models, such as 
\cite{bessaih2012invariant,matsumoto2014response}, but with different noise than ours and with very different goals. 
Our aim in this work  has been to use our stochastic shell model as a surrogate for  3D fluctuating 
Navier-Stokes equation, to assess the effects of thermal noise in the turbulent dissipation range of 
molecular fluids and to understand the interactions between thermal noise and turbulent intermittency. 
As we have argued in depth, the shell model is suitable for this purpose. It is theoretically interesting 
that truncated 3D Euler can mimic many of the features of 3D FNS. Not only does the bath of thermalized 
hydrodynamic modes at high wavenumbers create an ``efffective viscosity''  
\cite{kraichnan1975remarks,cichowlas2005effective,krstulovic2008two} but also,
following ideas of Kraichnan \cite{kraichnan1970convergents,kraichnan1975remarks},
it should create an ``effective noise'' satisfying a fluctuation-dissipation relation.  Nevertheless,
the predictions of 3D truncated Euler differ in several ways from those of 3D FNS, in particular lacking a viscous 
dissipation range at intermediate scales. Most significantly, 3D truncated Euler has never 
been proposed as a realistic model 
of the dissipation range of a molecular fluid, whereas 3D FNS is expected to be an accurate mesoscopic
model down to almost microscopic length scales.}

\subsection{\red{Prospects for Empirical Verification}}\lb{verify}  

\red{The most important question raised by our work
%Foremost among these questions 
is the existence of the predicted thermal equipartition range in the sub-Kolmogorov 
scales, which we argued supplants the traditional ``far-dissipation range'' of deterministic 
Navier-Stokes. We thus find that deterministic Navier-Stokes \eqref{NS} and 
fluctuating Navier-Stokes \eqref{FNS} make two radically different sets of predictions 
for the turbulent dissipation range, and it must now be determined which is correct.}  
Current numerical codes for solving the incompressible fluctuating hydrodynamic equations 
\cite{usabiaga2012staggered,donev2014low} are adequate to investigate 
turbulent flows at Taylor-scale Reynolds numbers up to 100 or so and 
such simulations should provide additional confirmation of our predictions. 
In fact, %we have been informed that a thermal $k^2$-spectrum does appear
% in preliminary computations of this type, at length-scales right around $\eta$
% (J. Bell, private communication).
 \red{since the original submission of this paper, a preprint  \cite{bell2021thermal}  has 
 appeared which reports on such a simulation, directly motivated by our work. Although 
 those simulations reached only $Re_\lambda=143,$ that suffices 
 to test our prediction of an equipartition $k^2$ energy spectrum appearing at the 
 Kolmogorov scale and Gaussian velocity statistics rather than strong intermittency 
 in the far-dissipation range. Both of these predictions were fully verified; see 
 \cite{bell2021thermal} for details.  This numerical confirmation gives strong {\it a posteriori} 
 validation to our methodology of using the stochastic shell model \eqref{nSabra} as 
 a surrogate for 3D FNS \eqref{FNS}. Our predictions for effects of inertial-range intermittency
 cannot yet be corroborated, because 3D FNS cannot currently be solved numerically at the high 
 Reynolds numbers required. Nevertheless, our work and that of \cite{bell2021thermal}
 set the stage for a clash of two competing physical theories.}
  
Ultimately, of course, the matter must be resolved by experiment. While 
the predictions of fluctuating hydrodynamics have been verified in many globally 
far-from-equilibrium flows \cite{dezarate2006hydrodynamic} and there is little 
doubt at all that thermal noise effects must be present at sub-Kolmogorov scales, 
the detailed predictions of fluctuating hydrodynamics can be legitimately 
questioned in turbulent flows where they have not yet been measured. 
\black{The possibility exists that the local equilibrium assumption underlying the 
fluctuation-dissipation relation \eqref{FDR} could break down, as already 
noted by Betchov \cite{betchov1961thermal} (see section II.D)}. This is 
especially true since extreme turbulent intermittency could threaten the validity 
of any hydrodynamic description at all, at least locally. 

\black{More than 60 years after the early experimental attempt of Betchov \cite{betchov1957fine},
it remains} a grand challenge to develop techniques which can measure the coarse-grained 
fluid velocities in Eq.\eqref{vel-coarse} at the relevant length-scales $\ell<\eta.$
All traditional fluid-velocity measurement techniques have well-known limitations in 
achieving such fine spatial resolution. \black{Betchov himself in his study \cite{betchov1957fine}
used a standard technique of} hot-wire anemometry \cite{sadeghi2018effects} \black{to measure 
turbulent velocity fields. He made special efforts to minimize the high-frequency noise 
in the wires in order to increase their resolution and sensitivity. Furthermore, he investigated 
a novel multi-jet configuration in a ``porcupine'' box designed to create a nearly isotropic flow of high 
turbulence intensity,  avoiding the weak electrical signal due to low turbulence intensity 
in grid-turbulence. Despite these efforts, the thermal noise spectrum of the fluid velocity predicted 
by Betchov remained about four orders of magnitude below the sensitivity of his measurements.
See his Fig.6, where the highest wavenumbers of his measured spectrum are also clearly contaminated 
by electrical noise in the wire. These limitations of hot-wire technology remain to the present day.}
Here we may note that a recent study of grid turbulence in the Modane wind 
tunnel by hot-wires \cite{gorbunova2020analysis} has remarked concerning the measured 
energy spectra that ``all of them appear to increase as functions of $k$ beyond a 
wave number $k_M$'' and that ``the value of $k_M$ depends on the spectra, but is 
found to be typically $k_M \eta\simeq  3$.'' This behavior, of course, na\"\i vely accords 
with our predictions. The authors explain these observations however ``as a contamination 
by the small-scale response of the hot wires'' and we have no reason to doubt this conclusion, 
but it underlines the essential limitations of the hot-wire technology. 

Another popular set of methods to  determine fluid velocity vectors  are those 
under the rubric of ``particle-image velocimetry'' or PIV, which do so by measuring 
the displacements of small, neutrally-bouyant solid beads that seed the flow 
\cite{raffel2018particle}. It has been argued that such particles are advected 
by a fluctuating fluid velocity coarse-grained over scales comparable to their radius 
\cite{donev2014low}, although the bead obviously displaces and distorts the flow 
in its near vicinity. \black{We may note that a new ``Giant von K\'arm\'an'' (GVK) 
experiment is currently underway at CEA in France which will attempt to measure 
turbulent velocities down to scales $\sim \eta/5$ using PIV with monodisperse 
polystyrene beads of diameter $~5\ \mu$m (B. Dubrulle, private communication) 
\cite{dubrulle2021giant}. It is not entirely clear that the velocity of even a single such particle
will correspond to the velocity relevant for fluctuating hydrodynamics, which corresponds 
to the local coarse-graining \eqref{vel-coarse} over individual molecules.}
The motion of such submerged beads is known to be sensitive 
to the local thermal fluctuations of the velocity, but the effects appear only in 
the long-time tails of the particle velocity auto-correlation \cite{donev2010hybrid}. 
Furthermore, while the instantaneous velocities of micron-scale Brownian particles
have been successfully measured in quiescent flows when confined by optical traps 
\cite{li2010measurement,kheifets2014observation}, it will be very difficult to measure 
the velocity of even one such freely-advected particle, let alone many.  
\black{The large number of particles that must seed the flow in PIV introduce 
``ghost particles'' that must be disambiguated by sophisticated post-processing techniques 
that introduce additional numerical noise in the inferred velocities
\cite{tan2020introducing}}. Another  velocity measurement technique exploiting tracer particles 
is ``laser doppler velocimetry'' or LDV, which has achieved micron-scale spatial resolution in turbulent boundary 
layers \cite{czarske2013micro}, but similar difficulties of interpretation and sensitivity appear. 
The sub-Kolmogorov scales of turbulent fluid flows remain a vast {\it terra incognita} of experimental science. 

It may be more reasonable to hope first for indirect evidence of the effects of thermal noise. 
There are, in fact, many physical processes in turbulent flows which are recognized 
to involve sub-Kolmogorov scales in a fundamental way but which in a stationary, laminar 
fluid are known to be strongly influenced by thermal noise. These include high 
Schmidt/Prandtl-number scalar mixing \cite{donev2014reversible}, 
droplet and bubble formation \cite{chaudhri2014modeling,gallo2020nucleation}, 
chemical reactions (combustion) \cite{lemarchand2004fluctuation,bhattacharjee2015fluctuating}
and locomotion of micro-organisms \cite{gotze2010mesoscale}, among others. 
Current theory and numerical modelling of all these processes in turbulent flows 
omit thermal noise completely, e.g. in the case of high Schmidt/Prandtl-number scalar mixing 
\cite{donzis2010batchelor,clay2018gpu,buaria2020turbulence}, 
dynamics of droplets and bubbles \cite{saito2018turbulence,elghobashi2019direct,milan2020sub}, 
chemical combustion \cite{sreenivasan2004possible,driscoll2008turbulent,echekki2010turbulent},  
and locomotion in turbulent flows \cite{durham2013turbulence,wheeler2019not}. 
The possibility exists in all of these cases of interesting interplay between turbulence 
and thermal effects, which might yield clear experimental signatures. These problems
are all ripe also for numerical investigation by fluctuating hydrodynamics, which 
should spur development of novel schemes which are more efficient at the high Reynolds 
numbers required. 

\subsection{\red{Validity of a Hydrodynamic Description}}\lb{validity} 

\red{A key question underlying the current intense interest in extreme events and 
smallest scales in a turbulent flow \cite{yeung2015extreme,yeung2020advancing, farazmand2017variational,
buaria2019extreme, buaria2020self,buaria2020vortex,nguyen2020characterizing} is whether 
such near-singular events may lead to a breakdown in the hydrodynamic approximation.
In our shell model simulation with parameters appropriate to the ABL, the most singular 
event that we observed in 300 large-eddy turnover times penetrated down to a 
length scale $8.4\ \mu$m, which is still 124 times greater than the mean-free path length of air.
The 3D FNS model should remain valid for such a singular event in the atmosphere. However, we cannot 
rule out that even more extreme events will occur if much longer times are considered or if 
the Reynolds number is further increased. As discussed in section \ref{intermittent:sec}, 
the Parisi-Frisch multifractal model predicts that the smallest length-scale $\eta_h$ for a 
zero H\"older singularity $h=0$ will reach down to a length-scale $\lambda_{micr}/Ma$
just marginally greater than $\lambda_{micr}$ for $Ma\ll 1,$  and there is some weak evidence 
that the smallest H\"older exponent in incompressible fluid turbulence is $h_{\min}=0$ \cite{iyer2020scaling}.
However, if negative H\"older singularities arise, then scales of order $\lambda_{micr}$ or even smaller 
could be excited. This possibility seems to us quite realistic and, if it should occur, the hydrodynamic 
description would break down in the vicinity of such an extreme singularity.}  

\red{Nevertheless, we argue that Landau-Lifschitz fluctuating hydrodynamics can still be employed, 
if supplemented with some finer-scale description near the singular set.}  
%The preceding considerations illustrate 
An important point, often not 
appreciated in discussions of the dynamical equations of turbulent flow, 
%which 
is that a hydrodynamic description can be valid even when the conditions 
of its applicability are not valid globally. A relevant example, discussed previously 
in \cite{eyink2007turbulence}, section III(e), is the singularities
speculated by Leray (\cite{leray1934mouvement}, section 3) to develop starting 
from smooth initial data for the incompressible Navier-Stokes dynamics above 
some critical Reynolds number $Re_c.$
It is known that the positions of such hypothetical singularities, if they exist, 
must be a set of space-time points $(\bx_*,t_*)$ of Hausdorff dimension $\leq 1$ where 
the fluid velocity itself blows up 
as $|\bu(x,t)|\geq C/r$ for $r\to 0$ as one approaches the singularity, 
with $r^2=|\bx-\bx_*|^2+\nu |t-t_*|.$ See \cite{caffarelli1982partial}, Corollary 1. 
In fact, this conclusion can be deduced heuristically from the multifractal 
result \eqref{PVlength} for the viscous cutoff length $\eta_h,$ if one 
takes into account a constant prefactor by replacing $Re\mapsto Re/Re_c.$ 
In that case one can see that $\eta_h>0$ for any finite value of $Re,$
except that $\eta_h\to 0$ when $Re>Re_c$ and $h\to -1.$ See \cite{eyink2007turbulence},
section III(e).

We therefore conclude (in disagreement with \cite{frisch1995turbulence}, 
section 8.3) that negative H\"older singularities and locally infinite fluid 
velocities are consistent with the incompressible fluid approximation, as long 
as those equations are interpreted in the integral or ``weak'' sense, i.e. as balances 
of momentum integrated over intervals of time and over control volumes in space. 
The local breakdown of the conditions of validity of the hydrodynamic 
approximation need not violate those equations, at least if the points 
of breakdown occur in a sufficiently low-dimensional subset of spacetime. 
In that case, a hybrid description should be possible with the incompressible 
fluid equations coupled to a more microscopic model (kinetic theory, molecular 
dynamics) in the region of singularity.  
For these same reasons, we believe that the fluctuating hydrodynamic description 
invoked in this work can be valid, even if negative H\"older singularities 
develop in the infinite-$Re$ limit or perhaps already for finite $Re>Re_c.$ 
Indeed, the rigorous Onsager-Machlup large-deviations theory derived in 
\cite{quastel1998lattice} for a stochastic lattice gas in the limit of (global) 
$Kn\ll 1$ and $Ma\ll1$ is valid at any fixed value of $Re$ even if Leray 
singularities appear in the solutions of incompressible Navier-Stokes equation.
It is worth observing that the fluctuating hydrodynamic equations can be 
derived microscopically with non-uniform spatial grids \cite{espanol2009microscopic} 
so that a global wavenumber cutoff $\Lambda$ need not be imposed 
and instead local cutoff lengths can be adapted to the particular solution. 

We may note that the 
%latter 
strong subsonic flow condition, $Ma\ll 1,$ does set 
some upper limit  on the Reynolds numbers that are achievable within an incompressible fluid 
approximation. In both nature and in the laboratory, $Re=UL/\nu$ is generally made larger 
by increasing $L$ and/or $U=(\varepsilon L)^{1/3},$ so that achieving very high $Re$ at 
fixed $\varepsilon$ requires $Ma\simeq 1$ or even $\gg 1$, as in astrophysical 
turbulence environments such as the molecular gas of the interstellar medium. 
In such compressible fluid turbulence strong shock discontinuities develop with $h=0$ 
and a shock-width of order $\sim \lambda_{mfp},$ so that the fluid approximation 
breaks down locally and a Boltzmann kinetic equation is required to describe the internal 
structure of the shock \cite{mott1951solution,liepmann1962structure,salomons1992usefulness}. 
For compressible fluid turbulence, the thermal effects will be described by the 
fluctuating hydrodynamics of the compressible Navier-Stokes equation 
\cite{dezarate2006hydrodynamic,schmitz1988fluctuations,eyink1990dissipation,
zubarev1983statistical,morozov1984langevin}
or for large-$Ma$ flows with strong shocks, by the nonlinear fluctuating Boltzmann equation \cite{bouchet2020boltzmann}. It should be noted that even in the 
latter case, a hydrodynamic description is still valid at length scales $\ell\gg \lambda_{mfp},$
because at scales much larger than the shock width the dynamics is accurately 
described by a (weak) Euler solution with an idealized discontinuity  
\cite{yu2005hydrodynamic}.  As we shall discuss in our following paper, the 
general description of turbulent inertial ranges by suitable (weak) solutions of 
the Euler equations \cite{eyink2018review,eyink2018cascades,drivas2018onsager}
is unchanged by thermal noise effects in the dissipation range.

New physical theory and new mathematical analysis are however
%also 
demanded by our results.
Existing derivations of the nonlinear fluctuating hydrodynamic equations 
\cite{zubarev1983statistical,morozov1984langevin,espanol2009microscopic}
are based upon the projection-operator methods of Zwanzig-Mori \cite{zwanzig1961memory,grabert2006projection}.
Although these methods are formally exact and have seen recent mathematical 
and computational development \cite{chorin2000optimal,hijon2010mori}, they have 
not been fully justified from the point of view of rigorous statistical mechanics. 
In fact, important questions exist regarding the ultimate limits of validity 
of the fluctuating hydrodynamic equations, since those equations often work quite well 
for micro- and nano-scale fluid systems without a clear separation of scales. 
This suggests that the existing rigorous framework of hydrodynamic scaling limits 
\cite{quastel1998lattice,spohn2012large} may be too restrictive. The resulting 
stochastic hydrodynamic equations present also some very challenging questions 
regarding their mathematical formulation. As the existing formal microscopic 
derivations \cite{zubarev1983statistical,morozov1984langevin,espanol2009microscopic}
make clear, the fluctuating hydrodynamics equations are {\it not} stochastic PDE's 
because they contain an explicit UV cutoff $\Lambda.$ Current mathematical theory 
(e.g. \cite{albeverio2008spde}) suffices to show that such cut-off models specify 
a well-posed dynamics for each finite value of $\Lambda,$ but the fundamental 
issue remains to be addressed that physical predictions should be $\Lambda$-independent. 
The reflexive response might be to attempt to show that a well-defined SPDE exists 
in the limit $\Lambda\to\infty,$ with a suitable choice of ``bare'' parameters, 
so that finite-$\Lambda$ models can be regarded as ``approximations'' to this idealized 
continuum limit. Based on our arguments in section \ref{formulation}, this point 
of view seems rather unphysical. It seems to us that a more natural goal is to establish some 
exact ``renormalization group invariance'' of the finite-$\Lambda$ effective field-theories 
which expresses invariance of their predictions to changes of UV cut-off $\Lambda$ and of other 
arbitrary features of the models, such as the numerical discretization.

\subsection{\red{Future Directions}}\lb{future} 

In this paper we have focused on the dissipation range of \black{fully-developed,} 
homogeneous turbulence, but there should be influences of thermal noise also on 
other turbulent \black{flows and processes. Several of these novel effects and 
new directions of research were suggested already by Betchov. For example, 
he argued that thermal noise could play an important role in triggering transition 
to turbulence (\cite{betchov1961thermal}, sections II.E-G), a possibility currently 
being actively explored \cite{luchini2017receptivity,fedorov2017receptivity}, and
that thermal noise could generate unpredictability in fully developed turbulence (\cite{betchov1961thermal},section II.H), anticipating modern 
ideas on spontaneous stochasticity \cite{mailybaev2016spontaneous,mailybaev2016spontaneously}. 
In \cite{betchov1964measure}, Betchov suggested information-theoretic approaches 
as a possible means to distinguish Gaussian thermal fluctuations from turbulent fluctuations, 
in line with recent research \cite{vladimirova2021second,shavit2020singular}. 
Betchov recognized as well that} thermal noise effects must occur not only in 
incompressible fluid turbulence but also in other turbulent systems, 
\black{e.g. in magnetized plasmas, where he extended magnetohydrodynamic equations to include 
stochastic electric fields from thermal fluctuations (\cite{betchov1961thermal}, Section III).} 
Such molecular noise characteristics have recently been derived for homogeneous plasma kinetics 
even at the level of large deviations \cite{feliachi2021dynamical} and at the linear level 
in the collisional two-fluid regime of magnetized plasmas \cite{krommes2018projectionA,krommes2018projectionB}. 
We expect that there will be signatures of thermal noise in astrophysical 
plasma turbulence at high magnetic Prandtl numbers, such as the partially 
ionized interstellar medium \cite{xu2017magnetohydrodynamic}. \black{In fact,  
thermal noise must be expected to act upon any} fluid modes which 
are strongly affected by \black{molecular} dissipation. \black{Another example is} 
the viscous sublayer eddies of wall-bounded turbulence, \black{where} additional fluctuating 
forces \black{are predicted to} appear in the vicinity of solid walls associated with 
dissipative, microscopic slip coefficients \cite{satten1984fluctuations,schmitz1988fluctuations}. 

\black{In consequence of these many important directions of research,} 
this paper which focuses on the dissipation range is \black{just} the 
first in a series to study the influence of thermal noise on turbulent flows. 
In \cite{bandak2020spontaneous} we shall discuss its more subtle 
effects on the turbulent inertial-range in the limit $Re\gg 1,$ in particular the role 
of thermal noise in triggering Eulerian spontaneous stochasticity \cite{mailybaev2016spontaneous,mailybaev2016spontaneously}. 
We plan to make also a parametric study of the Reynolds-number dependence of the 
dissipation-range intermittency discussed in this work, in order to explore 
possible limitations to the hydrodynamic description of small-scale fluid turbulence.
\red{In a work in preparation \cite{eyink2021high}, we study high Schmidt-number turbulent 
mixing and we show that the exponentially decaying scalar 
spectrum theoretically predicted for the viscous-diffusive range \cite{kraichnan1974convection} 
and verified numerically by deterministic Navier-Stokes simulations \cite{yeung2004simulations,clay2017strained} 
is erased by thermal noise and replaced by a $k^{-2}$ power-law spectrum 
associated to giant concentration fluctuations \cite{vailati1997giant,vailati2011fractal}.
Similar effects should be present also in the high magnetic Prandtl-number kinematic dynamo.} 
Because of the universality of the fluctuation-dissipation relation, thermal noise 
is inextricably linked to dissipation and the two effects must always appear together. 

\vspace{20pt}
{\bf Acknowledgements}: We acknowledge the Simons Foundation for support of this work through 
Targeted Grant in MPS-663054 at JHU and MPS-662985 at UIUC, ``Revisiting the Turbulence Problem 
Using Statistical Mechanics.'' We wish to thank also many colleagues for fruitful discussions of the science, 
especially Frank Alexander, John Bell, Freddy Bouchet, L\'{e}onie Canet, B\'{e}reng\`{e}re Dubrulle, Alej Garcia, Rui Ni and Tamer Zaki. 

\setcounter{section}{0} 
\renewcommand\thesection{\Alph{section}}
\renewcommand\thesubsection{\arabic{subsection}}

\section*{Appendices}

\section{Proof of the Fluctuation-Dissipation Relations} \lb{appA} 

We here derive the fluctuation-dissipation relation \eqref{FDR} for the 
incompressible Navier-Stokes equation and the corresponding result 
for the noisy Sabra model \eqref{nSabra}. The former result is well-known
\cite{dezarate2006hydrodynamic,schmitz1988fluctuations,forster1976long,forster1977large,
zubarev1983statistical,morozov1984langevin} but we give the proof here for 
completeness and, also, to stress the close parallels to the same result for 
the Sabra model. \red{A good general reference is \cite{ramshaw1986augmented}.} 

\subsection{FDR for Navier-Stokes}

We begin with the truncated fluctuating Navier-Stokes system \eqref{FNS-Lambda} 
which can be written as a system of SDE's for the Fourier modes 
\be \hat{\bu}_\bk=\int_\Omega d^3x \ e^{-i\bk\bdot\bx} \bu(\bx) \ee
of the velocity field satisfying $|\bk|<\Lambda.$ Because of the reality condition
\be \hat{\bu}_\bk^*=\hat{\bu}_{-\bk}\ee 
not all of these modes are independent. We take the modes whose wavevector lies 
in the half-set 
\be K^+= \left\{\bk:\ \begin{array}{ll}
                  k_x>0, & \mbox{or} \cr
                  k_y>0 & \mbox{if $k_x=0$, or}\cr 
                  k_z\geq 0 & \mbox{if $k_x=k_y=0$} 
\end{array}\right\} \ee 
as the independent complex modes. The fluctuating Navier-Stokes equation can then 
be written as  
\begin{eqnarray}
&& \partial_t\hat{u}_{\bk,m} + ik_n\left(\delta_{mp}-\frac{k_m k_p}{k^2}\right) 
 \sum_{\bp+\bq=\bk} \hat{u}_{\bp,n}\hat{u}_{\bq,p}  \cr
&& \hspace{60pt} +\nu k^2\hat{u}_{\bk,m} = 
 \hat{q}_{\bk,m} \lb{FNS-k} \end{eqnarray}
for $\bk\in K^+$ with $|\bk|<\Lambda,$ and where $\hat{\bq}_\bk$ is a 
suitable random force, further specified below, that represents the thermal noise. 
Here we note that the wavenumbers $\bp,$ $\bq$ which are summed over in the 
expression 
\be B_{\bk,m}(\hat{\bu},\hat{\bu}^*)= -i k_n\left(\delta_{mp}-\frac{k_m k_p}{k^2}\right) 
 \sum_{\bp+\bq=\bk} \hat{u}_{\bp,n}\hat{u}_{\bq,p} \lb{B-def} \ee 
may lie in the complementary set $K^-= -K^+$ and in the case that $\bp\in K^-,$ 
then $\hat{\bu}_\bp$ should be interpreted instead as $\hat{\bu}^*_{-\bp}.$
There is a corresponding equation of motion for the complex-conjugate variables
\be \partial_t\hat{u}_{\bk,m}^* = B_{\bk,m}^*[\hat{\bu},\hat{\bu}^*] -\nu k^2\hat{u}_{\bk,m}^* +
 \hat{{\rm q}}_{\bk,m}^* \ee 
with $B_{\bk,m}^*[\hat{\bu},\hat{\bu}^*] :=B_{\bk,m}[\hat{\bu},\hat{\bu}^*]^*$ when $\bk\in K^+$ and 
$|\bk|<\Lambda.$ Finally, we note that the random force $\hat{\bq}_\bk(t)$ is 
Gaussian with mean zero and covariance 
\be \langle \hat{{\rm q}}_{\bk,m}(t)\hat{{\rm q}}_{\bk',n}^*(t')\rangle
=\frac{2\nu k_BT}{\rho} V\delta_{\bk,\bk'}\delta(t-t') 
\left(k^2\delta_{mn}-k_mk_n\right) \lb{FDR-k} \ee 
with $\hat{\bq}^*_\bk=\hat{\bq}_{-\bk}$ and $\bk\bdot\hat{\bq}_\bk=0.$ To see
the equivalence with \eqref{FDR}, we note $\grad\bdot \tilde{\btau}$ in 
\eqref{FNS} can be written as 
\be \grad\bdot \tilde{\btau} = \tilde{\bq}+ \grad \tilde{\pi} \ee
with the random scalar field $\tilde{\pi}$ chosen so that $\grad\bdot\tilde{\bq}=0,$ 
and in that case $\tilde{\pi}$ can be absorbed into the pressure and 
$\tilde{\bq}$ is Gaussian with covariance 
%\newpage 
\begin{eqnarray} 
&&\langle \tilde{{\rm q}}_m(\bx,t) \tilde{{\rm q}}_n(\bx',t') \rangle \cr
&& =\frac{2\nu k_BT}{\rho}
\left(-\triangle_x\delta_{m,n}+\nabla_{x,m}\nabla_{x,n}\right)\delta^3(\bx-\bx')\delta(t-t') \cr
&& 
\end{eqnarray} 
The Fourier coefficient $\hat{\bq}_\bk(t)$ of $\tilde{\bq}(\bx,t)$ then satisfies \eqref{FDR-k}. 

We now wish to show that the noise covariance \eqref{FDR-k} is correctly 
chosen so that the Gibbs measure 
\be 
P_G[\hat{\bu},\hat{\bu}^*] =\frac{1}{Z} \exp\left[-{\mathcal E}/k_BT\right] \ee 
for kinetic energy 
\begin{eqnarray}
{\mathcal E}&=&\int_\Omega d^3x\ \frac{\rho}{2}|\bu(\bx)|^2\cr 
&=& \frac{\rho}{2V}\sum_{|\bk|<\Lambda} |\hat{\bu}_\bk|^2  
\ =\  \frac{\rho}{V}
\sum_{\bk\in K^+,|\bk|<\Lambda} |\hat{\bu}_\bk|^2 \cr
&& 
\end{eqnarray} 
is invariant under the stochastic dynamics \eqref{FNS-k} with no large-scale 
forcing. The proof is based
on the Fokker-Planck equation for the probability distribution $P[\hat{\bu},\hat{\bu}^*]$
of the Fourier modes. We employ the standard device of treating 
$\hat{\bu}_\bk$ and its complex conjugate $\hat{\bu}_\bk^*$ as independent variables in complex differential 
calculus (Wirtinger derivatives). The Fokker-Planck equation equivalent to the 
coupled Langevin equations (\ref{FNS-k}) is easily checked to be 
%\newpage 
\begin{eqnarray}
\partial_t P&=& \sum_{\bk\in K^+}\Bigg[ -\frac{\partial}{\partial \hat{\bu}_\bk} \bdot\Big(\bV_\bk[\hat{\bu},\hat{\bu}^*]P\Big) 
-\frac{\partial}{\partial \hat{\bu}_\bk^*}\bdot \Big(\bV^*_\bk[\hat{\bu},\hat{\bu}^*]P\Big) \cr 
&& \hspace{2pt} +\frac{2\nu k_BT}{\rho}V (k^2{\bf I}-\bk\bk): 
\frac{\partial^2}{\partial \hat{\bu}_\bk\partial \hat{\bu}_\bk^*}P\Bigg]
:={\mathcal L}^*P, \cr
&& 
\lb{FP-eq} \end{eqnarray}
where 
\be \bV_\bk[\hat{\bu},\hat{\bu}^*] = \bB_\bk[\hat{\bu},\hat{\bu}^*] - \nu k^2 \hat{\bu}_\bk \ee
and $\bV_\bk^*[\hat{\bu},\hat{\bu}^*] := \bV_\bk[\hat{\bu},\hat{\bu}^*]^*$ are the components
of the drift velocity in the phase-space. 

The proof further uses two fundamental properties 
of the truncated Euler dynamics: the 
{\it Liouville theorem} on conservation of phase-volume:
\be  \sum_{\bk\in K^+} 
\left( \frac{\partial}{\partial \hat{\bu}_\bk}\bdot \bB_\bk[\hat{\bu},\hat{\bu}^*] 
+ \frac{\partial}{\partial \hat{\bu}_\bk^*}\bdot \bB_\bk^*[\hat{\bu},\hat{\bu}^*] \right) =0 
\lb{liouville} \ee
and the {\it conservation of kinetic energy}:
\be  \sum_{\bk\in K^+} \left( \bB_\bk[\hat{\bu},\hat{\bu}^*]\bdot\frac{\partial {\mathcal E}}{\partial \hat{\bu}_\bk}
+ \bB_\bk^*[\hat{\bu},\hat{\bu}^*]\bdot\frac{\partial {\mathcal E}}{\partial \hat{\bu}_\bk^*}\right) =0. 
\lb{Econserve} \ee 
The Liouville theorem for truncated Euler has been well-known since the work of 
Lee \cite{lee1952some} and, in fact, follows easily from definition \eqref{B-def} by the observation that
\be \frac{\partial}{\partial \hat{\bu}_\bk}\bdot \bB_\bk[\hat{\bu},\hat{\bu}^*]=-2i\bk\bdot\hat{\bu}(\bzed), 
\quad \frac{\partial}{\partial \hat{\bu}_\bk^*}\bdot \bB_\bk^*[\hat{\bu},\hat{\bu}^*]=2i\bk\bdot\hat{\bu}(\bzed)\ee 
The conservation of kinetic energy \eqref{Econserve} follows from the detailed conservation for wavevector triads first 
noted by Onsager \cite{onsager1949statistical}. The consequence of (\ref{liouville}),(\ref{Econserve}) is that 
\begin{eqnarray}
&& \sum_{\bk\in K^+}
\left[ -\frac{\partial}{\partial \hat{\bu}_\bk}\bdot \Big(\bB_\bk[\hat{\bu},\hat{\bu}^*]P_G\Big) 
 -\frac{\partial}{\partial \hat{\bu}^*_\bk} \bdot \Big(\bB_\bk^*[\hat{\bu},\hat{\bu}^*]P_G\Big)\right]=0, \cr
&& \end{eqnarray} 
so that $P_G[\hat{\bu},\hat{\bu}^*]$ is an invariant measure for the truncated Euler system. 

The full stationarity condition $\partial_t P_G=0$ therefore simplifies to 
\begin{eqnarray} 
&&\sum_{\bk\in K^+} \nu k^2 \left[ \frac{\partial}{\partial \hat{\bu}_\bk}\bdot \Big(\hat{\bu}_\bk P_G\Big) 
+\frac{\partial}{\partial \hat{\bu}_\bk^*} \bdot\Big(\hat{\bu}_\bk^* P_G\Big) 
\right] \cr 
&& \hspace{10pt}=
-\frac{2\nu k_BT}{\rho}V \sum_{\bk\in K^+} (k^2{\bf I}-\bk\bk): 
\frac{\partial^2}{\partial \hat{\bu}_\bk\partial \hat{\bu}_\bk^*}P_G \cr
&& 
\lb{stat-cond} \end{eqnarray}
It is worth observing that this is exactly the stationarity condition 
for the Gibbs measure under the linear Ornstein-Uhlenbeck dynamics 
\be
\partial_t\hat{\bu}_{\bk} = -\nu k^2\hat{\bu}_{\bk} + \hat{\bq}_{\bk}, \quad 
\partial_t\hat{\bu}_{\bk}^* = -\nu k^2\hat{\bu}_{\bk}^* + \hat{\bq}_{\bk}^*
\ee
corresponding to the fluctuating Stokes equation. 
On the other hand, the elementary derivatives 
\begin{eqnarray}  
&& \frac{\partial P_G}{\partial \hat{\bu}_\bk} = -\frac{\rho \hat{\bu}_\bk^*}{Vk_BT}P_G, \quad 
\frac{\partial P_G}{\partial \hat{\bu}^*_\bk} = -\frac{\rho \hat{\bu}_\bk}{Vk_BT}P_G, \cr
&& \frac{\partial^2 P_G}{\partial \hat{\bu}_\bk\partial \hat{\bu}_\bk^*} = -\left[\frac{\rho }{Vk_BT}({\bI-\hat{\bk}\hat{\bk})}-\frac{\rho^2\hat{\bu}_\bk^*\hat{\bu}_\bk}{(Vk_BT)^2}\right]P_G \cr
&& 
\end{eqnarray}   
imply that both sides of (\ref{stat-cond}) are indeed equal to the same quantity 
\be \sum_{\bk\in K^+} \nu k^2 \left(4-\frac{2\rho|\hat{\bu}_\bk|^2}{Vk_BT}\right) P_G. \ee 
Thus, $P_G$ is an invariant measure, as claimed. Because of the non-degeneracy of the noise 
and the boundedness of the drift, this is in fact the unique invariant measure.  

It is furthermore easy to show by the same 
arguments that the backward Fokker-Planck operator 
\begin{eqnarray}
{\mathcal L} F&=& \sum_{\bk\in K^+}\Bigg[  \bV_\bk[\hat{\bu},\hat{\bu}^*]
\bdot \frac{\partial F}{\partial \hat{\bu}_\bk}
+\bV^*_\bk[\hat{\bu},\hat{\bu}^*] \frac{\partial F}{\partial \hat{\bu}_\bk^*}\bdot \cr 
&& \hspace{20pt} +\frac{2\nu k_BT}{\rho}V (k^2{\bf I}-\bk\bk): 
\frac{\partial^2}{\partial \hat{\bu}_\bk\partial \hat{\bu}_\bk^*}F\Bigg]
 \cr
&& 
\lb{BFP-op} \end{eqnarray}
for any real functions $F[\hat{\bu},\hat{\bu}^*],$ $G[\hat{\bu},\hat{\bu}^*]$ satisfies the following 
adjoint property with respect to the equilibrium Gibbs measure $P_G:$
\be \int G({\mathcal L}F)\,P_G \, d\hat{\bu}\,d\hat{\bu}^*= \int (\tilde{{\mathcal L}}G)F \, P_G \, d\hat{\bu}\,d\hat{\bu}^* 
\lb{adjoint} \ee 
where $d\hat{\bu}\,d\hat{\bu}^*=\prod_{\bk\in K^+} d\hat{\bu}_\bk\,d\hat{\bu}^*_\bk$ and where 
\begin{eqnarray}
\tilde{{\mathcal L}} F&=& \sum_{\bk\in K^+}\Bigg[  \tilde{\bV}_\bk[\hat{\bu},\hat{\bu}^*]
\bdot \frac{\partial F}{\partial \hat{\bu}_\bk}
+\tilde{\bV}^*_\bk[\hat{\bu},\hat{\bu}^*] \frac{\partial F}{\partial \hat{\bu}_\bk^*}\bdot \cr 
&& \hspace{20pt} +\frac{2\nu k_BT}{\rho}V (k^2{\bf I}-\bk\bk): 
\frac{\partial^2}{\partial \hat{\bu}_\bk\partial \hat{\bu}_\bk^*}F\Bigg]
 \cr
&& 
\lb{rev-BFP-op} \end{eqnarray}
with 
\be \tilde{\bV}_\bk[\hat{\bu},\hat{\bu}^*] = -\bB_\bk[\hat{\bu},\hat{\bu}^*] - \nu k^2 \hat{\bu}_\bk \ee
is the time-reversal $\tilde{\mathcal L}$ of the operator $\mathcal L$ under $\bu\to -\bu.$
The adjoint property \eqref{adjoint} is equivalent to the {\it detailed balance condition} 
\be P[\bu,t|\bu_0,0]P_G[\bu_0]=P[-\bu_0,t|-\bu,0]P_G[-\bu] \ee 
for the transition probability densities of the nonlinear diffusion \eqref{FP-eq}. Thus,  
the fluctuating Navier-Stokes dynamics is time-reversible in thermal equilibrium. 

Finally, we derive for the Gibbs distribution $P_G$ the spectrum of the mean energy per unit mass
\be E=\frac{\langle{\mathcal E}\rangle}{\rho V}=
\frac{1}{2V^2}\sum_{|\bk|<\Lambda} \langle |\hat{\bu}_\bk|^2 \rangle \ee 
which is conventionally considered for turbulence of an incompressible fluid. 
The variance of the Gaussian measure is given by 
\be \langle |\hat{\bu}_\bk|^2 \rangle = \frac{2Vk_BT}{\rho}, \ee 
which expresses energy equipartition, taking into account the two ``spin'' 
degrees of freedom for the solenoidal Fourier modes. Then taking the infinite-volume limit
\be \frac{1}{V}\sum_\bk \to \frac{1}{(2\pi)^3}\int d^3k \quad\mbox{for $V\to\infty$} \ee 
and we see that $E=\int_0^\Lambda dk\ E(k)$ with 
\be E(k)\sim \frac{k_BT}{\rho} \frac{4\pi k^2}{(2\pi)^3}.\ee
This is the $k^2$ equipartition spectrum first derived by Lee \cite{lee1952some} 
and Hopf \cite{hopf1952statistical} for truncated Euler dynamics. 

\subsection{FDR for Sabra Model} 

Corresponding results for the Sabra shell model are obtained by identical 
arguments. The Fokker-Planck equation for the probability distribution 
$P[u,u^*]$ of the vector $u=(u_0,u_1,...,u_N)$ of shell-variables is easily 
derived from the coupled Langevin equations (\ref{nSabra}) as: 
%\newpage 
\begin{eqnarray}
\partial_t P&=& \sum_{n=0}^N \Bigg[ -\frac{\partial}{\partial u_n} \Big(V_n[u,u^*]P\Big) 
-\frac{\partial}{\partial u_n^*} \Big(V^*_n[u,u^*]P\Big) \cr 
&& \hspace{65pt} +\frac{4\nu k_BT}{\varrho} k_n^2 \frac{\partial^2}{\partial u_n\partial u_n^*}P\Bigg]
:={\mathcal L}^*P, \cr
&& 
\lb{FPs-eq} \end{eqnarray}
where (now writing $B_n[u]$ as $B_n[u,u^*]$) 
\be V_n[u,u^*] = B_n[u,u^*] - \nu k_n^2 u_n \ee
and  $V_n^*[u,u^*] = V_n[u,u^*]^*.$
The inviscid Sabra model dynamics also satisfies a Liouville theorem
\be  \sum_{n=0}^N \left( \frac{\partial}{\partial u_n}B_n[u,u^*] + \frac{\partial}{\partial u_n^*} B^*_n[u,u^*]\right) =0, 
\lb{liouvilleS} \ee
which is direct by inspection of eq.\eqref{BS-def}, and conserves kinetic energy:
\be  \sum_{n=0}^N \left( B_n[u,u^*] \frac{\partial {\mathcal E}}{\partial u_n}
+ B^*_n[u,u^*]\frac{\partial {\mathcal E}}{\partial u_n^*} \right) =0 
\lb{EconserveS} \ee 
Consequently, the stationarity condition $\partial_t P_G=0$ for the 
measure $P_G[u,u^*] $ defined in \eqref{GibbsS} simplifies to
\begin{eqnarray} 
&&\sum_{n=0}^N \nu k_n^2 \left[ \frac{\partial}{\partial u_n} \Big(u_n P_G\Big) 
+\frac{\partial}{\partial u_n^*} \Big( u^*_n P_G\Big)\right] \cr 
&& \hspace{10pt}= 
-\left(\frac{4\nu k_BT}{\varrho}\right) \sum_{n=0}^N k_n^2 \frac{\partial^2}{\partial u_n\partial u_n^*} P_G 
\lb{stat-cond-S} 
\end{eqnarray}
On the other hand, the elementary derivatives 
\begin{eqnarray}  
&& \frac{\partial P_G}{\partial u_n} = -\frac{\varrho u_n^*}{2k_BT}P_G, \quad 
\frac{\partial P_G}{\partial u^*_n} = -\frac{\varrho u_n}{2k_BT}P_G, \cr
&& \frac{\partial^2 P_G}{\partial u_n\partial u_n^*} = -\left(\frac{\varrho }{2k_BT}-\frac{\varrho^2|u_n|^2}{(2k_BT)^2}\right)P_G 
\end{eqnarray}   
imply that both sides of \eqref{stat-cond-S} are equal to the same quantity 
\be \sum_{n=0}^N \nu k_n^2 \left(2-\frac{\varrho|u_n|^2}{k_BT }\right) P_G. \ee 
Thus, the Gibbs distribution $P_G$ is the unique invariant measure of the noisy Sabra model dynamics 
\eqref{nSabra} when the external driving force is set to zero, $f_n=0.$ Furthermore, the 
analogue of the adjoint property \eqref{adjoint} for fluctuating Navier-Stokes holds also 
for the noisy shell model, so that the dynamics is time-reversible under the transformation 
$u_n\to -u_n$ in thermal equilibrium. 

\section{Derivation of The Slaved Taylor-Ito\hspace{-8pt}$\bar{{\rm O}}$ Scheme}\lb{derive-slaved} 

For completeness we sketch here the derivation of the Taylor-It$\bar{{\rm o}}$ scheme 
from \cite{lord2004numerical} for our noisy Sabra model \eqref{nSabra}, written as  
\be 
du_n = a_n dt +b_n dW_n
\ee 
with
\be 
a_n = B_n[u]-\nu k_n^2u_n +f_n, \quad b_n=\left(\frac{2k_BT}{\varrho}\right)^{1/2}k_n
\ee 
Explicit integration of the linear viscous term gives 
\bea 
u_n(t_{k+1})&=& e^{-\nu k_n^2 \Delta t}\Big[u_n(t_k)+\int_{t_k}^{t_{k+1}} c_n(t,u(t))\ dt \cr
&& \hspace{40pt} +\int_{t_k}^{t_{k+1}} d_n(t) \ dW_n(t) \Big]\eea 
with 
\be c_n(t,u(t))=e^{\nu k_n^2 (t-t_k)}\big(B_n[u(t)]+f_n(t)\big) \lb{cn-def} \ee 
and 
\be d_n(t)= e^{\nu k_n^2 (t-t_k)} b_n. \ee 

Taylor-expanding $d_n(t)$ as 
\be d_n(t)= e^{\nu k_n^2 (t-t_k)} b_n= \left[1+\nu k_n^2 (t-t_k) +O((t-t_k)^2)\right] b_n \ee
and then substituting into the time-integral yields 
\bea
&& \int_{t_k}^{t_{k+1}} d_n(t) \ dW_n(t) \cr 
&& = b_n \Delta W_n(t_k) +\nu k_n^2 b_n\int_{t_k}^{t_{k+1}} (t-t_k)\ dW(t) + O((\Delta t)^{5/2}) \cr 
&& = b_n \Delta W_n(t_k) +\nu k_n^2 b_n \big(\Delta t\Delta W_n(t_k)-\Delta Z_n(t_k)\big) \cr
&& \hspace{150pt}  +  O((\Delta t)^{5/2}) \eea
by an integration by parts.

To similarly expand $c_n(t,u(t))$ we must use the It$\bar{{\rm o}}$ formula 
\bea 
&& c_n(t,u(t)) = c_n(t_k,u(t_k)) + \int_{t_k}^t ds\ (\partial_s +L) c_n(s,u(s)) \cr
&& + \sum_m \int_{t_k}^t ds\Big[b_m dW_m(s) \frac{\partial c_n}{\partial u_m}(s,u(s))+\cr
&& \hspace{120pt} + b_m dW_m^*(s) \frac{\partial c_n}{\partial u_m^*}(s,u(s))\Big] \cr
&& 
\eea 
where $L$ is the forward Kolmogorov operator 
\be L=\sum_m \left(a_m \frac{\partial}{\partial u_m}+
a_m^* \frac{\partial }{\partial u_m^*} +2b_m^2 \frac{\partial^2 }{\partial u_m\partial u_m^*}\right). \ee 
This implies that 
\bea 
&& \int_{t_k}^{t_{k+1}} dt\ c_n(t,u(t))\ dt 
= c_n(t_k,u(t_k)) \Delta t \cr
&& + \int_{t_k}^{t_{k+1}} dt \int_{t_k}^t ds\ (\partial_s +L) c_n(s,u(s)) \cr
&& + \sum_m \int_{t_k}^{t_{k+1}} d \int_{t_k}^t ds\ b_m\Big[dW_m(s) \frac{\partial c_n}{\partial u_m}(s,u(s)) \cr
&& \hspace{120pt} + dW_m^*(s) \frac{\partial c_n}{\partial u_m^*}(s,u(s))\Big]. \cr
&&
\eea
Substitution of the  It$\bar{{\rm o}}$ formulae for $(\partial_s +L) c_n(s,u(s))$
and $\frac{\partial c_n}{\partial u_m}(s,u(s))$ then gives the
It$\bar{{\rm o}}$-Taylor series approximation to the desired order 
\bea 
&& \int_{t_k}^{t_{k+1}} dt\ c_n(t,u(t))\ dt 
= c_n(t_k,u(t_k)) \Delta t \cr
&& + \int_{t_k}^{t_{k+1}} dt \int_{t_k}^t ds\ (\partial_t +L) c_n(t_k,u(t_k)) \cr
&& + \sum_m \int_{t_k}^{t_{k+1}} dt \int_{t_k}^t ds\ b_m\Big[dW_m(s) \frac{\partial c_n}{\partial u_m}(t_k,u(t_k)) \cr
&& \hspace{100pt} + dW_m^*(s) \frac{\partial c_n}{\partial u_m^*}(t_k,u(t_k))\Big] +R, \cr
&& 
\eea
where $R$ is a stochastic remainder term. Straightforward calculations using 
\eqref{cn-def} then give
\bea 
&& \int_{t_k}^{t_{k+1}} dt\ c_n(t,u(t)) = \Delta t [B_n(t_k,u(t_k))+f_n(t_k)] \cr
&&  +\frac{1}{2} (\Delta t)^2\left(\nu k_n^2[B_n(t_k,u(t_k))+f_n(t_k)]+\dot{f}_n(t_k)\right) \cr
&& + \frac{1}{2}(\Delta t)^2 \sum_m \Big[a_m \frac{\partial B_n}{\partial u_m}(t_k,u(t_k))+
               a_m^* \frac{\partial B_n}{\partial u_m^*}(t_k,u(t_k))\cr
&&   \hspace{128pt}  +2b_m^2 \frac{\partial^2 B_n}{\partial u_m\partial u_m^*}(t_k,u(t_k))\Big] \cr                
&& + \sum_m b_m\Big[\Delta Z_m(t_k) \frac{\partial B_n}{\partial u_m}(t_k,u(t_k)) \cr
&& \hspace{100pt} +\Delta Z_m^*(t_k) \frac{\partial B_n}{\partial u_m^*}(t_k,u(t_k))\Big] +R  \cr
&& 
\eea

Putting it all together gives the integration scheme 
\bea 
&& u_n(t_{k+1})= e^{-\nu k_n^2 \Delta t}\Bigg\{ 
u_n(t_k) +\Delta t[B_n(t_k,u(t_k))+f_n(t_k)]\cr
&& +\frac{1}{2} (\Delta t)^2 \left(\nu k_n^2[B_n(t_k,u(t_k))+f_n(t_k)]+\dot{f}_n(t_k)\right)\cr
&& + \frac{1}{2}(\Delta t)^2 \sum_m \Big[a_m \frac{\partial B_n}{\partial u_m}(t_k,u(t_k))+
               a_m^* \frac{\partial B_n}{\partial u_m^*}(t_k,u(t_k))\cr
&&   \hspace{128pt}  +2b_m^2 \frac{\partial^2 B_n}{\partial u_m\partial u_m^*}(t_k,u(t_k))\Big] \cr                
&& + \sum_m b_m\Big[\Delta Z_m(t_k) \frac{\partial B_n}{\partial u_m}(t_k,u(t_k)) \cr 
&& \hspace{120pt} +\Delta Z_m^*(t_k) \frac{\partial B_n}{\partial u_m^*}(t_k,u(t_k))\Big]  \cr
&& + b_n \left[(1 +\nu k_n^2 \Delta t)\Delta W_n(t_k)-\Delta Z_n(t_k)\right]\Bigg\} \cr
&& \lb{LR04} \eea
This result may be compared with equation (6.1) of \cite{lord2004numerical}. 

For the Sabra shell model, the only non-vanishing first-derivatives are 
\bea
\frac{\partial B_n}{\partial u_{n-2}} = \frac{1}{2}ik_{n-1}u_{n-1},&& \quad \frac{\partial B_n}{\partial u_{n+2}} 
= ik_{n+1}u^*_{n+1} \cr 
\frac{\partial B_n}{\partial u_{n-1}}  = \frac{1}{2}ik_{n-1}u_{n-2},&&\quad \frac{\partial B_n}{\partial u_{n+1}} = -\frac{1}{2}ik_n u^*_{n-1} \cr
\frac{\partial B_n}{\partial u^*_{n-1}}  = -\frac{1}{2}ik_n u_{n+1},&&\quad \frac{\partial B_n}{\partial u^*_{n+1}}  = ik_{n+1}u_{n+2} \cr 
&&
\eea
while for all $m$
\be \frac{\partial^2 B_n}{\partial u_m\partial u_m^*}=0. \ee
Substituting these results into \eqref{LR04} yields our numerical scheme \eqref{sabra-num}
for the noisy Sabra model. 

\section{Convergence of Steady-State Averages}\lb{converge-tests} 

Convergence of steady-state averages for our numerical study required a sufficiently large 
averaging time $T$ and proper resolution of the dynamics required a sufficiently large truncation 
wavenumber $N$ and sufficiently small time-step $\Delta t$. Here we describe the tests we have 
made that our choices of those parameters were sufficient. 

In the section \ref{N-independ} we already addressed at length the convergence of transition probabilities 
with respect to the high-wavenumber shell truncation $N$. Such convergence of transition probabilities 
implies convergence of the averages with respect to truncation wavenumber. The time-step was 
chosen as $\Delta t=10^{-5}$ so as to be smaller than the viscous time $t_{visc}=1/\nu k_N^2$ for the 
highest shellnumber $N.$ Setting the external forcing to zero, $f_n=0$ for all shells $n,$ guarantees
the gaussian Gibbs distribution \eqref{GibbsS} and we checked that the time-step $\Delta t=10^{-5}$ 
sufficed to reproduce that distribution to excellent accuracy for all shells, whereas reducing the step size 
to $\Delta t=1\times 10^{-4}\sim 3\times 10^{-4}$ introduced errors for $n$ near $N.$ In the forced 
simulation with $f_n$ chosen as in \eqref{stirring}, it was likewise found that the same same choice 
$\Delta t=10^{-5}$ sufficed to produce Gaussian thermal equilibrium very accurately at the highest 
two shells (see Fig.~\ref{energy-pdf}) and further produced accurately the known stretched-exponential 
decay in the deterministic model run (see insets in Fig.~\ref{strucfun:fig}). These consistency 
checks confirmed that our time-step $\Delta t=10^{-5}$ was sufficiently small.  

\textcolor{black}{
The total averaging time $T$ was taken to be 300 large eddy turnover times, based on convergence 
tests of the statistical averages presented in section \ref{num:sec}. Dividing the total time into ten 
subintervals of times $T/10$ and calculating averages separately over each led to negligible changes, which suggested that we had acceptable convergence for $-0.9<p<6. $ This was confirmed by the good agreement of our scaling exponents with the very accurate 
values obtained in \cite{lvov998improved}, as shown in Fig. \ref{sigexp:fig}.} 
Our work was not focused on high-precision of inertial-range scaling exponents, so that this was 
acceptable accuracy for our purposes. Crucially, the profound differences that we observed in the dissipation 
range between the statistics of the deterministic Sabra model \eqref{dSabra} and of the noisy Sabra model 
\eqref{nSabra} lie very far outside all error bars on the numerical calculations.

\bibliography{bibliography.bib}
 
\end{document}